\documentclass[11pt]{article}
\usepackage{palatino}
\usepackage{braket}
\usepackage{bm}
\usepackage{graphicx}
\usepackage{tikz}
\usepackage{marvosym}
\usepackage{ifthen}
  \usepackage{tikzsymbols}
\usetikzlibrary{patterns}
\usetikzlibrary{calc}
\usetikzlibrary{matrix}
\usetikzlibrary{tikzmark}
\usetikzlibrary{fit}
\usetikzlibrary{shadows.blur}
\usetikzlibrary{shapes.symbols}
\usetikzlibrary{shapes.geometric}
\usepackage[normalem]{ulem}

\usepackage{hyperref}

\usepackage{algorithm}
\usepackage[noend]{algpseudocode}

\usepackage{mdframed}

\usetikzlibrary{positioning}
\usepackage{cancel}
\usepackage{amssymb, amsmath}
\usepackage{amsthm}
\usepackage{thmtools}
\usepackage{epstopdf}
\newtheorem{lemma}{Lemma}
\newtheorem{theorem}{Theorem}
\newtheorem{corollary}{Corollary}
\newtheorem{remark}{Remark}
\declaretheorem[style=definition,numbered=yes]{definition}

\let\oldsqcap\sqcap
\renewcommand{\sqcap}{\scalebox{1}[1.25]{$\oldsqcap$}}

\newcommand\blfootnote[1]{%
  \begingroup
  \renewcommand\thefootnote{}\footnote{#1}%
  \addtocounter{footnote}{-1}%
  \endgroup
}

\begin{document}

\title{
On the Capacity of Distributed Quantum Storage}
\author{Hua Sun, Syed A. Jafar}
\date{}                                           
\maketitle

\blfootnote{Hua Sun (email: hua.sun@unt.edu) is with the Department of Electrical Engineering at the University of North Texas. Syed A. Jafar (email: syed@uci.edu) is with the  Department of Electrical Engineering and Computer Science (EECS) at the University of California Irvine. }

\begin{abstract}
A distributed quantum storage code maps a quantum message to $N$ storage nodes, of arbitrary specified sizes, such that the stored message is robust to an arbitrary specified set of erasure patterns. The sizes of the storage nodes, and erasure patterns may not be homogeneous. The capacity of distributed quantum storage is the maximum feasible size of the quantum message (relative to the sizes of the storage nodes), when the scaling of the  size of the message and all storage nodes by the same scaling factor is allowed. Representing the decoding sets as hyperedges in a storage graph, the capacity is characterized for various graphs, including MDS graph, wheel graph, Fano graph, and intersection graph. The achievability is related via quantum CSS codes to a classical secure storage problem. Remarkably, our coding schemes utilize non-trivial alignment structures to ensure recovery and security in the corresponding classical secure storage problem, which leads to similarly non-trivial quantum codes. The converse is based on quantum information inequalities, e.g., strong sub-additivity and weak monotonicity of quantum entropy, tailored to the topology of the storage graphs.
\end{abstract}

\newpage

\allowdisplaybreaks

\section{Introduction}
Robust storage of quantum information is a central technical challenge that stands in the way of the highly anticipated quantum technologies of the future. Efficient distributed quantum computing for instance requires efficient distributed storage of quantum information along with efficient recovery from  failures of subsystems. There is a  rich history of progress  in quantum coding theory leading to a variety of efficient  constructions of quantum error correction codes (QECCs) and bounds on optimal code parameters \cite{Grassl_Bounds,Grassl_Popovski}. Recently there is also interest in applying Shannon theoretic reasoning (based on properties of von Neumann entropies)  to discover the fundamental limits of quantum storage \cite{Grassl_Huber_Winter,Mani_Winter}. 

Distinct from coding theory which prioritizes practical code constructions with (relatively) short codelengths and/or low dimensional systems, a Shannon theoretic approach may prioritize asymptotic optimality guarantees, seeking tractability in the expanded space obtained by allowing arbitrarily long codelengths and/or arbitrarily large local dimensions while constraining only the \emph{relative} values (i.e., ratios) of key parameters. The two distinct perspectives lead to different challenges and produce complementary insights. For example, consider the quantum Singleton bound, $n\geq k+2(d-1)$ that is satisfied by any $[[n,k,d]]_q$ quantum code \cite{Knill_Laflamme, Rains, Cerf_Cleve, Huber_Grassl}, where $n,k,d,q$ represent the number of physical qudits, the number of logical qudits, the minimum distance, and the local dimension of each qudit, respectively. Given a local dimension $q$, codes achieving this bound, called quantum maximum distance separable (QMDS) codes, may not exist. For instance if $q=2$ (qubit systems) then there is no quantum code for one logical qubit $(k=1)$, that tolerates the erasure of any one ($d-1=1$) physical qubit, and needs only three ($n=3$) physical qubits (the minimum feasible value is $n=4$ in this case \cite{Grassl_4qubits}). This is reflected in the QMDS conjecture \cite{Huber_Grassl, Ketkar_Klappenecker_Kumar_Sarvepalli, Grassl_Rotteler} that remains a prominent open problem in coding theory. On the other hand, if  the local dimension $q$ is allowed to be arbitrarily large, then QMDS codes always exist and need only generic structures.\footnote{This is also the case in classical coding theory, where the corresponding classical MDS conjecture remains open, but over large alphabet almost any randomly generated code is MDS.} Remarkably, the Singleton bound is also a Shannon theoretic bound \cite{Grassl_Huber_Winter,Mani_Winter}, and once the local dimension constraint is relaxed (allowing arbitrarily large $q$) it immediately provides a \emph{tight} information theoretic characterization of the optimal code parameters. 
The improved tractability afforded by relaxed local dimension constraints makes a Shannon theoretic approach suitable for exploring (asymptotically) optimal codes for more general settings, e.g., for \emph{heterogeneous} storage systems. 

A common simplifying assumption in studies of QECCs is that the subsystems  comprising the quantum storage are homogeneous, i.e., they have the same size, and are equally likely to be impacted by noise. It is known, however,  that noise characteristics depend very much on the hardware implementation, varying significantly from one physical qudit to another, and failures can be highly correlated. Noting that the performance of a QECC is strongly impacted by such disparities, recent works have emphasized the critical importance of studying optimal code designs for heterogeneous noise structures \cite{heterogeneouserrors}. 

The concerns are  further amplified if the storage is broadly  \emph{distributed}. Distributed  storage systems are likely to employ a variety of quantum subsystems, differing in size and robustness level from one location to another. Furthermore, if these distributed storage systems are to some day mature to larger scales (paralleling existing classical datacenters), then the constituent subsystems will themselves need to approach large sizes, e.g., each constituent subsystem in a large scale distributed storage system may itself be a  storage system employing an internal QECC. A Shannon theoretic approach may be especially well suited for the study of such systems. Last but not the least,  since larger subsystems may be able to employ internal consistency checks within each location, the dominant failure mode in large scale distributed quantum storage systems (similar to classical datacenters) is likely to be errors in \emph{known} locations, also known as erasures.

Motivated by these observations, and drawing inspiration from the formulations in \cite{Grassl_Huber_Winter, Mani_Winter}, in this work we undertake a Shannon theoretic study of the fundamental limits of heterogeneous distributed quantum storage systems  subject to erasures. We model the distributed quantum storage as an $N$-partite quantum system $Q_1 \cdots Q_N$ where each $Q_n, n \in [N]$ is viewed as a storage node. The \emph{relative sizes}  of the storage nodes are specified via the  parameters $\lambda_n, n\in[N]$. Since only the relative values are of interest it will be convenient to set $\min_{n\in[N]}\lambda_n=1$.  The set of erasure patterns that must be protected against is also specified. The sizes and the erasure patterns need not be homogeneous. We explore the \emph{capacity} of such a distributed storage system, defined informally (see Section \ref{sec:defs} for formal definitions) as the maximum (relative) amount ($\lambda_0$) of arbitrary quantum information ($Q_0$) that can be stored in the $N$ storage nodes while respecting all specified storage size constraints and ensuring robustness against all specified erasure patterns. 

\begin{figure}[h]
\begin{center}
\begin{tikzpicture}[scale=0.7, thick]
\begin{scope}[scale=1.2]
\node (N1) at (0cm,0cm) [draw, circle,  inner sep=0.1cm] {$1$};
\draw [line width=0.25cm, black!20] (N1) circle [radius=1.45cm];
\draw [black] (N1) circle [radius=1.45cm];
\node (N2) at ({1.5cm*cos(150)},{1.5cm*sin(150)}) [draw, circle, inner sep=0.1cm, fill=white] {$2$};
\node (N3) at ({1.5cm*cos(30)},{1.5cm*sin(30)}) [draw, circle, inner sep=0.1cm, fill=white] {$3$};
\node (N4) at ({1.5cm*cos(270)},{1.5cm*sin(270)}) [draw, circle, inner sep=0.1cm, fill=white] {$4$};
\draw [line width=0.25cm, red!20](N1)--(N2);
\draw [black](N1)--(N2);
\draw [line width=0.25cm,blue!20](N1)--(N3);
\draw [black](N1)--(N3);
\draw [line width=0.25cm,green!40!white](N1)--(N4);
\draw [black](N1)--(N4);
\end{scope}
\begin{scope}[shift={(7,-1.75)},yscale=2.5]
\draw[thick,-latex] (0,0)--(6,0);
\draw[thick,-latex] (0,0)--(0,1.2);
\draw[thick] (0,2/3)--(2,1)--(6,1);
\node at (0.1,-0.2){$1$};
\node at (2,-0.2){$2$};
\node at (4.75,-0.2){$\max(\lambda_1,\lambda_2)\rightarrow$};
\node at (-0.5,2/3-0.1){$2/3$};
\node at (-0.5,1){$1$};
\node at (-0.5,1.4){$C(\mathcal{W}_4)$};
\draw[dashed](0,1)--(2,1);
\draw[dashed](2,1)--(2,0);
\end{scope}
\end{tikzpicture}
\caption{(Left) The `wheel graph $\mathcal{W}_4$' representing a $4$-partite  quantum storage system $Q_1Q_2Q_3Q_4$, where each  storage node $Q_n$ is represented by a vertex $n$, $n\in[4]$, and decoding sets $\{Q_1,Q_2\}$, $\{Q_1,Q_3\}$, $\{Q_1,Q_4\}$, $\{Q_2,Q_3,Q_4\}$ are represented by the red,  blue, green, and black hyperedges, respectively. (Right) The storage capacity $C(\mathcal{W}_4)=\min(\lambda_1,\lambda_2,(\lambda_1+\lambda_2)/3)$ is shown as a function of $\max(\lambda_1,\lambda_2)$ assuming without loss of generality that $\min(\lambda_1,\lambda_2)=1$ and $\lambda_2\leq \lambda_3\leq \lambda_4$.}\label{fig:w4graph}
\end{center}
\end{figure}
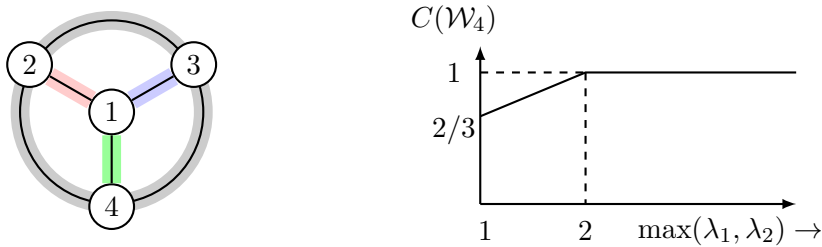

Let us illustrate the problem with an example. Consider a distributed storage system with $4$ storage nodes,  $Q_1,Q_2,Q_3,Q_4$, of relative sizes $\lambda_1,\lambda_2,\lambda_3,\lambda_4$ (these are arbitrary values specified by the problem, e.g., $\lambda_1=1, \lambda_2=2,\lambda_3=3,\lambda_4=4$). Say the erasure patterns are specified such that the nodes in any one of subsets $\{Q_1\}$, $\{Q_2,Q_3\}$, $\{Q_2,Q_4\}$, $\{Q_3,Q_4\}$ may be erased. Equivalently, the quantum information stored in this distributed storage system must be perfectly decodable from the surviving (unerased) set of nodes $\mathcal{D}(e)=\{Q_n: n\in e\}$, for any $e\in\mathcal{E}=\{\{1,2\}$, $\{1,3\}$, $\{1,4\}$, $\{2,3,4\}\}$. Assume, without loss of generality in this case, that $\lambda_2\leq\lambda_3\leq\lambda_4$. A graphical representation of this storage system (labeled the `wheel graph $\mathcal{W}_4$') appears in Figure \ref{fig:w4graph}.

\vspace{0.1in}
\vspace{-0.1in}

The capacity of this storage system is the largest (relative) amount $(\lambda_0)$ of quantum information $Q_0$ that can be stored in $Q_1Q_2Q_3Q_4$ such that $Q_0$ can be perfectly recovered from any $\mathcal{D}(e)$, $e\in\mathcal{E}$.  For this storage system, we show (special case of Theorem \ref{thm:wheel} in Section \ref{sec:wheel}) that the storage capacity is $C(\mathcal{W}_4)\triangleq\min\left(\lambda_1,\lambda_2,\frac{\lambda_1+\lambda_2}{3}\right)$. A capacity characterization requires matching achievability and converse arguments. For the converse we provide a Shannon theoretic  proof that if there exists \emph{any} storage scheme (not limited to any special class of codes, e.g., CSS codes \cite{Calderbank_Shor, Steane})  capable of storing arbitrary quantum information $Q_0$ of size $\log_2|Q_0|=\kappa\lambda_0$ qubits into a $4$-partite quantum system $Q_1Q_2Q_3Q_4$ with sizes $\log_2|Q_n| \leq \kappa \lambda_n$ qubits, $n\in[4]$, for some scaling factor $\kappa>0$ that can be chosen freely by the coding scheme, such that $Q_0$ can be recovered perfectly from any $\mathcal{D}(e)$, $e\in\mathcal{E}$, then we must have $\lambda_0\leq C(\mathcal{W}_4)=\min\left(\lambda_1,\lambda_2,\frac{\lambda_1+\lambda_2}{3}\right)$. For achievability we provide a constructive proof that given $\lambda_1, \lambda_2,\lambda_3,\lambda_4$, and any $\epsilon>0$, there exists a storage scheme capable of storing arbitrary quantum information $Q_0$ of size $\log_2|Q_0|\geq \kappa(C(\mathcal{W}_4)-\epsilon)$ qubits into a $4$-partite quantum system $Q_1Q_2Q_3Q_4$ with sizes $\log_2|Q_n| \leq \kappa \lambda_n$ qubits, $n\in[4]$, for some scaling factor $\kappa>0$, such that $Q_0$ can be recovered perfectly from $\mathcal{D}(e)$, $e\in\mathcal{E}$.

As evident from the example, distributed quantum storage systems can be conveniently represented as graphs with vertices identifying storage nodes along with their relative size parameters, and hyperedges identifying the decoding sets. Given a storage graph, in order to find a Shannon theoretic converse we apply quantum information inequalities, e.g., strong sub-additivity and weak monotonicity of quantum entropy, tailored to the specified storage structure. The converse proof for the wheel graph $\mathcal{W}_4$ example is presented in  Section \ref{ex:wheel}.

To find achievable coding schemes, we take advantage of an important  connection previously noted in literature in a few slightly different forms \cite{Smith, Hayashi_Song}. This connection, via CSS codes, is essentially between our quantum distributed storage coding problem, and a corresponding classical secure storage problem. 
Through this connection (formalized in Theorem \ref{thm:css} and Corollary \ref{cor:css} in Section \ref{sec:genbounds}), quantum systems $Q_0, Q_1,\cdots, Q_N$ with sizes $\kappa\lambda_0,\kappa\lambda_1,\cdots,\kappa\lambda_N$ qudits (a qudit represents a $q$-dimensional quantum system) are mapped to classical symbols $Y_{Q_0}, Y_{Q_1},\cdots, Y_{Q_N}$ in $\mathbb{F}_q^{\kappa\lambda_0}$, $ \mathbb{F}_q^{\kappa\lambda_1}$, $\cdots,\mathbb{F}_q^{\kappa\lambda_N}$, respectively (provided $q$ is a prime power so that the finite field $\mathbb{F}_q$ exists, and $\kappa\lambda_n\in\mathbb{N}$, $n\in\{0,\cdots,N\}$). $Y_{Q_0}$ now represents as a classical secret, whereas $Y_{Q_1},\cdots,Y_{Q_N}$ are the classical secret shares. The secret $Y_{Q_0}$ (comprising $\kappa\lambda_0$ i.i.d. uniform symbols from $\mathbb{F}_q$) must be decodable from any set of secret shares $\mathcal{Y}_{\mathcal{D}(e)}=\{Y_{Q_n}: n\in e\}$, where $e\in\mathcal{E}$, and $\mathcal{E}$ is the set of decoding-set-indices inherited from the quantum distributed storage problem. Also, quite importantly, any set of secret shares corresponding to erased systems, i.e., $\{Y_{Q_n}: n\in [N]\setminus e\}$ must not reveal any information about the secret. The latter constraint is a consequence of the quantum no-cloning theorem. The key observation is the following. If there exists an $\mathbb{F}_q$-linear solution to this classical problem, then by utilizing the connection between the two problems, the solution translates into a quantum distributed storage code for our original problem.

\begin{figure}[t]
\begin{tikzpicture}[scale=0.81]
\node (W4) at (0,0) [draw, rectangle, anchor=west]{\begin{tikzpicture}
\begin{scope}[scale=0.55,xshift=0.9cm, yshift=-0.5cm]
\node (N1) at (0cm,0cm) [draw, circle,  inner sep=0.1cm] {$1$};
\draw [line width=0.25cm, black!20] (N1) circle [radius=1.45cm];
\draw [black] (N1) circle [radius=1.45cm];
\node (N2) at ({1.5cm*cos(150)},{1.5cm*sin(150)}) [draw, circle, inner sep=0.1cm, fill=white] {$2$};
\node (N3) at ({1.5cm*cos(30)},{1.5cm*sin(30)}) [draw, circle, inner sep=0.1cm, fill=white] {$3$};
\node (N4) at ({1.5cm*cos(270)},{1.5cm*sin(270)}) [draw, circle, inner sep=0.1cm, fill=white] {$4$};
\draw [line width=0.25cm, red!20](N1)--(N2);
\draw [black](N1)--(N2);
\draw [line width=0.25cm,blue!20](N1)--(N3);
\draw [black](N1)--(N3);
\draw [line width=0.25cm,green!40!white](N1)--(N4);
\draw [black](N1)--(N4);
\node at (-1.25,2.5) [align=left]{\footnotesize $(\lambda_1,\lambda_2,\lambda_3,\lambda_4)$\\\footnotesize $=(1,2,2,2)$};
\end{scope}
\end{tikzpicture}};
\def\gapa{0.3cm}
\node (A) at (4.5,0)[draw,rectangle, anchor=west]{\scalebox{0.9}{\small 
$\begin{array}{l}
Q_0\rightarrow Y_{Q_0}\in\mathbb{F}_q^{\kappa}\mbox{ (secret)}\\[\gapa]
Q_1\rightarrow Y_{Q_1}\in\mathbb{F}_q^{\kappa}\mbox{ (share)}\\[\gapa]
Q_2\rightarrow Y_{Q_2}\in\mathbb{F}_q^{2\kappa}\mbox{ (share)}\\[\gapa]
Q_3\rightarrow Y_{Q_3}\in\mathbb{F}_q^{2\kappa}\mbox{ (share)}\\[\gapa]
Q_4\rightarrow Y_{Q_4}\in\mathbb{F}_q^{2\kappa}\mbox{ (share)}\\
\end{array}
$}
};
\def\gapb{0.1cm}
\node (B) at (20,-0)[draw, rectangle, anchor=east]{\scalebox{0.9}{\small
$\begin{array}{l}
~H(Y_{Q_0})=\kappa\log_2(q)\mbox{ bits} \mbox{ (uniform)}\\[0.36cm]
\begin{array}{l|l}
\mbox{Decodability:}&\mbox{Security:}\\[\gapb]
H(Y_{Q_0}\mid Y_{Q_1},Y_{Q_2})=0&I(Y_{Q_0}; Y_{Q_3},Y_{Q_4})=0\\[\gapb]
H(Y_{Q_0}\mid Y_{Q_1},Y_{Q_3})=0&I(Y_{Q_0}; Y_{Q_2},Y_{Q_4})=0\\[\gapb]
H(Y_{Q_0}\mid Y_{Q_1},Y_{Q_4})=0&I(Y_{Q_0}; Y_{Q_2},Y_{Q_3})=0\\[\gapb]
H(Y_{Q_0}\mid Y_{Q_2},Y_{Q_3},Y_{Q_4})=0&I(Y_{Q_0}; Y_{Q_1})=0\\[\gapb]
\end{array}
\end{array}
$}
};
\node (C) at (20,-5)[anchor=east, draw,rectangle]{\scalebox{0.9}{\small
$\begin{array}{l}
\mbox{$\mathbb{F}_q$-linear solution ($\kappa=1$):}\\
a,b_1,b_2,b_3: \mbox{i.i.d. unif. in $\mathbb{F}_q$}\\
\mbox{secret: } a, \mbox{ noise: } b_1,b_2,b_3\\
Y_{Q_0}=(a)\\
Y_{Q_1}=(b_1)\\
Y_{Q_2}=(a+b_1,~ b_2)\\
Y_{Q_3}=(a+b_1, ~b_3)\\
Y_{Q_4}=(a+b_1, ~a+b_2+b_3)\\
\end{array}
$}
};
\node (D) at (0,-5)[anchor=west, draw,rectangle]{\scalebox{0.9}{\small
$\begin{array}{l}
\mbox{Quantum System $\leftrightarrow$ Hilbert space: }\\(Q_0\leftrightarrow \mathbb{C}^q),\\
(Q_1\leftrightarrow \mathbb{C}^q), (Q_2\leftrightarrow \mathbb{C}^q\otimes \mathbb{C}^q), (Q_3\leftrightarrow \mathbb{C}^q\otimes \mathbb{C}^q), (Q_4\leftrightarrow \mathbb{C}^q\otimes \mathbb{C}^q)\\[0.2cm]
\mbox{Orthonormal basis for $\mathbb{C}^q$: $\{\ket{a}: a\in\mathbb{F}_q\}$}\\[0.1cm]
\mbox{Quantum Code: For all $a\in\mathbb{F}_q$, map input $\ket{a}_{Q_0}$ to,}\\
\displaystyle \frac{1}{\sqrt{q^{3}}}\sum_{b_1,b_2,b_3\in\mathbb{F}_q}\ket{b_1}_{Q_1}\ket{a+b_1,b_2}_{Q_2}\ket{a+b_1,b_3}_{Q_3}\ket{a+b_1,a+b_2+b_3}_{Q_4}\\
\end{array}
$}
};
\draw [-latex] (W4)--(A);
\draw [-latex] (A)--(B);
\draw [-latex] (C.north|-B.south) -- (C.north|-C.north);
\draw [-latex] (C)--(D);
\end{tikzpicture}
\caption{Sketch of achievability argument $(\lambda_0=1)$ for a particular wheel graph $\mathcal{W}_4$ setting, with size constraints $(\lambda_1,\lambda_2,\lambda_3,\lambda_4)=(1,2,2,2)$. Quantum nodes are  mapped to classical nodes of corresponding sizes over $\mathbb{F}_q$, decodability and security constraints (stated here in entropic terms) define the classical secure storage problem, and an $\mathbb{F}_q$-linear solution to the classical problem is mapped to a quantum distributed storage code via a CSS code construction.}\label{fig:connection}
\end{figure}
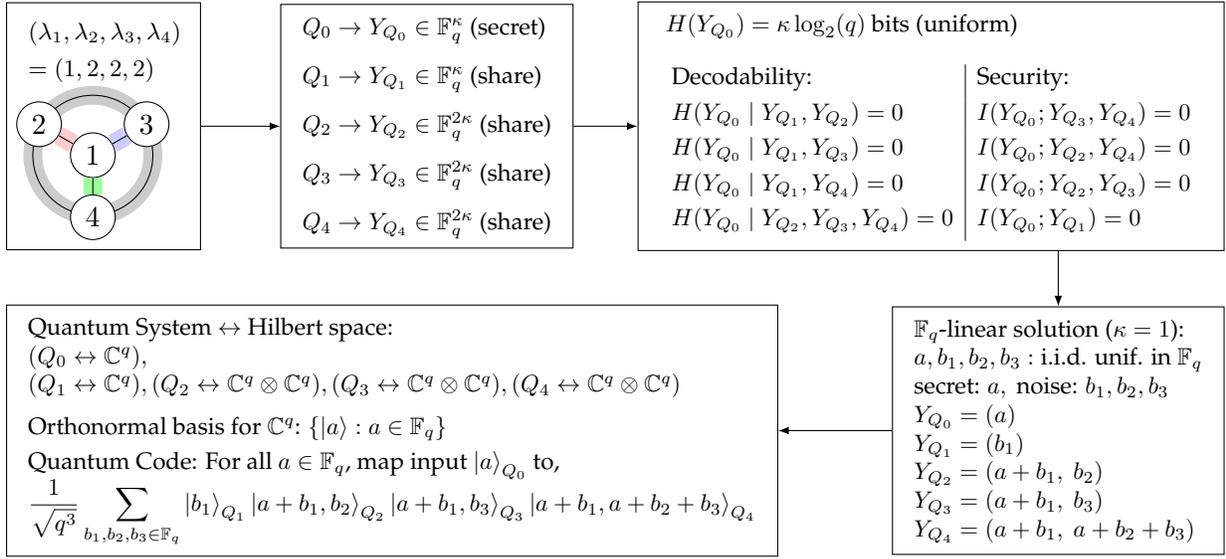
For an illustration via our wheel graph $\mathcal{W}_4$ example, consider the size constraints, say $\lambda_1=1,\lambda_2=\lambda_3=\lambda_4=2$, for which the capacity $C(\mathcal{W}_4)=1$. The mapping to the classical secure storage problem is shown in Figure \ref{fig:connection}, along with the optimal solution. The general solution for the $\mathcal{W}_4$ example with arbitrary size constraints $\lambda_n$ is presented in Section \ref{ex:wheel}.

The storage graph structures that emerge out of the heterogeneous aspects of distributed quantum storage, lead to rather non-trivial classical secure storage problems that have not been explored in the classical secret sharing and secure storage literature. Unlike homogeneous settings where generic structures tend to be optimal, these problems in general require \emph{interference alignment} principles for optimal code design in order to efficiently satisfy the simultaneous decodability and security constraints. Even in the simple example of the wheel graph $\mathcal{W}_4$,  alignment can be seen in the mirroring of the $a+b_1$ symbol in $Y_{Q_2},Y_{Q_3},Y_{Q_4}$. This is needed to ensure that any two of these shares (because the corresponding quantum storage nodes can be erased) do not reveal anything about the secret. Gaining insight into the role that interference alignment might play in quantum storage codes is a key motivation for the present work.

\bigskip
{\it Notation:} $[N]$ denotes the set $\{1,2,\cdots, N\}$ and $[i:j]$ denotes the set $\{i, i+1, \cdots, j\}$ for $i \leq j, i,j,N \in \mathbb{N}$.
$|X|$ denotes the cardinality of $X$ if $X$ is a set, the dimension of $X$ if $X$ is a vector, and the dimension
of the Hilbert space associated with $X$ if $X$ is a quantum system. For a set $\mathcal{X}$, the set of its cardinality-$k$ subsets is denoted as $\binom{\mathcal{X}}{k} \triangleq \{ \mathcal{Y} : \mathcal{Y} \subset \mathcal{X}, |\mathcal{Y}| = k \}$ and the set of all subsets of $\mathcal{X}$ is denoted as $2^\mathcal{X}$.
For two sets $\mathcal{X}, \mathcal{Y}$, the notation $\mathcal{X} \setminus \mathcal{Y}$ denotes the set of elements that are in $\mathcal{X}$ but not in $\mathcal{Y}$. For matrices ${\bf M}_1, {\bf M}_2$ of compatible dimensions, $({\bf M}_1, {\bf M}_2)$ and $({\bf M}_1; {\bf M}_2)$ represent their horizontal and vertical concatenations, respectively. 
For a matrix ${\bf M}$, the notation $\langle {\bf M} \rangle$ denotes the vector space spanned by the columns of ${\bf M}$ (defined over the same field as the elements of ${\bf M}$).
${\bf I}_{k\times k}$ represents the $k\times k$ identity matrix and ${\bf 0}_{k_1\times k_2}$ represents the $k_1\times k_2$ matrix with all its elements equal to $0$; the subscript that indicates the matrix dimension may be omitted when it is clear from the context. For quantum systems $X, Y$, the notation $X \rightsquigarrow Y$ denotes that a unitary (isometric) map is applied to $X$ so that it is transformed to $Y$. 
When $X$ is a set of classical random variables or quantum systems, $H(X)$ denotes the joint Shannon or von Neumann (quantum) entropy of the elements in $X$.
$\mathbb{F}_q$ denotes the finite field with $q$ elements where $q$ is a power of a prime.

\section{Problem Statement}
\subsection{Preliminaries}
Before we formally introduce our problem statement, let us recall some elementary aspects of modeling of quantum systems.  A quantum system $Q$ is associated with a Hilbert space $\mathcal{H}_Q$. The dimension of a quantum system $Q$, denoted $|Q|$, corresponds to the dimension of the Hilbert space $\mathcal{H}_Q$.   In this work we only consider finite dimensional Hilbert spaces.  Arbitrary states of $Q$ are represented by density matrices, i.e., unit trace positive semidefinite matrices over $\mathcal{H}_Q$. Define $S(\mathcal{H}_Q)$  as the set of density matrices
over $\mathcal{H}_Q$. 
Define the set of unit rank density matrices, corresponding to pure states, as $S_1(\mathcal{H}_Q)$. Pure states can also be represented as unit vectors in $\mathcal{H}_Q$.
For a composite system comprised of $N$ quantum subsystems, if the subsystem $Q_n$ is associated with the Hilbert space $\mathcal{H}_{Q_n}$, $n\in[N]$, then the composite system $Q_1Q_2\cdots Q_N$ is associated with the Hilbert space $\mathcal{H}_{Q_1}\otimes \mathcal{H}_{Q_2}\otimes\cdots\otimes\mathcal{H}_{Q_N}$.
A \emph{qudit}, representing a $q$-dimensional quantum system, is a conventional unit for expressing the size of a quantum system. When $q=2$ the qudit is called a qubit.\footnote{The choice of units is a cosmetic issue, since translation from one unit to another is trivial, but if the quantum systems involved are naturally composed of $q$-dimensional subsystems for some $q\neq 2$ then using a qudit (instead of a qubit) as the unit may produce cleaner expressions.} The size of a quantum system $Q$ is  $\log_2|Q|$ qubits, which is the same as $\log_q|Q|$ qudits.  For the composite system $Q_1Q_2\cdots Q_N$, where each $Q_n$ has size $\log_2|Q_n|$ qubits, $n\in[N]$,  the composite system has size $\log_2|Q_1Q_2\cdots Q_N|=\log_2|Q_1|+\log_2|Q_2|+\cdots+\log_2|Q_N|$ qubits. 

\subsection{Distributed Quantum Storage Capacity Formulation}\label{sec:defs}
A quantum {\em storage structure} is specified as a graph $\mathcal{G}=((\lambda_1,\cdots,\lambda_N),\mathcal{E})$. The graph $\mathcal{G}$ has $N$ vertices representing quantum storage nodes, $$\mathcal{Q}(\mathcal{G})=\{Q_1,Q_2,\cdots, Q_N\},$$ and a set of hyperedges, $\mathcal{E}\subset 2^{[N]}$, that defines the corresponding decoding sets, $$\mathcal{D}(e)\triangleq\{Q_n: n\in e\}, ~~~e\in\mathcal{E},$$ and their complements,
$$\mathcal{D}^c(e)\triangleq \{Q_n: n\in [N]\setminus e\},~~~e\in\mathcal{E}.$$
A storage node is \emph{redundant} if it is not included in any decoding set. A decoding set is \emph{redundant} if it includes a smaller decoding set as a proper subset. 
We assume that the storage graphs have no redundant storage nodes, and no redundant decoding sets, i.e., for every $Q_n, n\in[N]$, there exists some $e \in \mathcal{E}$ such that $Q_n \in e$, and there are no $e, e' \in \mathcal{E}$ such that $e \subset e'$.

Each storage node $Q_n$ has size constrained as $\log_q|Q_n|\leq\kappa\lambda_n$ qudits,  for all $n\in[N]$. Here $\kappa$ is a scaling factor, and a qudit represents a $q$-dimensional quantum system.

A quantum {\em message}, $Q_0$, is a quantum system with size $\log_q|Q_0|=\kappa\lambda_0$ qudits. Along with $Q_0$, define $R$, without loss of generality also of the same size $\log_q|R|=\kappa\lambda_0$ qudits, as a {\em reference} quantum system such that $RQ_0$ is an arbitrary pure state $\ket{\varphi}\bra{\varphi}_{RQ_0}\in S_1(\mathcal{H}_R\otimes \mathcal{H}_{Q_0})=S_1(\mathbb{C}^{q^{\kappa\lambda_0}}\otimes \mathbb{C}^{q^{\kappa\lambda_0}})$.

We say that storage $(\mathcal{G},q,\kappa)$ {\em fits} $Q_0$  if and only if there exists a CPTP (completely positive trace preserving) encoding map,\footnote{While the map is specified between two sets of density matrices, it can be readily extended to all linear operators (matrices). Refer to Section 5.1 of \cite{Hayashi} or Appendix B of \cite{Wilde}.} $$\mbox{ENC}: S(\mathcal{H}_{Q_0})\rightarrow S(\mathcal{H}_{Q_1}\otimes\cdots\otimes\mathcal{H}_{Q_N}),$$ that maps the quantum message $Q_0$ to the storage nodes $Q_1Q_2\cdots Q_N$, and for each  $e\in\mathcal{E}$, a CPTP decoding map,\footnote{Note that we adopt a compound channel model, where there are $|\mathcal{E}|$ decoders, one for each erasure pattern.} $$\mbox{DEC}^{e}: S(\otimes_{n\in e} \mathcal{H}_{Q_n})\rightarrow S(\mathcal{H}_{Q_0}),$$ 
that maps the storage nodes in the corresponding decoding set $\mathcal{D}(e)$ to an output $\widehat{Q}_{0}$ such that $R\widehat{Q}_{0}$ is in state $\ket{\varphi}$, i.e., the original message $Q_0$ and any entanglement with the reference system are perfectly recovered.
\begin{align}
&I_{R}\otimes\left(\mbox{DEC}^e_{(Q_n: n\in e)\rightarrow \widehat{Q}_0}\circ \mathcal{N}^e_{(Q_n:n\in[N])\rightarrow (Q_n: n\in e)}\circ \mbox{ENC}_{Q_0\rightarrow (Q_n: n\in[N])}\right)_{Q_0\rightarrow \widehat{Q}_0}\left(\ket{\varphi}\bra{\varphi}\right)_{RQ_0}=\left(\ket{\varphi}\bra{\varphi}\right)_{R\widehat{Q}_0}, \notag\\
&\forall \ket{\varphi}\bra{\varphi}\in S_1(\mathcal{H}_R\otimes \mathcal{H}_{Q_0})=S_1(\mathbb{C}^{q^{\kappa\lambda_0}}\otimes \mathbb{C}^{q^{\kappa\lambda_0}}), \forall e\in\mathcal{E}.\label{eq:allstates}
\end{align}
Here the channel $\mathcal{N}^e(\cdot) \triangleq \mbox{Tr}_{(Q_n:n\in[N]\setminus e)}(\cdot)$ simply erases the quantum systems $Q_n, n\in[N]\setminus e$.

Given storage $(\mathcal{G},q,\kappa)$, we wish to determine how much quantum information can be stored, i.e., the largest $\lambda_0$ such that $(\mathcal{G},q,\kappa)$ fits $Q_0$. While it is important from a coding theoretic perspective to answer the question for any given $\mathcal{G},q,\kappa$, we will adopt a Shannon theoretic perspective and  focus instead on a coarser (but more tractable) objective, the \emph{storage capacity} (defined next) which relaxes the dependence on  $\kappa$ and  $q$, by optimizing over these parameters.

The {\em storage capacity} of a storage graph $\mathcal{G}$ is defined as
\begin{align}
C(\mathcal{G})&\triangleq \sup_{q,\kappa\in\mathbb{N}}\sup\{\lambda_0: (\mathcal{G},q,\kappa)\mbox{ fits } Q_0 \}.
\end{align}
\noindent A subscript `$u$'  is added ($C_u(\mathcal{G})$) if the storage is  uniform, i.e., $\lambda_1 = \lambda_2 = \cdots = \lambda_N = 1$. 

Since the sizes of $Q_0,Q_1,\cdots,Q_N$ are allowed to scale proportionately, only the relative values of $\lambda_0,\lambda_1,\cdots,\lambda_N$ are important for storage capacity. Without loss of generality we will assume,
\begin{eqnarray}
\min_{n\in[N]}\lambda_n=1.\label{eq:lambdamin1}
\end{eqnarray}

\begin{figure}[h]
\begin{tikzpicture}
\node (ENC) at (0,0) [draw, thick, rectangle, minimum width=1cm, minimum height=2cm]{\footnotesize ENC};
\node (QR) at ($(-2.5cm,1cm)+(ENC.west)$) {\footnotesize $\ket{\varphi}$};
\node (QRhat) at ($(12.5cm,1cm)+(ENC.west)$) {\footnotesize $\ket{\varphi}$};
\draw [-] (QR.east)--($(-1cm,0cm)+(ENC.west)$)node[midway,below=0.1cm]{\footnotesize $Q_0$}-- (ENC.west);

\draw [-] (QR.east)--($(-1cm,2cm)+(ENC.west)$)node[midway,above=0.1cm]{\footnotesize $R$}-- ($(11cm,2cm)+(ENC.west)$)--(QRhat.west);
\node (Ne) at (4,0) [draw, thick, rectangle, minimum width=1cm, minimum height=2cm, align=center]{\footnotesize $\mathcal{N}^e$};

\node (De) at (9,0) [draw, thick, rectangle, minimum width=1cm, minimum height=2cm]{\footnotesize $\mbox{DEC}^e$};
\draw [-] (QRhat.west)--($(1cm,0cm)+(De.east)$)node[midway,below=0.1cm]{\footnotesize $\widehat{Q}_0$}-- (De.east);

\node (Q1) at (2,0.75) {\footnotesize $Q_1$};
\node (Q2) at (2,0.25) {\footnotesize $Q_2$};
\node (vdots) at (2,-0.15) {\footnotesize $\vdots$};
\node (QN) at (2,-0.75) {\footnotesize $Q_N$};
\draw (ENC.east|-Q1) -- (Q1)--(Q1-|Ne.west);
\draw (ENC.east|-Q2) -- (Q2)--(Q2-|Ne.west);
\draw (ENC.east|-QN) -- (QN)--(QN-|Ne.west);

\node (Qerased) at (6.5,0.5) {\footnotesize $(Q_n: n\in [N]\setminus e)$};
\draw (Ne.east|-Qerased)-- (Qerased);
\node (Qunerased) at (6.5,-0.5) {\footnotesize $(Q_n: n\in e)$};
\draw (Ne.east|-Qunerased) -- (Qunerased)--(Qunerased-|De.west);
\end{tikzpicture}
\caption{Erasure channel corresponding to the storage capacity problem. Given any $e\in\mathcal{E}$, the channel $\mathcal{N}^e$ erases the storage nodes $Q_i$ for all $i\in[N]\setminus e$, leaving only storage nodes $Q_j, j\in e$ for the decoder $\mbox{DEC}^e$. Note that the encoder does not depend on $e$, because the positions of erasures are not known at the time of encoding, but the decoder is chosen based on the realization of $e$ because the decoding is done with the knowledge of which storage nodes are erased. }\label{fig:channel}
\end{figure}
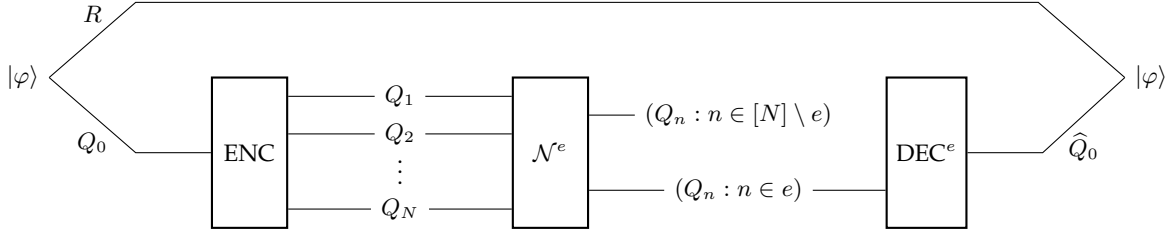

\begin{remark} Similar to the observation in \cite{Grassl_Huber_Winter}, the storage capacity, as defined above corresponds to the Shannon theoretic capacity of the quantum erasure channel shown in Figure \ref{fig:channel}. The  scaling factor $\kappa$ corresponds in this case to the  number of channel uses. Over $\kappa$ channel uses, each $Q_n$ represents the composite system $Q_n(1)Q_n(2)\cdots Q_n(\kappa)\equiv Q_n$ which has size $\kappa\lambda_n$ qudits, i.e., $\log_q|Q_n|=\kappa\lambda_n$, whereas each subsystem $Q_n(k)$ corresponding to channel use $k\in[\kappa]$ has size fixed as $\log_q|Q_n(k)|=\lambda_n$ qudits. Recall that in the Shannon theoretic formulation the number of channel uses $\kappa$ can scale arbitrarily. The alphabet  $q$ can also be chosen arbitrarily in this channel as it only amounts to a choice of `sub-packetization' of the composite quantum systems which is inconsequential for  a Shannon theoretic capacity formulation (see e.g., Theorem VI.5 of \cite{Cannons_Dougherty_Freiling_Zeger}). The setting in Figure \ref{fig:channel} corresponds to a `compound channel' setting in the information theory literature (cf. \cite{loyka2016general} and the references therein). Here `$e$' is a channel state that is chosen from a set of possible states $\mathcal{E}$, and then held fixed across all channel uses. The  state $(e)$ is known to the receiver. The transmitter knows only the set of possible realizations, $\mathcal{E}$, but not the actual realization $(e)$.
\end{remark}

The following notion of a maximal storage graph will be useful.
\begin{definition}[Maximal Storage Graph] A quantum storage graph $\mathcal{G}=((\lambda_1, \cdots, \lambda_N),\mathcal{E})$ is said to be \emph{strongly maximal} if its capacity $C(\mathcal{G})> 0$, but for all $\mathcal{G}'=((\lambda_1', \cdots, \lambda'_{N'}),\mathcal{E}')$ such that $\mathcal{G}$ is a proper\footnote{$\mathcal{G}=((\lambda_1, \cdots, \lambda_N),\mathcal{E})$ is said to be a proper subgraph of $\mathcal{G}'=((\lambda_1', \cdots, \lambda'_{N'}),\mathcal{E}')$ if either $N = N'$, $\mathcal{E} \subsetneq \mathcal{E'}$ (i.e., the vertex set is identical while the edge set is a proper subset) or $N < N'$, $\mathcal{E} \subsetneq \mathcal{E'}$ (note that if the vertex set is a proper subset, then it implies that the edge set must also be a proper subset as we assume there is no redundant storage node).}  subgraph of $\mathcal{G}'$, the capacity $C(\mathcal{G}')=0$. A quantum storage graph $\mathcal{G}$ is said to be \emph{weakly maximal} if  for all $\mathcal{G}'$ such that $\mathcal{G}$ is a proper subgraph of $\mathcal{G}'$, the capacity $C(\mathcal{G}')<C(\mathcal{G})$.
\end{definition}

\section{Results}
\subsection{Capacity Bounds for Arbitrary Storage Graphs}\label{sec:genbounds}

We first present upper and lower bounds on the capacity of arbitrary storage graphs in the following two theorems, respectively. For a set of quantum storage nodes $\mathcal{S} \subset \mathcal{Q}(\mathcal{G}) = \{Q_1, Q_2, \cdots, Q_N\}$,  define the compact notation $\Lambda( \mathcal{S}) = \sum_{i: Q_i \in \mathcal{S}} \lambda_i$.

\begin{theorem} \label{thm:genupbound} 
The quantum storage capacity of  $\mathcal{G}=((\lambda_1,\cdots,\lambda_N),\mathcal{E})$ is bounded from above as,
\begin{align}
\mbox{Intersection Bound: }&&C(\mathcal{G})&\leq \Lambda\Big( \mathcal{D}(e_i)\cap \mathcal{D}(e_j) \Big),&&\forall e_i, e_j\in\mathcal{E}, e_i \neq e_j, \label{up:intersection}\\
\mbox{Wheel Bound: }&&C(\mathcal{G})&\leq \frac{\Lambda(\mathcal{Q}_1)+\Lambda(\mathcal{Q}_2)+\cdots+\Lambda(\mathcal{Q}_{k})}{2k-1},&&\forall  k\in[2:n-2], \label{up:wheel}
\end{align}
where $\{\mathcal{Q}_1,\mathcal{Q}_2,\cdots,\mathcal{Q}_n\}$ is a partition of $\mathcal{Q}(\mathcal{G})$ such that $\Lambda(\mathcal{Q}_2)\leq \Lambda(\mathcal{Q}_3)\leq \cdots \leq \Lambda(\mathcal{Q}_n)$ and each of $\mathcal{Q}_1\cup\mathcal{Q}_2$, $\mathcal{Q}_1\cup\mathcal{Q}_3$, $\cdots$, $\mathcal{Q}_1\cup\mathcal{Q}_n$, and $\mathcal{Q}_2\cup\mathcal{Q}_3\cup\cdots\cup\mathcal{Q}_n$ contains a decoding set of $\mathcal{G}$ as a subset.
\end{theorem}

The proof of Theorem \ref{thm:genupbound} is presented in Section \ref{sec:genupbound}. The intersection bound (\ref{up:intersection}) is a consequence of the no-cloning theorem \cite{Wootters_Zurek, Dieks} while the wheel bound (\ref{up:wheel}) requires the use of the strong sub-additivity (sub-modularity)  and  weak monotonicity properties of quantum entropy \cite{Lieb_Ruskai, Pippenger}.

\begin{theorem} \label{thm:css}
(CSS Bound) The quantum storage capacity of $\mathcal{G}=((\lambda_1,\cdots,\lambda_N),\mathcal{E})$
is bounded from below as $C(\mathcal{G}) \geq k/\kappa$ if there exist $k,\delta, n, \kappa \in\mathbb{N}$, a finite field $\mathbb{F}_q$ and matrices ${\bf A}\in\mathbb{F}_q^{k\times n}, {\bf B}\in\mathbb{F}_q^{\delta\times n}$ with rank$({\bf A})=k$, rank$({\bf A};{\bf B})=k+\delta$, such that for all $e\in\mathcal{E}$,
\begin{align}
0&=H(\bm{a}\mid \mathcal{Y}_{\mathcal{D}(e)}),\label{css:dec}\\
0&=I(\bm{a};\mathcal{Y}_{\mathcal{D}^c(e)}).\label{css:sec}
\end{align}
The following notation is used here.
\begin{enumerate}
\item $\mathcal{Y}_\mathcal{S}\triangleq\{Y_{Q}: Q\in \mathcal{S}\}$ for all $\mathcal{S}\subset\mathcal{Q}(\mathcal{G})$,  
\item $(Y_{Q_1},Y_{Q_2},\cdots,Y_{Q_N})\triangleq \bm{a}{\bf A}+\bm{b}{\bf B}\in\mathbb{F}_q^{1\times n}$,
where
\begin{eqnarray}
&&n = \kappa (\lambda_1 + \cdots + \lambda_N),\\
&& {\bf A} = ({\bf A}_1, \cdots, {\bf A}_N), ~{\bf A}_i \in \mathbb{F}_q^{k \times \kappa \lambda_i}, \\
&& {\bf B} = ({\bf B}_1, \cdots, {\bf B}_N), ~{\bf B}_i \in \mathbb{F}_q^{\delta \times \kappa \lambda_i}, \\
&& Y_{Q_i} = \bm{a}{\bf A}_i+\bm{b}{\bf B}_i \in\mathbb{F}_q^{1\times |Y_{Q_i}|}, |Y_{Q_i}| = \log_q |Q_i| = \kappa \lambda_i, i \in [N],
\end{eqnarray}
\item $\bm{a},\bm{b}$ are $1\times k$ and $1\times \delta$ random vectors, respectively, with i.i.d. uniform elements in $\mathbb{F}_q$. 
\end{enumerate}
\end{theorem}

Theorem \ref{thm:css} allows the construction of an achievable scheme for the quantum storage problem via a  classical secure storage problem, i.e., $k$ classical symbols $\bm{a}$ (comprising a `secret' $Y_{Q_0}$) are stored over classical storage nodes $Y_{Q_1}, \cdots, Y_{Q_N}$ such that from any decoding set $\mathcal{Y}_{\mathcal{D}(e)}$, we may recover the classical symbols $\bm{a}$ (refer to (\ref{css:dec})) while from the complement of any decoding set $\mathcal{Y}_{\mathcal{D}^c(e)}$, nothing is revealed about $\bm{a}$ (refer to (\ref{css:sec})). Security is guaranteed with the help of independent classical uniform noise symbols $\bm{b}$. The classical decoding constraint (\ref{css:dec}) and the classical security constraint (\ref{css:sec}) (the random variables are classical so only Shannon entropies are involved) can be equivalently stated as rank constraints on matrices ${\bf A}, {\bf B}$, presented in the following corollary.

\begin{corollary}\label{cor:css}
For any $\mathcal{S} = \{Q_{i_1}, \cdots, Q_{i_{|\mathcal{S}|}}\} \subset \mathcal{Q}(\mathcal{G})$, define
\begin{align}
{\bf A}_\mathcal{S}& \triangleq ( {\bf A}_{i_1}, {\bf A}_{i_2}, \cdots, {\bf A}_{i_{|\mathcal{S}|}} ),\\
{\bf B}_\mathcal{S} &\triangleq ( {\bf B}_{i_1}, {\bf B}_{i_2}, \cdots, {\bf B}_{i_{|\mathcal{S}|}} ).
\end{align}
Then the following equivalence relations hold.
\begin{eqnarray}
(\ref{css:dec}) &\Leftrightarrow& \mbox{rank}({\bf A}_{\mathcal{D}(e)}; {\bf B}_{\mathcal{D}(e)}) - \mbox{rank}({\bf B}_{\mathcal{D}(e)}) = k \label{css:rank1}\\
&\Leftrightarrow& \langle {\bf I}_{k\times k}; {\bf 0}_{\delta\times k} \rangle \subset \langle {\bf A}_{\mathcal{D}(e)}; {\bf B}_{\mathcal{D}(e)} \rangle, \label{css:space1}\\
(\ref{css:sec}) &\Leftrightarrow& \mbox{rank}({\bf A}_{\mathcal{D}^c(e)}; {\bf B}_{\mathcal{D}^c(e)}) = \mbox{rank}({\bf B}_{\mathcal{D}^c(e)}) \label{css:rank2}\\
&\Leftrightarrow& \langle {\bf I}_{k\times k}; {\bf 0}_{\delta\times k} \rangle \cap \langle {\bf A}_{\mathcal{D}^c(e)}; {\bf B}_{\mathcal{D}^c(e)} \rangle = \{{\bf 0}\}. \label{css:space2}
\end{eqnarray}
\end{corollary}

Interestingly, the code construction for the classical secure storage problem (\ref{css:dec}), (\ref{css:sec}) can be translated into a quantum storage code, which turns out to be equivalent to the canonical CSS quantum code (thus Theorem \ref{thm:css} is called the CSS bound). We note that Theorem \ref{thm:css} (and its connection to CSS codes) is not new and it has appeared in similar (sometimes equivalent or more general) forms in the literature (see e.g., \cite{Smith}, Theorem 1 of  \cite{Hayashi_Song}, and \cite{Senthoor_Sarvepalli}). For the sake of completeness, a proof of Theorem \ref{thm:css} and Corollary \ref{cor:css} is included in Section \ref{sec:css} and Section \ref{sec:csscor}, respectively.

\bigskip

Equipped with the converse bound in Theorem \ref{thm:genupbound} and the achievability argument in Theorem \ref{thm:css}, we are able to characterize the class of all storage graphs for which $C(\mathcal{G}) > 0$, and the exact capacity of small graphs with either $N \leq 4$ nodes or $|\mathcal{E}| \leq 3$ decoding sets. The result is stated in the following corollary.

\begin{corollary} \label{thm:small}
For a storage  graph  $\mathcal{G}=((\lambda_1,\cdots,\lambda_N),\mathcal{E})$, 
\begin{enumerate}
\item the capacity $C(\mathcal{G})=0$, if and only if there exist $e_i, e_j\in\mathcal{E}$ such that $e_i\cap e_j=\emptyset$; 
\item if $N \leq 4$ or $|\mathcal{E}| \leq 3$, then 
Theorem \ref{thm:genupbound} provides a tight bound on the capacity $C(\mathcal{G})$.
\end{enumerate}
\end{corollary}
Note that the tight capacity characterization for small graphs ($N \leq 4$ or $|\mathcal{E}| \leq 3$) holds whether the storage is uniform or non-uniform. The proof of Corollary \ref{thm:small} is presented in Section \ref{sec:small}.

\subsection{Capacities of Certain Storage Graphs}

\subsubsection{MDS Graph $\mathcal{M}_{N,K}$}

The MDS graph, defined as $\mathcal{M}_{N,K}=\left((\lambda_1, \cdots, \lambda_N),\binom{[N]}{K}\right)$, represents a code structure comprising $N$ storage nodes, $\mathcal{Q}(\mathcal{M}_{N,K})=\{Q_1,Q_2,\cdots,Q_N\}$, such that any $K$ storage nodes form a decoding set, i.e., $\mathcal{E}=\binom{[N]}{K}$. The storage need not be uniform. Let us assume, without loss of generality, that $\lambda_1\leq\lambda_2 \leq \cdots\leq\lambda_N$.

\begin{theorem}\label{thm:mds}
The capacity of the MDS graph $\mathcal{M}_{N,K}, 2K > N$ is 
\begin{eqnarray}
C(\mathcal{M}_{N,K})=\min_{\mathcal{I}\in\binom{ \{Q_1,\cdots, Q_N\} }{2K-N}} \Lambda(\mathcal{I}) = \lambda_1 + \lambda_2 + \cdots + \lambda_{2K-N}.
\end{eqnarray}
For uniform storage, the capacity is $C_u(\mathcal{M}_{N,K})=2K-N$.
\end{theorem}

When $2K \leq N$, the capacity is $C(\mathcal{M}_{N,K})$ = 0 due to Corollary \ref{thm:small}, because there exist decoding sets that have no intersection. Thus, we only need to consider the cases where $2K > N$. The converse follows from the intersection bound (\ref{up:intersection}). The achievability for uniform storage is already well known, and can be shown by using random (generic) linear codes in the classical secure storage problem with constraints (\ref{css:dec}), (\ref{css:sec}) and then applying Theorem \ref{thm:css} to translate into quantum codes; the achievability for non-uniform storage is based on `\emph{space sharing}' over the optimal codes of a number of distinct MDS graphs with uniform storage. An example for $\mathcal{M}_{4,3}$, i.e., with $N=4, K=3$,  is given in Section \ref{ex:mds}. The proof of Theorem \ref{thm:mds} is presented in Section \ref{sec:mds}.

\subsubsection{Wheel Graph $\mathcal{W}_N$}\label{sec:wheel}
The wheel graph, defined as $\mathcal{W}_N=\left((\lambda_1, \cdots, \lambda_N),\mathcal{E} \right)$, has $N \geq 4$ storage nodes and $N$ decoding sets, as follows. 
\begin{align}
\mathcal{E}&=\big\{ \{1,2\}, \{1,3\}, \cdots, \{1,N\}, \{2,3,\cdots,N\} \big\}.
\end{align}
Assume, without loss of generality, that $\lambda_2 \leq \lambda_3 \leq \cdots \leq \lambda_N$. 

$\mathcal{W}_5$ is illustrated below.
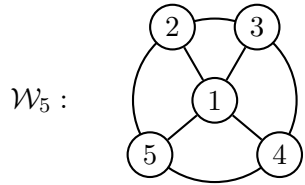
\begin{center}
\begin{tikzpicture}[scale=0.75, thick]
\begin{scope}[shift={(7,0)}]
\node (N1) at (0cm,0cm) [draw, circle, inner sep=0.1cm] {$1$};
\draw (N1) circle [radius=1.45cm];
\node (N2) at ({1.5cm*cos(120)},{1.5cm*sin(120)}) [draw, circle, inner sep=0.1cm, fill=white] {$2$};
\node (N3) at ({1.5cm*cos(60)},{1.5cm*sin(60)}) [draw, circle, inner sep=0.1cm, fill=white] {$3$};
\node (N4) at ({1.5cm*cos(-40)},{1.5cm*sin(-40)}) [draw, circle, inner sep=0.1cm, fill=white] {$4$};
\node (N5) at ({1.5cm*cos(220)},{1.5cm*sin(220)}) [draw, circle, inner sep=0.1cm, fill=white] {$5$};
\draw (N1)--(N2);
\draw (N1)--(N3);
\draw (N1)--(N4);
\draw (N1)--(N5);
\node (W) [left= 1.5cm of N1]{$\mathcal{W}_5:$};
\end{scope}
\end{tikzpicture}
\end{center}

\begin{theorem} \label{thm:wheel}
The storage capacity of the wheel graph $\mathcal{W}_N, N \geq 4$ is bounded from above as
\begin{align}
C(\mathcal{W}_N) \leq \min\left( \lambda_1, \lambda_2, ~\frac{\lambda_1 + \lambda_2}{3},~\frac{\lambda_1 + \lambda_2 + \lambda_3}{5},\cdots, \frac{\lambda_1+\lambda_2+\cdots+\lambda_{N-2}}{2N-5}\right),
\label{cap:wheel}
\end{align}
and the upper bound is tight in each of the following cases.
\begin{enumerate}
\item $C(\mathcal{W}_4)=\min\left(\lambda_1,\lambda_2,\frac{\lambda_1+\lambda_2}{3}\right)$.
\item $C(\mathcal{W}_5)=\min\left(\lambda_1,\lambda_2,\frac{\lambda_1+\lambda_2}{3},\frac{\lambda_1+\lambda_2+\lambda_3}{5}\right)$.
\item If $\lambda_2 = \lambda_3 = \cdots = \lambda_N$, then $C(\mathcal{W}_N) = \min(\lambda_1, \lambda_2, \frac{\lambda_1 + (N-3)\lambda_2}{2N-5})$. In particular, for uniform storage, $C_u(\mathcal{W}_N)= \frac{N-2}{2N-5}$.
\end{enumerate}
\end{theorem}

In the upper bound of $C(\mathcal{W}_N)$, the first two terms $\lambda_1, \lambda_2$ follow from the intersection bound (\ref{up:intersection}) while the remaining bounds follow from the wheel bound (\ref{up:wheel}). When the upper bound is achievable, the code construction requires structured  (instead of random) linear codes inspired by interference alignment principles. Whether the upper bound (\ref{cap:wheel}) is tight for $N \geq 6$ is an open problem. 
The case of $N=4$, i.e., the wheel graph $\mathcal{W}_4$ is further illustrated in Section \ref{ex:wheel}. The proof of Theorem \ref{thm:wheel} is presented in Section \ref{sec:wheel_proof}.

\subsubsection{Fano Graph $\mathcal{F}_7$}
The Fano graph, $\mathcal{F}_7=( (\lambda_1, \cdots, \lambda_7),\mathcal{E})$, has $7$ storage nodes and $7$ decoding sets, as follows.
\begin{align}
\mathcal{E}&=\{\{1,2,4\},\{4,5,6\},\{1,3,6\},\{2,6,7\},\{3,4,7\},\{1,5,7\},\{2,3,5\}\}.
\end{align}

\vspace{-0.1in}
\begin{figure}[h]
\begin{center}
\begin{tikzpicture}[scale=0.75, thick]
\node (N5) at (0cm,0cm) [draw, circle, inner sep=0.1cm] {$4$};
\node (N7) at (5cm,0) [draw, circle, inner sep=0.1cm] {$6$};
\node (N1) at ({5cm*cos(60)},{5cm*sin(60)}) [draw, circle, inner sep=0.1cm] {$1$};
\node (N4) at ({(5cm+5cm*cos(60))/3},{5cm*sin(60)/3}) [draw, circle, inner sep=0.1cm] {$7$};
\draw (N4) circle [radius=1.45];
\node (N2) at ($(N1)! 0.5!(N5)$) [draw, circle, inner sep=0.1cm, fill=white] {$2$};
\node (N3) at ($(N1)! 0.5!(N7)$) [draw, circle, inner sep=0.1cm, fill=white] {$3$};
\node (N6) at ($(N5)! 0.5!(N7)$) [draw, circle, inner sep=0.1cm, fill=white] {$5$};
\draw (N1)--(N2)--(N5)--(N6)--(N7)--(N3)--(N1);
\draw (N1)--(N4)--(N6);
\draw (N3)--(N4)--(N5);
\draw (N2)--(N4)--(N7);
\node [right =0cm of N4]{\colorbox{red!10}{\footnotesize $a+b_1+b_2+b_3$}};
\node [right =0cm of N1]{\colorbox{red!10}{\footnotesize $a+b_1$}};
\node [left =0cm of N2]{\colorbox{red!10}{\footnotesize $a+b_2$}};
\node [right =0cm of N3]{\colorbox{red!10}{\footnotesize $a+b_3$}};
\node [below =0cm of N5]{\colorbox{red!10}{\footnotesize $a+b_1+b_2$}};
\node [below =0cm of N6]{\colorbox{red!10}{\footnotesize $a+b_2+b_3$}};
\node [below =0cm of N7]{\colorbox{red!10}{\footnotesize $a+b_1+b_3$}};
\end{tikzpicture}
\end{center}
\vspace{-0.2in}
\caption{The Fano graph is shown. There are $7$ storage nodes and $7$ decoding sets. The decoding sets are the hyperedges corresponding to the $6$ straight lines and the circle. The  solution for the corresponding classical secure storage problem is shown with secret `$a$' and noise `$b_1,b_2,b_3$,' all i.i.d. uniform in $\mathbb{F}_q$ where  $\mathbb{F}_q$ is any finite field with an even characteristic.}\label{fig:Fano}
\end{figure}
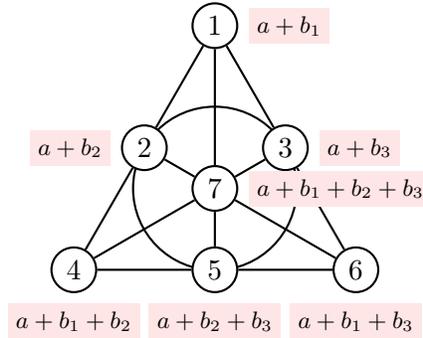

\begin{theorem}
The capacity of the $7$-node Fano  graph $\mathcal{F}_7$ is  $C(\mathcal{F}_7)= 1$. 
\end{theorem}
Note that the capacity does not depend on $\lambda_1,\cdots,\lambda_7$, so the capacity is the same for uniform as well as non-uniform storage. Recall the assumption  \eqref{eq:lambdamin1}  that $\min_{n\in[7]}\lambda_n=1$.

\proof  Without loss of generality, suppose $\min_{n\in[7]}\lambda_n=\lambda_1=1$. The converse follows from the intersection bound (\ref{up:intersection}) as $$C(\mathcal{F}_7) \leq \Lambda( \{Q_1,Q_2,Q_4\} \cap \{Q_1,Q_3,Q_6\}) = \lambda_1 = 1.$$ 

The achievability follows from Theorem \ref{thm:css}. Since reducing the sizes of quantum systems cannot increase the storage capacity, it suffices to show achievability for $\lambda_1=\lambda_2=\cdots=\lambda_7=1$. The corresponding classical secure storage code is shown in Figure \ref{fig:Fano} where $k = \kappa = \lambda_1 = \cdots = \lambda_7 = 1$. It is not difficult to verify that the decoding constraint (\ref{css:dec}) and the security constraint (\ref{css:sec}) hold over any finite field $\mathbb{F}_q$ with even characteristic. For example, consider the decoding set $\{Q_4,Q_5,Q_6\}$ for which the corresponding classical decoding constraint is satisfied because of the following equation,
$$a = (a+b_1+b_2) + (a+b_2+b_3)  + (a+b_1+b_3),$$ 
which holds only over the binary (extension) field. Similarly, consider the classical security condition corresponding to the decoding set $\{Q_4,Q_5,Q_6\}$. Its complement is the set $\{Q_1,Q_2,Q_3, Q_7\}$, and the corresponding classical storage $a+b_1, a+b_2, a+b_3, a+b_1+b_2+b_3$ reveals nothing about the secret $a$, i.e.,
\begin{align}
&I(a; a+b_1, a+b_2, a+b_3, a+b_1+b_2+b_3)\notag\\
&=I(a;a+b_1,a+b_2,a+b_3)+I(a; a+b_1+b_2+b_3\mid a+b_1, a+b_2, a+b_3)\\
&=0+H(a+b_1+b_2+b_3\mid a+b_1, a+b_2, a+b_3),\\
&=0,\label{eq:useeven}
\end{align}
 only if $\mathbb{F}_q$ has an even characteristic. This is important in the last step \eqref{eq:useeven}, because over a field of even characteristic, $a+b_1+b_2+b_3$ is determined by (is the sum of) $a+b_1, a+b_2, a+b_3$. Thus the classical code has characteristic dependent alignment structure. 
 
\hfill\qed

\subsubsection{Intersection Graph $\sqcap_{\Delta,m}$}
Defined for $\Delta > m$, the intersection graph $\sqcap_{\Delta,m} = ( (\lambda_1, \cdots, \lambda_N), \mathcal{E} )$, has $N = \binom{\Delta}{m}$ storage nodes and $\Delta$ decoding sets. Let $\pi:\binom{[\Delta]}{m}\rightarrow[N]=\{1,2,\cdots,\binom{\Delta}{m}\}$ be an aribtrary bijection, that determines the labeling of the storage nodes. For any set $\mathcal{S}\in\binom{[\Delta]}{m}$, define the compact notation $\overline{\mathcal{S}}\triangleq \pi(\mathcal{S})$. Thus, the set of storage nodes is identified as, 
\begin{align}
\mathcal{Q}(\sqcap_{\Delta,m}) = \left\{Q_{\overline{\mathcal{S}}} : \mathcal{S} \in \binom{[\Delta]}{m} \right\}.
\end{align} 
We will refer to $\mathcal{S}\in\binom{[\Delta]}{m}$ as the `label' of the quantum storage node $Q_{\overline{\mathcal{S}}}$. 
Define 
\begin{align}
e_i&=\left\{\overline{\mathcal{S}}: i \in \mathcal{S},~~ \mathcal{S}\in \binom{[\Delta]}{m}\right\},~~~~~~i\in[\Delta],\\
\mathcal{E}&=\{e_1,e_2,\cdots, e_\Delta\},\\
\intertext{so that for each $e_i$, $i\in[\Delta]$, the decoding set $\mathcal{D}(e_i)$ is defined as}
\mathcal{D}(e_i) &=\left\{Q_n: n\in e_i\right\}= \left\{Q_{\overline{\mathcal{S}}}: i \in \mathcal{S}, \mathcal{S} \in \binom{[\Delta]}{m}  \right\}.
\end{align}
In  words, the decoding set $\mathcal{D}(e_i)$ contains all $\binom{\Delta-1}{m-1}$ storage nodes $Q_{\overline{\mathcal{S}}}$ whose label $\mathcal{S}$ contains $i$. Note that the intersection of any two decoding sets, $\mathcal{D}(e_i), \mathcal{D}(e_j)$  comprises all $\binom{\Delta-2}{m-2}$ storage nodes whose labels contain both $i$ and $j$.

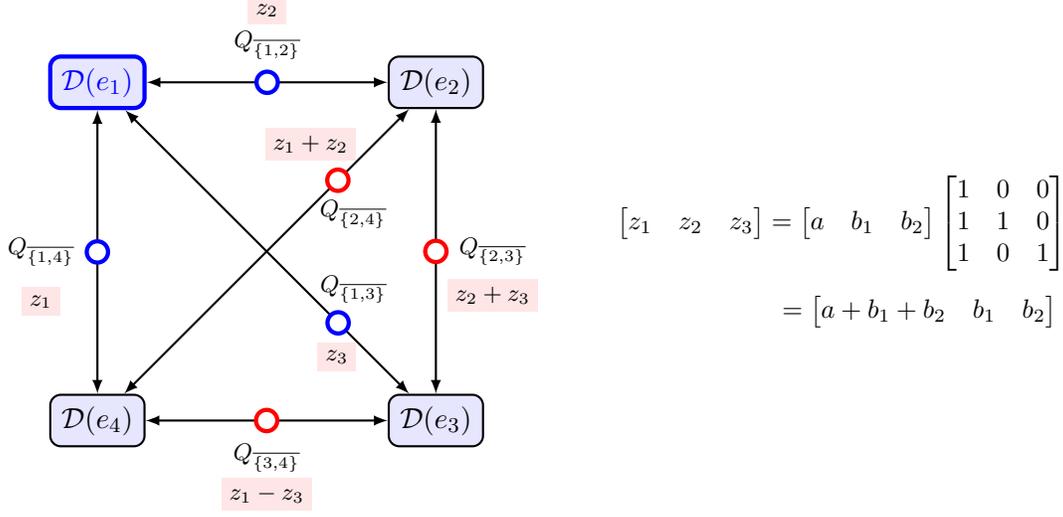
\begin{figure}[h]
\center
\begin{tikzpicture}[scale=0.9,thick]
\node (name1) at (11cm,2.5cm)[align=left]{\small $\begin{bmatrix}z_1&z_2&z_3\end{bmatrix}=\begin{bmatrix}a&b_1&b_2\end{bmatrix}
\begin{bmatrix}
1 & 0 & 0 \\
1 & 1 & 0 \\
1 & 0 & 1 
\end{bmatrix}$\\[0.3cm]
\small ~~~~~~~~~~~~~~~~~~~~~~~~ $= \begin{bmatrix} a+b_1+b_2 & b_1 & b_2 \end{bmatrix}$
};
\node (N247) at (0cm,5cm) [draw, ultra thick, blue, fill=blue!10, rectangle, rounded corners, inner sep=0.15cm] { $\mathcal{D}(e_1)$};
\node (N125) at (0cm,0cm) [draw, fill=blue!10, rectangle, rounded corners,  inner sep=0.15cm] { $\mathcal{D}(e_4)$};

\node (N345) at (5cm,0cm) [draw, fill=blue!10, rectangle, rounded corners, inner sep=0.15cm] { $\mathcal{D}(e_3)$};
\node (N137) at (5cm,5cm) [draw, fill=blue!10, rectangle, rounded corners,  inner sep=0.15cm] { $\mathcal{D}(e_2)$};

\draw [latex-latex](N137)--(N247) node (N7) [midway, circle, draw, blue, ultra thick, fill=white, inner sep=0.1cm]{};
\node (N7label)[above=0cm of N7]{\footnotesize $Q_{\overline{\{1,2\}}}$};
\node [above =-0.2cm of N7label]{\colorbox{red!10}{\footnotesize $z_2$}};

\draw [latex-latex](N125)--(N247) node (N2) [midway, circle, draw, ultra thick, blue, fill=white,  inner sep=0.1cm]{};
\node (N2label) [left =0cm of N2]{\footnotesize $Q_{\overline{\{1,4\}}}$};
\node [below =0cm of N2label]{\colorbox{red!10}{\footnotesize $z_1$}};

\draw [latex-latex](N345)--(N247) node (N4) [pos=0.25, circle, draw, blue, fill=white, ultra thick, inner sep=0.1cm]{};
\node [below left = 0cm and -0.5cm of N4]{\colorbox{red!10}{\footnotesize $z_3$}};
\node [above right = 0cm and -0.5cm of N4]{\footnotesize $Q_{\overline{\{1,3\}}}$};

\draw [latex-latex](N125)--(N345) node (N5) [pos=0.5, circle, draw, red, fill=white, ultra thick, inner sep=0.1cm]{};
\node (N5label)[below =0cm of N5]{\footnotesize $Q_{\overline{\{3,4\}}}$};
\node [below =-0.2cm of N5label]{\colorbox{red!10}{\footnotesize $z_1-z_3$}};

\draw [latex-latex](N125)--(N137) node (N1) [pos=0.75, circle, draw, red, fill=white, ultra thick, inner sep=0.1cm]{};
\node [below right =0cm and -0.5cm of N1]{\footnotesize $Q_{\overline{\{2,4\}}}$};
\node [above left =0cm and -0.5cm of N1]{\colorbox{red!10}{\footnotesize $z_1+z_2$}};

\draw [latex-latex](N137)--(N345) node (N3) [pos=0.5, circle, draw, red, fill=white, ultra thick, inner sep=0.1cm]{};
\node (N3label) [right =0cm of N3]{\footnotesize $Q_{\overline{\{2,3\}}}$};
\node [below =-0.1cm of N3label]{\colorbox{red!10}{\footnotesize $z_2+z_3$}};
\end{tikzpicture}
\caption{The intersection graph $\sqcap_{4,2}$ is shown. Storage nodes $\{Q_{\overline{\mathcal{S}}}\}$ appear as empty circles, decoding sets $\{\mathcal{D}(e_i)\}$ as filled rectangles. Arrows from storage nodes to decoding sets indicate membership of the decoding set. Decoding set $\mathcal{D}(e_1)$ and its participating nodes are shown in blue, and the nodes in the complement $\mathcal{D}^c(e_1)$ are shown in red. Also shown is the optimal classical secure storage code over any finite field $\mathbb{F}_q$,  with secret $(a\in\mathbb{F}_q)$ and noise $(b_1,b_2\in\mathbb{F}_q)$, respectively. The optimal classical code translates into an optimal quantum code via Theorem \ref{thm:css}. 
}\label{fig:int42}
\end{figure}
 
For intersection graphs we consider only uniform storage, as this already leads to non-trivial code structures. The intersection graph $\sqcap_{\Delta,m}$ is interesting because on the one hand, the storage structure is `smooth,' thus somewhat homogeneous. Here by smoothness we mean that any storage node belongs to {\it exactly} $m$ decoding sets and all storage nodes have the same size.
On the other hand, despite this smoothness of the intersection graph, its capacity characterization involves non-trivial structured alignment of the code spaces. Figure \ref{fig:int42} illustrates the intersection graph $\sqcap_{4,2}$ along with an optimal solution to the corresponding classical secure storage problem. 

We are able to characterize the uniform storage capacity of intersection graphs in the following theorem.

\begin{theorem}\label{thm:int}
The capacity of the intersection graph $\sqcap_{\Delta,m}$ with uniform storage is,  
\begin{align}
C_u(\sqcap_{\Delta,m})=\binom{\Delta-2}{m-2}.
\end{align}
\end{theorem}

The converse follows from the intersection bound (\ref{up:intersection}). The achievability relies on a rather delicate \emph{alignment} based classical secure storage code, in conjunction with Theorem \ref{thm:css}.  The proof of Theorem \ref{thm:int} is presented in Section \ref{sec:int}. 

As noted, optimal solutions to intersection graphs involve non-trivial alignment structures. Let us briefly preview this aspect. Consider the simple example of the  intersection graph $\sqcap_{4,2}$ in Figure \ref{fig:int42}. Note that the classical secret $(a)$ and noise $(b_1,b_2)$ symbols are first linearly `\emph{precoded}' into  the symbols $(z_1,z_2,z_3)$ and then the coding is performed on the $z_i$ symbols. It is not difficult to verify that with the `precoding' from the secret and noise symbols to $z_i$ symbols, shown explicitly in Figure \ref{fig:int42}, and the storage code over $z_i$ shown in the graph, all decoding and security constraints are satisfied. It is also worthwhile to note that the precoding aspect is not particularly interesting from a Shannon theoretic perspective. This is because, as it turns out, the precoding requires no special structure over  large $q$. Indeed, for large $q$, almost any random choice of the linear precoding matrix will turn out to be sufficient. The critical aspect of the solution is the required non-trivial alignment structure, which is revealed in the next step following the precoding, i.e., in the coding that is performed over the $z_i$ symbols.

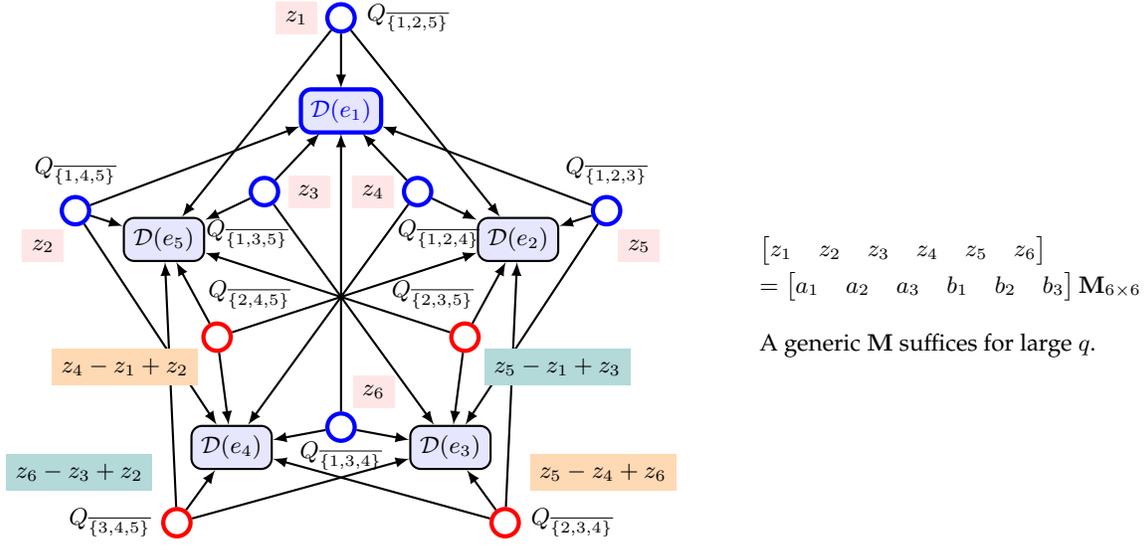
\begin{figure}[!t]
\center
\begin{tikzpicture}[scale=0.9]
\node (name1) at (9cm,0cm)[align=left, black]
{\footnotesize $\begin{bmatrix}z_1&z_2&z_3&z_4&z_5&z_6\end{bmatrix}$\\
\footnotesize $=\begin{bmatrix}a_1&a_2&a_3&b_1&b_2&b_3\end{bmatrix}{\bf M}_{6\times 6}$\\[0.3cm]
\footnotesize A generic ${\bf M}$ suffices for large $q$.
};

\def\rad{2.75}
\foreach \i in {1,...,5}
	{\node (D\i) at ({\rad*cos(deg(pi/2-(\i-1)*2*pi/5))},{\rad*sin(deg(pi/2-(\i-1)*2*pi/5))}) [draw, black, thick, rectangle, rounded corners, fill=blue!10]{\footnotesize $\mathcal{D}(e_\i)$};

\ifthenelse{\i = 3 \OR \i=4}
	{\node (Q\i) at ({1.5*\rad*cos(deg(pi/2-(\i-1)*2*pi/5))},{1.5*\rad*sin(deg(pi/2-(\i-1)*2*pi/5))}) [draw, red, ultra thick, circle, fill=white]{};}
	{\node (Q\i) at ({1.5*\rad*cos(deg(pi/2-(\i-1)*2*pi/5))},{1.5*\rad*sin(deg(pi/2-(\i-1)*2*pi/5))}) [draw, blue, ultra thick, circle, fill=white]{};}
	
	\ifthenelse{\i = 2 \OR \i=4}
		{\node (Qin\i) at ({0.7*\rad*cos(deg(pi/2-pi/5-(\i-1)*2*pi/5))},{0.7*\rad*sin(deg(pi/2-pi/5-(\i-1)*2*pi/5))}) [draw, red, ultra thick, circle, fill=white]{};}
		{\node (Qin\i) at ({0.7*\rad*cos(deg(pi/2-pi/5-(\i-1)*2*pi/5))},{0.7*\rad*sin(deg(pi/2-pi/5-(\i-1)*2*pi/5))}) [draw, blue, ultra thick, circle, fill=white]{};}
		
	}
\node at ({\rad*cos(deg(pi/2-(1-1)*2*pi/5))},{\rad*sin(deg(pi/2-(1-1)*2*pi/5))}) [draw, blue, ultra thick, rectangle, rounded corners, fill=blue!10]{\footnotesize $\mathcal{D}(e_1)$};	
	
\foreach \i/\j/\k in {1/2/5,2/1/3,3/2/4,4/3/5,5/4/1}
	{\draw[thick,black,-latex] (Q\i)--(D\i);
	\draw[thick,black,-latex] (Q\i)--(D\j);
	\draw[thick,black,-latex] (Q\i)--(D\k);}
\foreach \i/\j/\k in {1/2/4, 2/3/5, 3/4/1, 4/5/2, 5/1/3}
	{\draw[thick,black,-latex] (Qin\i)--(D\i);
	\draw[thick,black,-latex] (Qin\i)--(D\j);
	\draw[thick,black,-latex] (Qin\i)--(D\k);}	
	
\node [right =0cm of Q1] {\footnotesize $Q_{\overline{\{1,2,5\}}}$};
\node [left =0.1cm of Q1, fill=red!10] {\footnotesize $z_1$};
\node [above =0cm of Q2] {\footnotesize $Q_{\overline{\{1,2,3\}}}$};
\node [below right =0.1cm and -0cm of Q2, fill=red!10] {\footnotesize $z_5$};
\node [above =0cm of Q5] {\footnotesize $Q_{\overline{\{1,4,5\}}}$};
\node [below left =0.1cm and -0cm of Q5, fill=red!10] {\footnotesize $z_2$};

\node (Q4label) [left =0cm of Q4] {\footnotesize $Q_{\overline{\{3,4,5\}}}$};
\node [above left = 0.1cm and -1.25cm of  Q4label, fill=teal!30]{\footnotesize $z_6-z_3+z_2$};

\node (Q3label) [right =0cm of Q3] {\footnotesize $Q_{\overline{\{2,3,4\}}}$};
\node [above right = 0.1cm and -1.25cm of  Q3label, fill=orange!30]{\footnotesize $z_5-z_4+z_6$};

\node [below right =0.1cm and -0.55cm of Qin1] {\footnotesize $Q_{\overline{\{1,2,4\}}}$};
\node [left=0.1cm of  Qin1, fill=red!10]{\footnotesize $z_4$};

\node [above left =0.1cm and -0.4cm of Qin2] {\footnotesize $Q_{\overline{\{2,3,5\}}}$};
\node [below right=-0cm and 0.1cm of  Qin2, fill=teal!30]{\footnotesize $z_5-z_1+z_3$};

\node [above right =0.1cm and -0.4cm of Qin4] {\footnotesize $Q_{\overline{\{2,4,5\}}}$};
\node [below left=-0cm and 0.1cm of  Qin4, fill=orange!30]{\footnotesize $z_4-z_1+z_2$};

\node [below left =0.1cm and -0.6cm of Qin5] {\footnotesize $Q_{\overline{\{1,3,5\}}}$};
\node [right=0.1cm of  Qin5, fill=red!10]{\footnotesize $z_3$};
\node [below=-0.15cm of Qin3] {\footnotesize $Q_{\overline{\{1,3,4\}}}$};
\node [above right = 0.1cm and 0cm of  Qin3, fill=red!10]{\footnotesize $z_6$};
\end{tikzpicture}
\caption{The intersection graph $\sqcap_{5,3}$ is illustrated, along with an optimal solution to the corresponding classical secure storage problem.}
\label{fig:intgraphs}
\end{figure}

 A bit more insightful example  is illustrated in Figure \ref{fig:intgraphs} which illustrates the intersection graph $\sqcap_{5,3}$ along with the solution to its corresponding secure storage problem. Leaving a more detailed discussion of this example to Section \ref{ex:int}, let us note here that the solution shown in Figure \ref{fig:intgraphs} assumes large $q$ so that generic precoding suffices. Recall that alphabet sizes are not constrained under the Shannon theoretic formulation. However, finding optimal solutions over smaller finite fields may be an interesting open problem from a coding theoretic perspective.

An intuitive sketch of the solution to the classical secure storage problem for intersection graphs is as follows. A generic precoding step maps $k=\binom{\Delta-2}{m-2}$ (capacity of the code, $\kappa=1$) secrets $(a_1,\cdots, a_k)$ along with $n-k$ noise symbols $(b_1,\cdots, b_{n-k})$ to the $n=\binom{\Delta-1}{m-1}$ (size of a decoding set) symbols $z_1,\cdots,z_n$. A classical code over these $z_i$ symbols determines the storage associated with each node such that the nodes in any decoding set can recover all $n$ of $z_i$ symbols. To guarantee security it is important to ensure that the $N-n$ nodes outside the decoding set  remain secured by the $n-k$ noise dimensions. This is a challenging requirement because  the complement of a decoding set, $\mathcal{D}^c(e_i)$ has $N-n=\binom{\Delta-1}{m}$ nodes, which is \emph{greater} than the number of noise symbols $n-k=\binom{\Delta-2}{m-1}$. Yet, in the optimal solution, any $\mathcal{D}^c(e_i)$ leaks no information about the secrets $a_1,\cdots, a_k$, precisely due to the  crucial \emph{alignment} structure of the classical code. For example in Figure \ref{fig:intgraphs}, $\mathcal{D}^c(e_1)$ contains $4$ nodes (the red nodes), and there are only $3$ independent noise symbols $(b_1,b_2,b_3)$. Yet, the code is able to satisfy the security requirement because only $3$ of the $4$ symbols in $\mathcal{D}^c(e_1)$ carry independent information, whereas the $4^{th}$ symbol is simply a linear combination of the other $3$. In other words,  the information contained in the $4^{th}$ symbol appears along a dimension that \emph{aligns} within the $3$ dimensions already occupied by the other $3$ symbols, and the $3$ independent noise symbols suffice to protect those $3$ information dimensions from leaking any secrets. Further details are left to  Section \ref{ex:int}.

\subsection{Maximal Storage Graphs}
In this section, we briefly explore the maximality of storage graphs. Strongly maximal storage graphs are completely characterized in the following theorem.

\begin{theorem}\label{thm:strong}
A storage graph $\mathcal{G}=((\lambda_1, \cdots, \lambda_N),\mathcal{E})$ is strongly maximal if and only if for every $\mathcal{S} \subset \mathcal{Q}(\mathcal{G}) = \{Q_1,\cdots, Q_N\}$, there exists at least one $\mathcal{D}(e), e \in\mathcal{E}$ 
that is either completely included or completely excluded by $\mathcal{S}$, i.e., $\mathcal{S} \cap \mathcal{D}(e) \in\{\emptyset, \mathcal{D}(e)\}$. 
\end{theorem}

Theorem \ref{thm:strong} is proved through a more refined treatment of the `if and only if' condition for $C(\mathcal{G}) > 0$ in Corollary \ref{thm:small}. The details are presented in Section \ref{sec:strong}.

\begin{remark}\label{cor:uniform}
If we replace the non-uniform storage capacity $C(\mathcal{G})$ by the uniform storage capacity $C_u(\mathcal{G})$ in the  definition of strongly maximal storage graph, the `if and only if' condition in Theorem \ref{thm:strong} remains the same. This is seen by checking that every step in the proof of Theorem \ref{thm:strong} (see Section \ref{sec:strong}) holds for $C_u(\mathcal{G})$ as well, so the same proof applies.
\end{remark}

Let us characterize the maximality of the storage graphs that we have considered.
\begin{corollary} \label{cor:max}
The following maximality properties are noted.
\begin{enumerate}
\item The MDS graph $\mathcal{M}_{N,K}, 2K > N$ is weakly maximal for all $N,K$, and strongly maximal iff $2K-N=1$.
\item The wheel graph $\mathcal{W}_N, N \geq 4$ is strongly maximal. 
\item The Fano graph $\mathcal{F}_7$ is strongly maximal. 
\item The intersection graph $\sqcap_{\Delta,m}, \Delta > m$ is strongly (weakly) maximal iff $\Delta=3,m=2$.
\end{enumerate}
\end{corollary}

For strong maximality, we simply apply the `if and only if' condition in Theorem \ref{thm:strong}. Weak maximality requires  more specialized treatment (e.g., carefully chosen additional decoding sets to test if the capacity decreases). The proof of Corollary \ref{cor:max} is presented in Section \ref{sec:max}.

\section{Examples}
\subsection{MDS Graph $\mathcal{M}_{4,3}$} \label{ex:mds}
The converse follows immediately from the intersection bound (\ref{up:intersection}). Let us consider the achievability. First, consider the uniform storage case. Let us start from the MDS graph $\mathcal{M}_{3,2}$ and show that $C_u(\mathcal{M}_{3,2}) = 1$ is achievable, which will be used in characterizing $C(\mathcal{M}_{4,3})$. We set
\begin{eqnarray}
Y_{Q_1} = b, Y_{Q_2} = a + b, Y_{Q_3} = 2a + b,
\end{eqnarray}
where $a, b$ are i.i.d. uniform elements in $\mathbb{F}_3$, so from any two $Y_Q$, we may recover $a$ and from any single $Y_Q$, nothing is revealed about $a$. Noting that $k = \kappa = 1, n = 3, \lambda_1 = \lambda_2 = \lambda_3 = 1$, we  apply Theorem \ref{thm:css} to obtain the desired quantum code. Thus, $C_u(\mathcal{M}_{3,2}) \geq 1$.

Next we proceed to $\mathcal{M}_{4,3}$ and show that $C_u(\mathcal{M}_{4,3}) = 2$ is achievable. We set
\begin{eqnarray}
Y_{Q_1} = b, Y_{Q_2} = a_1 + b, Y_{Q_3} = a_2 + b, Y_{Q_4} = a_1 + a_2 + b,
\end{eqnarray}
where $a_1, a_2, b$ are i.i.d. uniform elements in $\mathbb{F}_3$. Note that from any three $Y_Q$, we can recover $\bm{a} = (a_1, a_2)$ and from any single $Y_Q$, nothing is revealed about $\bm{a}=(a_1,a_2)$. Noting that $k = 2, \kappa = 1$, and applying Theorem \ref{thm:css}, produces the quantum code. Thus,  $C_u(\mathcal{M}_{4,3}) \geq 2$.

Next, consider the non-uniform storage case. We show that $C(\mathcal{M}_{4,3}) = \lambda_1 + \lambda_2$ is achievable. 
Assume $\lambda_1, \lambda_2 $ are positive integers (the case of arbitrary real numbers will be treated in the general proof in Section \ref{sec:mds}).
As $C_u(\mathcal{M}_{4,3}) \geq 2$,  
we are able to store a quantum message $Q_0(1)$ of size $\lambda_0(1)=2 \lambda_1 $ qudits in $\mathcal{M}_{4,3}$ while utilizing $\lambda_1$ qudits of storage from each $Q_i, i \in [4]$. Now there remain unused $0$ qudits in $Q_1$, $\lambda_2 - \lambda_1$ qudits in $Q_2$, $\lambda_3 - \lambda_1$ qudits in $Q_3$, $\lambda_4 - \lambda_1$ qudits in $Q_4$. Recall that $\lambda_1 \leq \lambda_2 \leq \lambda_3 \leq \lambda_4$, and the storage in $Q_1$ is fully utilized. We  now view $Q_2, Q_3, Q_4$ as an $\mathcal{M}_{3,2}$ MDS graph in order to store another quantum message $Q_0(2)$. It is important to note that the original $\mathcal{M}_{4,3}$ MDS constraint with storage node set $\{Q_1, Q_2, Q_3, Q_4\}$ ensures that any $3$ surviving (unerased) $Q_i$ from $Q_1, Q_2, Q_3, Q_4$ will contain at least two of $Q_2, Q_3, Q_4$, and the $\mathcal{M}_{3,2}$ MDS graph with storage node set $\{Q_2, Q_3, Q_4\}$ guarantees the quantum message $Q_0(2)$ can be perfectly recovered. 
As the uniform storage capacity $C_u(\mathcal{M}_{3,2}) \geq 1$, we are able to store the quantum message $Q_0(2)$ of size $\lambda_0(2)=\lambda_2 - \lambda_1$ qudits in $\mathcal{M}_{3,2}$ by utilizing $\lambda_2 - \lambda_1$ qudits of storage from each of $Q_2, Q_3, Q_4$. After this step, $Q_2$ is fully utilized, $Q_3$ is left with $\lambda_3 - \lambda_2$ qudits, and $Q_4$ is left with $\lambda_4 - \lambda_2$ qudits. These remaining qudits are not used. Overall we have stored quantum message $Q_0=Q_0(1)Q_0(2)$ of size $\lambda_0 = \lambda_0(1)+\lambda_0(2)=(2\lambda_1) + (\lambda_2 - \lambda_1) = \lambda_1 + \lambda_2$ qudits over the MDS graph $\mathcal{M}_{4,3}$ where each storage node $Q_i, i \in [4]$ has size $\lambda_i$ qudits. Thus, we have shown $C(\mathcal{M}_{4,3}) \geq \lambda_1 + \lambda_2$, as desired.

\subsection{Wheel Graph $\mathcal{W}_4$}\label{ex:wheel}
We show that $C(\mathcal{W}_4) = \min\left( \lambda_1, \lambda_2 , \frac{\lambda_1 + \lambda_2}{3} \right)$. From this $C_u(\mathcal{W}_4) = \frac{2}{3}$ follows because  uniform storage corresponds to $\lambda_1 = \lambda_2 = 1$. The converse and achievability proofs are presented sequentially.

\paragraph {\it Converse:} 
Since the problem formulation \eqref{eq:allstates} requires perfect recovery for any pure state of $RQ_0$, for the converse bound let us consider the case where $Q_0$ is maximally entangled with the reference $R$. As every CPTP map has an isometric (unitary) extension, let us assume without loss of generality that the encoder is a \emph{unitary} transformation that maps $Q_0$ and ancilla qudits to $Q_1Q_2Q_3Q_4Q_5$. The artificial system $Q_5$ is included as the purifying system. This allows the possibility that $RQ_1Q_2Q_3Q_4$ is mixed, while $RQ_1Q_2Q_3Q_4Q_5$ is pure. Note that $Q_5$ is not included in any decoding set.  

Since $RQ_1Q_2Q_3Q_4Q_5$ is pure, and $I(R; Q_3,Q_4,Q_5)=0$ (because $\{Q_1,Q_2\}$ is a decoding set, refer to (\ref{eq:sec}) for a proof), we have
\begin{align}
H(R,Q_3,Q_4,Q_5)&=H(Q_1,Q_2)\label{step:5}\\
\overset{(\ref{eq:sec})}{\implies} H(R)+H(Q_3,Q_4,Q_5)&\leq H(Q_1)+H(Q_2).\label{step:6}\\
\intertext{Similarly,}
H(R)+H(Q_2,Q_4,Q_5)&\leq H(Q_1)+H(Q_3) \label{step:7}\\
\overset{(\ref{step:6})+(\ref{step:7})}{\implies} 2H(R)+H(Q_2,Q_3,Q_4,Q_5)+H(Q_4,Q_5)&\leq 2H(Q_1)+H(Q_2)+H(Q_3) \label{step:8}\\
\implies 2H(R)+H(R,Q_1)+H(Q_4)+H(Q_5\mid Q_4)&\leq 2H(Q_1)+H(Q_2)+H(Q_3) \label{step:9}\\
\implies 3H(R)+H(Q_4)+H(Q_5\mid Q_4)&\leq H(Q_1)+H(Q_2)+H(Q_3), \label{step:10}\\
\intertext{and similarly, } 
3H(R)+H(Q_3)+H(Q_5\mid Q_3)&\leq H(Q_1)+H(Q_2)+H(Q_4). \label{step:11}
\end{align}
Here (\ref{step:7}) follows symmetrically from (\ref{step:6}) as $\{Q_1, Q_3\}$ is also a decoding set; strong sub-additivity is used to obtain (\ref{step:8}); (\ref{step:9}) follows from the fact that $RQ_1Q_2Q_3Q_4Q_5$ is pure so that $H(Q_2,Q_3,Q_4,Q_5)$ $= H(R,Q_1)$; (\ref{step:10}) is due to the fact that $H(R, Q_1) = H(R) + H(Q_1)$ as $\{Q_2, Q_3, Q_4\}$ is a decoding set so that from (\ref{eq:sec}), $I(R; Q_1)$ = 0; and (\ref{step:11}) follows similarly from (\ref{step:10}) as we may switch $Q_3$ and $Q_4$. 

Adding (\ref{step:10}), (\ref{step:11}) and applying weak monotonicity, $H(A\mid B)+H(A\mid C)\geq 0$ with $(A,B,C)=(Q_5,Q_4,Q_3)$, we obtain the following bound (where $Q_5$ disappears).
\begin{eqnarray}
6H(R)+H(Q_3)+H(Q_4) &\leq& 2H(Q_1)+2H(Q_2)+H(Q_3)+H(Q_4)\\
\implies H(R) &\leq& \frac{H(Q_1)+H(Q_2)}{3} \label{eq:stancon}\\
\implies  \kappa \lambda_0 = H(R)\leq \frac{H(Q_1)+H(Q_2)}{3}  &\leq& \frac{\log_q |Q_1| + \log_q |Q_2|}{3} = \kappa \frac{\lambda_1 + \lambda_2}{3} \\
\implies C(\mathcal{W}_4) = \sup \lambda_0 &\leq& \frac{\lambda_1 + \lambda_2}{3}.
\end{eqnarray}

\paragraph{\it Achievability:} 
We first present two component codes that will be used later.
\begin{enumerate}
\item $\lambda_0=1$ code for storage graph $\mathcal{W}_4$ with $(\lambda_1, \lambda_2, \lambda_3, \lambda_4) = (2, 1, 1, 1)$.  

Let $\kappa = 1$ and $q > 9$ be any prime number. Use the following classical  secure storage code.
\begin{eqnarray}
&& Y_{Q_1} = (b_1, b_2), ~Y_{Q_2} = a + b_1 + b_2, \\
&& Y_{Q_3} = a + 2b_1 + 3b_2, ~Y_{Q_4} = a + 4b_1 + 9b_2,
\end{eqnarray}
where $a, b_1, b_2$ are i.i.d. uniform elements in $\mathbb{F}_{q}$. It is readily verified that the decoding constraint (\ref{css:dec}) and the security constraint (\ref{css:sec}) are satisfied, so Theorem \ref{thm:css} produces the claimed quantum code.

\item $\lambda_0=1$ code  for storage graph $\mathcal{W}_4$ with $(\lambda_1, \lambda_2, \lambda_3, \lambda_4) = (1, 2, 2, 2)$.

Let $\kappa = 1$ and $q$ be any prime. Use the following classical (structured) secure storage code.
\begin{eqnarray}
&& Y_{Q_1} = b_1, ~Y_{Q_2} = (a+b_1, b_2), \label{ex:code1} \\
&& Y_{Q_3} = (a+b_1, b_3), ~Y_{Q_4} = (a+b_1, a+b_2+b_3), \label{ex:code2} 
\end{eqnarray}
where $a, b_1, b_2, b_3$ are i.i.d. uniform elements in $\mathbb{F}_q$. It is straightforward to verify (\ref{css:dec}) and  (\ref{css:sec}) hold. An interesting aspect of this code is that $Y_{Q_2}, Y_{Q_3}, Y_{Q_4}$ contain the same element $a+b_1$, inspired by noise alignment (same noise as in $Y_{Q_1}$) and signal alignment principles \cite{Li_Sun_CDS, Li_Sun_CDS2, Li_Sun_Extremal, Li_Sun_ExtremalS}.
\end{enumerate}

Next we show how to use the above two component codes to achieve $C(\mathcal{W}_4)$ $=$ $\min\left( \lambda_1, \lambda_2 , \frac{\lambda_1 + \lambda_2}{3} \right)$. There are three cases. Choose a prime $q>9$ so that both component codes are defined. Since rationals are dense over the reals, it suffices to consider rational $\lambda_1, \lambda_2$. Moreover, since  the scaling factor $\kappa$ can be chosen freely, there is no loss of generality in the assumption that $\lambda_1, \lambda_2$, and the number of qudits used below (i.e., $\lambda_0(1) = (2\lambda_1 - \lambda_2)/3$ in (\ref{eq:qd}), $\lambda_0(2) = (2\lambda_2-\lambda_1)/3$ in (\ref{eq:qdd})) are \emph{integers}.
\begin{enumerate}
\item When $\lambda_1 \leq \lambda_2/2$, $\min\left( \lambda_1, \lambda_2 , \frac{\lambda_1 + \lambda_2}{3} \right) = \lambda_1$. 
Use the second component code to store $\lambda_1$ qudits of $Q_0$ in ($Q_1, Q_2, Q_3, Q_4$) while utilizing storage in the amounts $(\lambda_1, 2\lambda_1, 2\lambda_1, 2\lambda_1)$ qudits, respectively. Let us verify that the number of qudits utilized in $Q_i$ is  not greater than the size of $Q_i$. This is so because $\lambda_1 \leq \lambda_2/2$ and $\lambda_2 \leq \lambda_3 \leq \lambda_4$. So the scheme works.
\item When $\lambda_2/2 \leq \lambda_1 \leq 2\lambda_2$, $\min\left( \lambda_1, \lambda_2 , \frac{\lambda_1 + \lambda_2}{3} \right) = \frac{\lambda_1 + \lambda_2}{3}$.
\begin{align}
& \mbox{Use the first component code to store $\lambda_0(1) = (2\lambda_1 - \lambda_2)/3 \geq 0$ qudits of $Q_0$} \notag\\
& ~~~~\mbox{in $(Q_1, Q_2, Q_3, Q_4)$ by utilizing $(2\lambda_0(1), \lambda_0(1), \lambda_0(1), \lambda_0(1))$ qudits of storage.} \label{eq:qd} \\
& \mbox{Use the second component code to store $\lambda_0(2) = (2\lambda_2-\lambda_1)/3 \geq 0$ qudits of $Q_0$} \notag\\
& ~~~~\mbox{in $(Q_1, Q_2, Q_3, Q_4)$ by utilizing $(\lambda_0(2), 2\lambda_0(2), 2\lambda_0(2), 2\lambda_0(2))$ qudits of storage.}\label{eq:qdd}
\end{align}
In total, we have stored $\lambda_0 = \lambda_0(1) + \lambda_0(2) = (\lambda_1 + \lambda_2)/3$ qudits of $Q_0$ in $(Q_1, Q_2, Q_3, Q_4)$ while utilizing $(2\lambda_0(1) + \lambda_0(2), \lambda_0(1) + 2\lambda_0(2), \lambda_0(1) + 2\lambda_0(2), \lambda_0(1) + 2\lambda_0(2))$ qudits of storage. We verify that the amount of storage utilized by the scheme from the storage node $Q_i$ is not greater than the size of $Q_i$, because,
\begin{align}
2\lambda_0(1) + \lambda_0(2) &= \lambda_1, \\
\lambda_0(1) + 2\lambda_0(2) &= \lambda_2 \leq \lambda_3 \leq \lambda_4.
\end{align}
So the scheme works.
\item When $2\lambda_2 \leq \lambda_1$, $\min\left( \lambda_1, \lambda_2 , \frac{\lambda_1 + \lambda_2}{3} \right) = \lambda_2$. 
Use the first component code to store $\lambda_2$ qudits of $Q_0$ in $(Q_1, Q_2, Q_3, Q_4)$ by utilizing storage in the amounts $(2\lambda_2, \lambda_2, \lambda_2, \lambda_2)$ qudits, respectively. The utilized storage in $Q_i$ is not greater than the size of $Q_i$ because $2\lambda_2 \leq \lambda_1$ and $\lambda_2 \leq \lambda_3 \leq \lambda_4$. So the scheme works.
\end{enumerate}

\subsubsection{From Classical to Quantum: Illustration of Theorem \ref{thm:css}}
We take the classical code (\ref{ex:code1}), (\ref{ex:code2}) as an example to illustrate the translation to a quantum code in Theorem \ref{thm:css}.

First, consider the quantum encoding procedure.
Set $\lambda_0 = \kappa = 1$ and $q=2$, i.e., $Q_0$ has $1$ qubit. We only consider the encoding of $Q_0$ and defer the inclusion of the entangled reference system to the general proof of Theorem \ref{thm:css} in Section \ref{sec:css}. It suffices to consider the encoding and decoding of the computational basis $\ket{a}, a = 0, 1$ (as the coding operations are linear, the code for any superposition of $\ket{0}$, $\ket{1}$ is also implied).

\begin{eqnarray}
\ket{a} &\rightsquigarrow& \sum_{\bm{b}} \ket{a {\bf A} + \bm{b} {\bf B}} =  \frac{1}{\sqrt{2^3}} \sum_{b_1, b_2, b_3 \in \mathbb{F}_2}  \ket{b_1}_{Q_1} \ket{a+b_1, b_2}_{Q_2} \ket{a+b_1, b_3}_{Q_3} \ket{a+b_1, a+b_2+b_3}_{Q_4}, \notag\\
\end{eqnarray}
where $(\lambda_1, \lambda_2, \lambda_3, \lambda_4) = (1,2,2,2)$, i.e., $Q_1, Q_2, Q_3, Q_4$ contain $1$ qubit, $2$ qubits, $2$ qubits, and $2$ qubits respectively in the final output (corresponding to the $4$ kets above).  

Second, consider the quantum decoding procedure. 
\begin{enumerate}
\item Consider decoding set $\{Q_1, Q_2\}$. The decoding sets $\{Q_1, Q_3\}, \{Q_1, Q_4\}$ are similar.
\begin{eqnarray}
&&  \frac{1}{\sqrt{2^3}} \sum_{b_1, b_2, b_3 \in \mathbb{F}_2}  \ket{b_1} \ket{a+b_1, b_2} \ket{a+b_1, b_3} \ket{a+b_1, a+b_2+b_3} \\
&\rightsquigarrow&   \frac{1}{\sqrt{2^3}} \sum_{b_1, b_2, b_3 \in \mathbb{F}_2} \ket{a}  \ket{a+b_1, a+b_2} \ket{a+b_1, b_3} \ket{a+b_1, a+b_2+b_3} \\
&& (\mbox{define}~a+b_1 = c_1, a+b_2 = c_2, b_3 = c_3) \notag \\
&=&   \frac{1}{\sqrt{2^3}} \sum_{b_1, b_2, b_3 \in \mathbb{F}_2}  \ket{ a }  \ket{ c_1,  c_2}  \ket{c_1, c_3} \ket{ c_1, c_2+c_3}  \\
&& (\mbox{$c_1, c_2, c_3$ also takes all values in $\mathbb{F}_2$}) \notag \\
&=&  \frac{1}{\sqrt{2^3}} \sum_{c_1, c_2, c_3 \in \mathbb{F}_2}  \ket{ a }  \ket{ c_1,  c_2}  \ket{c_1, c_3} \ket{ c_1, c_2+c_3}  \\
&=&  \ket{ a }   \frac{1}{\sqrt{2^3}} \sum_{c_1, c_2, c_3 \in \mathbb{F}_2}  \ket{ c_1,  c_2}  \ket{c_1, c_3} \ket{ c_1, c_2+c_3} 
\end{eqnarray}
so that $\ket{a}$ is unentangled from the rest of the codeword and recovered with no error. Note that the storage nodes $Q_3, Q_4$ that are outside the considered decoding set are not touched in decoding.

\item Consider decoding set $\{Q_2, Q_3, Q_4\}$.
\begin{eqnarray}
&&  \frac{1}{\sqrt{2^3}} \sum_{b_1, b_2, b_3 \in \mathbb{F}_2}  \ket{b_1} \ket{a+b_1, b_2} \ket{a+b_1, b_3} \ket{a+b_1, a+b_2+b_3} \\
&\rightsquigarrow&  \frac{1}{\sqrt{2^3}}  \sum_{b_1, b_2, b_3 \in \mathbb{F}_2} \ket{b_1}  \ket{a+b_1, b_2} \ket{a+b_1, b_3} \ket{a+b_1, a} \\
&\rightsquigarrow&  \frac{1}{\sqrt{2^3}}  \sum_{b_1, b_2, b_3 \in \mathbb{F}_2} \ket{b_1}  \ket{b_1, b_2} \ket{b_1, b_3} \ket{b_1, a} \\
&=&   \left(  \frac{1}{\sqrt{2^3}} \sum_{b_1, b_2, b_3 \in \mathbb{F}_2} \ket{b_1}  \ket{b_1, b_2} \ket{b_1, b_3} \ket{b_1} \right) \ket{a}
\end{eqnarray}
so that $\ket{a}$ is recovered.
\end{enumerate}

\subsection{Intersection Graph $\sqcap_{5,3}$}\label{ex:int}
Consider the uniform storage case where $\lambda_i = 1, \forall i$. For the converse, consider the intersection bound (\ref{up:intersection}). 
\begin{eqnarray}
C_u(\sqcap_{5,3}) &\leq& \Lambda\Big(\mathcal{D}(e_1) \cap \mathcal{D}(e_2)\Big) = \Lambda\left(\left\{Q_{\overline{\mathcal{S}}}: 1 \in \mathcal{S}, 2 \in \mathcal{S}, \mathcal{S} \in \binom{[5]}{3} \right\} \right) \\
&=& \Lambda\left( \{Q_{123}, Q_{124}, Q_{125} \} \right) = \binom{5-2}{3-2} = 3,
\end{eqnarray}
where $Q_{\overline{\{i,j,k\}}}$ is abbreviated as $Q_{ijk}$ to simplify the notation.

For the achievability, we use Theorem \ref{thm:css} and translate from the following classical (structured) secure storage code. Set $k = 3, \kappa = 1$ and for this example, think of $q$ as a sufficiently large prime power (explicit choice will be given in the general proof in Section \ref{sec:int}). 
Set
\begin{eqnarray}
&& Y_{Q_{123}}, Y_{Q_{124}}, Y_{Q_{125}}, Y_{Q_{134}}, Y_{Q_{135}}, Y_{Q_{145}} ~\mbox{each as a generic (random linear)} \notag\\
&&~~~~~\mbox{combination of}~a_1, a_2, a_3, b_1, b_2, b_3, \label{eq:d1} \\
&& Y_{Q_{234}} = Y_{Q_{123}} - Y_{Q_{124}}  + Y_{Q_{134}}, \label{eq:d2}\\
&& Y_{Q_{235}} = Y_{Q_{123}} - Y_{Q_{125}}  + Y_{Q_{135}}, \label{eq:d5}\\
&& Y_{Q_{245}} = Y_{Q_{124}} - Y_{Q_{125}}  + Y_{Q_{145}}, \label{eq:d3} \\
&& Y_{Q_{345}} = Y_{Q_{134}} - Y_{Q_{135}}  + Y_{Q_{145}}, \label{eq:d4}
\end{eqnarray}
where $a_1, a_2, a_3, b_1, b_2, b_3$ are i.i.d. uniform elements in $\mathbb{F}_q$. The storage nodes in $\mathcal{D}(e_1)$ (i.e., storage nodes whose label contains `$1$') are each assigned a generic linear combination of $a_1, a_2, a_3, b_1, b_2, b_3$ (refer to (\ref{eq:d1})). The  storage nodes that are not in $\mathcal{D}(e_1)$ (i.e., those whose label does not contain `$1$') are assigned linear combinations of those in $\mathcal{D}(e_1)$ according to an alternating sign structure as specified above. For example, consider $Y_{Q_{234}}$ in (\ref{eq:d2}), which is set as a linear combination of $Y_{Q_{\mathcal{S}}}$ where $\mathcal{S}$ are all sets obtained by replacing one of the elements of $\{2,3,4\}$ with `$1$'.

Next, we use a dimension counting argument to show that the decoding constraint (\ref{css:dec}) and the security constraint (\ref{css:sec}) are satisfied.
First, consider (\ref{css:dec}). From $\mathcal{D}(e_1)$, (\ref{eq:d1}) shows that $\{Y_Q: Q \in \mathcal{D}(e_1)\}$ contains $6$ generic linear combinations of $a_1, a_2, a_3, b_1, b_2, b_3$ and thus may be set so that $a_1, a_2, a_3$ can be recovered. $\mathcal{D}(e_i), i \in [2:5]$ are similar, so let us  consider $\mathcal{D}(e_2)$. From (\ref{eq:d2}) - (\ref{eq:d3}),
\begin{eqnarray}
\left( Y_{Q_{123}}, Y_{Q_{124}}, Y_{Q_{125}}, Y_{Q_{234}}, Y_{Q_{235}}, Y_{Q_{245}} \right) \overset{\mbox{\scriptsize invertible}}{\longleftrightarrow} \left( Y_{Q_{123}}, Y_{Q_{124}}, Y_{Q_{125}}, Y_{Q_{134}}, Y_{Q_{135}}, Y_{Q_{145}} \right)
\end{eqnarray}
so that $\mathcal{D}(e_2)$  recovers $a_1, a_2, a_3$ as well.

Second, consider (\ref{css:sec}). We wish to show that nothing is revealed about $a_1, a_2, a_3$ from $\mathcal{Y}_{\mathcal{D}^c(e_i)}$ where $\mathcal{D}^c(e_i)$ contains $\binom{5}{3} - \binom{4}{2} = 4$ linear combinations of $a_1, a_2, a_3, b_1, b_2, b_3$. This is guaranteed by ensuring the $4$ linear combinations in $\mathcal{Y}_{\mathcal{D}^c(e_i)}$ occupy $3$ dimensions and the $3$ dimensions are fully covered by $b_1, b_2, b_3$ (so that $a_1, a_2, a_3$ are perfectly protected). We discuss next how this is guaranteed by the design in (\ref{eq:d1}) - (\ref{eq:d4}). 

Let us start from the simpler case for $\mathcal{D}^c(e_i)$ when $i \in [2:5]$. As the cases are similar, we consider $\mathcal{D}^c(e_2) = \{Q_{134}, Q_{135}, Q_{145}, Q_{345} \}$ for concreteness. From (\ref{eq:d4}), we see that $Y_{Q_{345}}$ is a function of $Y_{Q_{134}}, Y_{Q_{135}}, Y_{Q_{145}}$ so that indeed $\mathcal{Y}_{\mathcal{D}^c(e_2)} = \{Y_{Q_{134}}, Y_{Q_{135}}, Y_{Q_{145}}, Y_{Q_{345}}\}$ contains at most $3$ dimensions, $Y_{Q_{134}}, Y_{Q_{135}}, Y_{Q_{145}}$, wherein the $b$ symbols lie in generic spaces so that indeed (\ref{css:sec}) can be satisfied (a detailed proof is deferred to Section \ref{sec:int}).

We are now left with only $\mathcal{D}^c(e_1) = \{Q_{234}, Q_{235}, Q_{245}, Q_{345} \}$. Indeed, this is the most interesting case. From the surface, it might seem that $\mathcal{Y}_{\mathcal{D}^c(e_1)} = \{ Y_{Q_{234}}, Y_{Q_{235}}, Y_{Q_{245}}, Y_{Q_{345}} \}$ contains $4$ independent dimensions. However, this is not the case, as we show that, perhaps surprisingly, $Y_{Q_{345}}$ is a linear combination of $Y_{Q_{234}}, Y_{Q_{235}}, Y_{Q_{245}}$ as follows,
\begin{eqnarray}
&& Y_{Q_{234}} - Y_{Q_{235}} + Y_{Q_{245}} \notag\\
&\overset{(\ref{eq:d2}) - (\ref{eq:d3})}{=}& (Y_{Q_{123}} - Y_{Q_{124}}  + Y_{Q_{134}}) - (Y_{Q_{123}} - Y_{Q_{125}}  + Y_{Q_{135}}) + (Y_{Q_{124}} - Y_{Q_{125}}  + Y_{Q_{145}}) \\
&=& Y_{Q_{134}} - Y_{Q_{135}}  + Y_{Q_{145}} ~ \overset{(\ref{eq:d4})}{=} ~ Y_{Q_{345}}.
\end{eqnarray}
Thus, $\mathcal{Y}_{\mathcal{D}^c(e_1)}$ has only $3$ dimensions as desired (as a result of `alignment') and $Y_{Q_{234}}, Y_{Q_{235}}, Y_{Q_{245}}$ contain generic linear combinations of $b_1, b_2, b_3$ so that (\ref{css:sec}) can be satisfied.

To summarize,  non-trivial linear dependence exists by design in the above code construction. It is noteworthy that related alignment structures have appeared in the context of private information retrieval (private computation) \cite{Sun_Jafar_PC}, coded caching (for function retrieval) \cite{WSJTC_Scalar}, and secure aggregation \cite{WSJC_Groupwise, WSJC_GroupwiseR}.

\section{Proofs}
\subsection{Proof of Theorem \ref{thm:genupbound}}\label{sec:genupbound}

Let us start with a  lemma on the properties of quantum entropy and mutual information that will be used in the proof of Theorem \ref{thm:genupbound}. Unless stated explicitly otherwise, the quantum information measure terms are with respect to the joint state of the reference system $R$ and all quantum storage nodes $Q_1,\cdots, Q_N$, denoted as $\rho_{RQ_1\cdots Q_N}$. 
Recall from \eqref{eq:allstates} that our problem formulation requires perfect recovery for \emph{every} pure state of $RQ_0$, so in particular perfect recovery is required for a maximally entangled state $R Q_0$ where $Q_0$ is maximally mixed, i.e., $H(R, Q_0)_{\ket{\varphi}_{RQ_0}}=0$ and $H(R) = H(Q_0)_{\ket{\varphi}_{RQ_0}} = \log_q |Q_0| = \kappa \lambda_0$. In what follows, we assume $R Q_0$ is maximally entangled.

\begin{lemma}\label{lemma:genupbound}
For a storage graph $\mathcal{G}=((\lambda_1,\cdots,\lambda_N),\mathcal{E})$, and maximally entangled $RQ_0$, we have 
\begin{align}
\mbox{(Perfect recovery)} && I(R; \mathcal{D}(e)) &= 2\kappa \lambda_0, && \forall e \in \mathcal{E}, \label{eq:dec}\\
\mbox{(Monogamy of entanglement) } && I(R; \mathcal{Q}_1) + I(R; \mathcal{Q}_2) &\leq 2\kappa \lambda_0,  
&& \forall \mathcal{Q}_1 \cap \mathcal{Q}_2 = \emptyset, \mathcal{Q}_1, \mathcal{Q}_2 \subset \mathcal{Q}(\mathcal{G}), \label{eq:monogamy} \\
\mbox{(No-cloning) } && I(R; \mathcal{D}^c(e)) &= 0, && \forall e \in \mathcal{E}. \label{eq:sec}
\end{align}
\end{lemma}

{\it Proof:} (\ref{eq:dec}) was first proved by Schumacher and Nielsen \cite{Schumacher_Nielsen}. Intuitively, it says that as the original quantum message $Q_0$ can be perfectly recovered, then $\mathcal{D}(e)$ must contain all entanglement with the reference system $R$ (note that $I(R; Q_0)_{\ket{\varphi}_{RQ_0}} = 2\kappa \lambda_0$). A short proof based on quantum data processing inequality is as follows. For any $e \in \mathcal{E}$, after passing through the decoding mapping, the mutual information cannot increase, i.e.,
\begin{eqnarray}
I(R; \mathcal{D}(e)) \geq I(R; \widehat{Q}_0)_{\ket{\varphi}_{R \widehat{Q}_0}} = I(R; Q_0)_{\ket{\varphi}_{R Q_0}} = 2 H(R),
\end{eqnarray}
as decoding is perfect, i.e., $\ket{\varphi}_{R \widehat{Q}_0} = \ket{\varphi}_{R {Q}_0}$. Combining with the fact that $I(R; \mathcal{D}(e)) \leq 2 H(R) = 2\kappa \lambda_0$ completes the proof of (\ref{eq:dec}).

(\ref{eq:monogamy}), which we refer to as monogamy of entanglement, is a simple consequence of weak monotonicity.
\begin{align}
 I(R; \mathcal{Q}_1) + I(R; \mathcal{Q}_2) &= 2H(R) - \big( \underbrace{ H(R \mid \mathcal{Q}_1) + H(R \mid \mathcal{Q}_2) }_{\geq 0} \big) \\
 &\leq 2H(R) = 2 \kappa \lambda_0.
\end{align}

(\ref{eq:sec}), which we refer to as no-cloning, follows by combining (\ref{eq:dec}) and (\ref{eq:monogamy}) (setting $\mathcal{Q}_1 = \mathcal{D}(e)$ and $\mathcal{Q}_2 = \mathcal{D}^c(e)$) and using the property that quantum mutual information is non-negative. Intuitively, (\ref{eq:sec}) says that as the decoding set $\mathcal{D}(e)$ contains all entanglement with $R$, the complement $\mathcal{D}^c(e)$ and $R$ must be in a product state so that cloning is not possible.

\hfill\qed

Equipped with Lemma \ref{lemma:genupbound}, we now prove the intersection bound (\ref{up:intersection}). Consider two decoding sets $\mathcal{D}(e_i), \mathcal{D}(e_j)$, where $e_i, e_j \in \mathcal{E}$.
\begin{eqnarray}
2\kappa \lambda_0 &\overset{(\ref{eq:dec})}{=}& I(R; \mathcal{D}(e_i)) \\
&=& I(R; \mathcal{D}(e_i) \cap \mathcal{D}(e_j), \mathcal{D}(e_i) \setminus \mathcal{D}(e_j)) \\
&=& I(R; \mathcal{D}(e_i) \setminus \mathcal{D}(e_j)) +  I(R; \mathcal{D}(e_i) \cap \mathcal{D}(e_j)  \mid \mathcal{D}(e_i) \setminus \mathcal{D}(e_j)) \\
&=& \underbrace{ I(R ; \mathcal{D}^c(e_j) )}_{\overset{(\ref{eq:sec})}{=} 0}  - \underbrace{ I(R ; \mathcal{D}^c(e_j) \setminus (\mathcal{D}(e_i) \setminus \mathcal{D}(e_j)) \mid \mathcal{D}(e_i) \setminus \mathcal{D}(e_j)) }_{\geq 0} \notag\\
&&~+ \underbrace{ I(R; \mathcal{D}(e_i) \cap \mathcal{D}(e_j) \mid \mathcal{D}(e_i) \setminus \mathcal{D}(e_j)) }_{\leq 2 H( \mathcal{D}(e_i) \cap \mathcal{D}(e_j))}\\
&\leq& 2 H( \mathcal{D}(e_i) \cap \mathcal{D}(e_j) ) \\
&\leq& 2 \log_q \Pi_{Q \in \mathcal{D}(e_i) \cap \mathcal{D}(e_j)} |Q| \\
&=& 2 \sum_{k: Q_k \in \mathcal{D}(e_i) \cap \mathcal{D}(e_j)} \kappa \lambda_k =2 \kappa \Lambda\Big( \mathcal{D}(e_i) \cap \mathcal{D}(e_j) \Big) \label{eq:int}\\
\implies C(\mathcal{G}) &=& \sup {\lambda_0} \leq  \Lambda\Big( \mathcal{D}(e_i) \cap \mathcal{D}(e_j) \Big). 
\end{eqnarray}

Next we prove the wheel bound (\ref{up:wheel}) as a generalization of the converse bound for the wheel graph $\mathcal{W}_4$ presented in Section \ref{ex:wheel}.

Consider the maximally entangled pure state $RQ_0$, mapped through an isometric extension of the encoding CPTP map to pure state $RQ_1\cdots Q_NQ_{N+1}$, where $Q_{N+1}$ is an artificial  purifying system and is not in any decoding set.  

Consider any $k \in [2:n-2]$ and $\{\mathcal{Q}_1, \mathcal{Q}_2\}, \cdots, \{\mathcal{Q}_1, \mathcal{Q}_k\}, \{\mathcal{Q}_1, \mathcal{Q}_{k+1}\}, \{\mathcal{Q}_1, \mathcal{Q}_{k+2}, \cdots, \mathcal{Q}_n\}$ that each contains a decoding set, so from (\ref{eq:sec}), $I(R; \{\mathcal{Q}_1, \mathcal{Q}_i\}^c) = 0, i \in [2:k+2]$. As $RQ_1\cdots Q_NQ_{N+1}$ is pure,
\begin{eqnarray}
H(R) + H(\mathcal{Q}_3,  \mathcal{Q}_4, \cdots, \mathcal{Q}_k, \mathcal{Q}_{k+1}, \cdots, \mathcal{Q}_n, Q_{N+1}) &=& H(\mathcal{Q}_1, \mathcal{Q}_2) ~\leq~ H(\mathcal{Q}_1) + H(\mathcal{Q}_2), \label{eq:t1}\\
H(R) + H(\mathcal{Q}_2, \mathcal{Q}_4, \cdots, \mathcal{Q}_k, \mathcal{Q}_{k+1}, \cdots, \mathcal{Q}_n, Q_{N+1}) &=& H(\mathcal{Q}_1, \mathcal{Q}_3) ~\leq~ H(\mathcal{Q}_1) + H(\mathcal{Q}_3), \label{eq:t2}\\
&\vdots& \notag \\
H(R) + H(\mathcal{Q}_2, \mathcal{Q}_3, \cdots, \mathcal{Q}_{k-1}, \mathcal{Q}_{k+1}, \cdots, \mathcal{Q}_n, Q_{N+1}) &=& H(\mathcal{Q}_1, \mathcal{Q}_k) ~\leq~ H(\mathcal{Q}_1) + H(\mathcal{Q}_k), \label{eq:t3}\\
H(R) + H(\mathcal{Q}_2, \mathcal{Q}_3, \cdots, \mathcal{Q}_{k}, \mathcal{Q}_{k+2}, \cdots, \mathcal{Q}_n, Q_{N+1}) &=& H(\mathcal{Q}_1, \mathcal{Q}_{k+1}) ~\leq~ H(\mathcal{Q}_1) + H(\mathcal{Q}_{k+1}), \notag\\
\label{eq:t4}\\
H(R) + H( \mathcal{Q}_{2}, \mathcal{Q}_{3}, \cdots, \mathcal{Q}_{k}, \mathcal{Q}_{k+1}, Q_{N+1}) &=& H(\mathcal{Q}_1, \mathcal{Q}_{k+2}, \cdots, \mathcal{Q}_n)  \notag\\
&\leq& H(\mathcal{Q}_1) + H(\mathcal{Q}_{k+2}, \cdots, \mathcal{Q}_n). \label{eq:t5}
\end{eqnarray}
Adding (\ref{eq:t1}) - (\ref{eq:t4}) and applying strong sub-additivity repeatedly, we obtain
\begin{eqnarray}
&& k H(R) + (k-1) \underbrace{ H(\mathcal{Q}_2, \mathcal{Q}_3, \cdots, \mathcal{Q}_{k}, \mathcal{Q}_{k+1}, \cdots, \mathcal{Q}_n, Q_{N+1})}_{=H(R, \mathcal{Q}_1) \overset{(\ref{eq:sec})}{=} H(R) + H(\mathcal{Q}_1) } + H(\mathcal{Q}_{k+2}, \cdots, \mathcal{Q}_n, Q_{N+1})\notag\\
&\leq& k H(\mathcal{Q}_1) + H(\mathcal{Q}_2) + \cdots + H(\mathcal{Q}_k) + H(\mathcal{Q}_{k+1}) \label{eq:tt} \\
\implies && (2k-1) H(R) + H(\mathcal{Q}_{k+2}, \cdots, \mathcal{Q}_n) + H(Q_{N+1} \mid \mathcal{Q}_{k+2}, \cdots, \mathcal{Q}_n) \notag\\
&\leq& H(\mathcal{Q}_1) + H(\mathcal{Q}_2) + \cdots + H(\mathcal{Q}_k) + H(\mathcal{Q}_{k+1}). \label{eq:t6}
\end{eqnarray}
Here (\ref{eq:tt}) follows from the fact that $\mathcal{Q}_2 \cup \cdots \cup \mathcal{Q}_n$ contains a decoding set, so $I(R; \mathcal{Q}_1) = 0$ by (\ref{eq:sec}).
Adding (\ref{eq:t1}) - (\ref{eq:t3}), (\ref{eq:t5}) and applying strong sub-additivity repeatedly, we have
\begin{eqnarray}
&& k H(R) + (k-1) \underbrace{ H(\mathcal{Q}_2, \mathcal{Q}_3, \cdots, \mathcal{Q}_{k}, \mathcal{Q}_{k+1}, \cdots, \mathcal{Q}_n, Q_{N+1})}_{=H(R, \mathcal{Q}_1) \overset{(\ref{eq:sec})}{=} H(R) + H(\mathcal{Q}_1) } + H(\mathcal{Q}_{k+1}, Q_{N+1})\notag\\
&\leq& k H(\mathcal{Q}_1) + H(\mathcal{Q}_2) + \cdots + H(\mathcal{Q}_k) + H(\mathcal{Q}_{k+2}, \cdots, \mathcal{Q}_n)\\
\implies && (2k-1) H(R) + H(\mathcal{Q}_{k+1}) + H( Q_{N+1} \mid \mathcal{Q}_{k+1}) \notag\\
&\leq& H(\mathcal{Q}_1) + H(\mathcal{Q}_2) + \cdots + H(\mathcal{Q}_k) + H(\mathcal{Q}_{k+2}, \cdots, \mathcal{Q}_n). \label{eq:t7}
\end{eqnarray}
Adding (\ref{eq:t6}) and (\ref{eq:t7}) and applying weak monotonicity, we have
\begin{eqnarray}
2(2k-1) H(R) &\leq& 2( H(\mathcal{Q}_1) + H(\mathcal{Q}_2) + \cdots + H(\mathcal{Q}_k) ) \\
\implies  \kappa \lambda_0 = H(R) &\leq&  \kappa \frac{\Lambda(\mathcal{Q}_1) + \Lambda(\mathcal{Q}_2) + \cdots + \Lambda(\mathcal{Q}_k)}{2k-1} \\
\implies C(\mathcal{G}) = \sup \lambda_0 &\leq&  \frac{\Lambda(\mathcal{Q}_1) + \Lambda(\mathcal{Q}_2) + \cdots + \Lambda(\mathcal{Q}_k)}{2k-1}. 
\end{eqnarray}

\subsection{Proof of Theorem \ref{thm:css}}\label{sec:css}
We are given matrices ${\bf A}, {\bf B}$ with elements in finite field $\mathbb{F}_q$ that satisfy the conditions in Theorem \ref{thm:css}. We show how to use them to design a quantum encoding and decoding scheme as follows.

\paragraph{\it Quantum Encoding:} 
Set $\kappa \lambda_0 = k$ so that $Q_0$ has $k$ qudits where each qudit is $q$-dimensional. Suppose $Q_0$ is in an arbitrary state with density matrix $\omega$. Without loss of generality, suppose $\omega$ has a spectral decomposition $\omega = \sum_i p_i \ket{i} \bra{i}$ and a purification $\ket{\varphi}_{RQ_0} = \sum_{i} \sqrt{p_i} \ket{i} \ket{i}$. The encoding proceeds as follows.
\begin{eqnarray}
\ket{\varphi}_{R Q_0} &=& \sum_{i} \sqrt{p_i} \ket{i} \ket{i} \\
&=& \sum_{a_1, \cdots, a_k \in \mathbb{F}_q} \sqrt{p_{a_1,\cdots, a_k}} \ket{a_1, \cdots, a_k} \ket{a_1,\cdots, a_k} \\
&=& \sum_{\bm{a} \in \mathbb{F}_q^{1\times k}} \sqrt{p_{\bm{a}}} \ket{\bm{a}} \ket{\bm{a}} \\
&\rightsquigarrow&  \sum_{\bm{a}} \sqrt{p_{\bm{a}}}  \ket{\bm{a}}  \ket{\bm{a} {\bf A}} \label{eq:enc1} \\
&\rightsquigarrow&  \sum_{\bm{a}} \sqrt{p_{\bm{a}}}  \ket{\bm{a}}  \ket{\bm{a} {\bf A}} \frac{1}{\sqrt{q^\delta}} \sum_{\bm{b} \in \mathbb{F}_q^{1\times \delta}} \ket{\bm{b}} \label{eq:enc2} \\
&\rightsquigarrow&  \sum_{\bm{a}} \sqrt{p_{\bm{a}}}  \ket{\bm{a}}_R \frac{1}{\sqrt{q^\delta}}  \sum_{\bm{b}}  \ket{\bm{a} {\bf A} + \bm{b} {\bf B}}_{Q_1 \cdots Q_N} \label{eq:enc3}
\end{eqnarray}
where (\ref{eq:enc1}) is an isometric map (which can be extended to unitary) because rank$({\bf A}) = k$, i.e., ${\bf A}$ has full rank; (\ref{eq:enc3}) is an isometric map because rank$({\bf A};{\bf B}) = k+\delta$ so that from $\bm{a}{\bf A}+\bm{b}{\bf B}$, one can recover $\bm{a}, \bm{b}$.
Note that the encoding of $Q_0$ to $Q_1, \cdots Q_N$ does not touch $R$, i.e., $R$ goes through an identity map and $RQ_1\cdots Q_N$ ends up in a pure state. The encoding is now complete.

\paragraph{\it Quantum Decoding:} 
Consider any decoding set $\mathcal{D}(e)$, $e \in \mathcal{E}$ (abbreviated as $\mathcal{D}$ in this section where $e$ is fixed and omitted). For the quantum decoding procedure, we will perform a change of basis operation for $({\bf A}; {\bf B})$. 
To this end, suppose $\langle ({\bf A}_\mathcal{D}; {\bf B}_\mathcal{D}) \rangle \cap \langle ({\bf A}_{\mathcal{D}^c}; {\bf B}_{\mathcal{D}^c}) \rangle$ has dimension $\delta_1$.
Suppose $\mbox{rank}({\bf A}_\mathcal{D}; {\bf B}_\mathcal{D}) = k + \delta_1 + \delta_2$ where $\delta_1 \geq 0, \delta_2 \geq 0$ as $\langle {\bf I}_{k\times k}; {\bf 0}_{\delta\times k} \rangle$ is a subspace of $\langle {\bf A}_\mathcal{D}; {\bf B}_\mathcal{D} \rangle$ (refer to (\ref{css:space1})) and $\langle {\bf I}_{k\times k}; {\bf 0}_{\delta\times k} \rangle$ has no intersection with $\langle {\bf A}_{\mathcal{D}^c}; {\bf B}_{\mathcal{D}^c} \rangle$ (refer to (\ref{css:space2})); and  $\mbox{rank}({\bf A}_{\mathcal{D}^c}; {\bf B}_{\mathcal{D}^c}) = \delta_1 + \delta_3$ where $\delta_1 + \delta_2 + \delta_3 = \delta$ as $\mbox{rank}({\bf A};{\bf B}) = k+\delta$. We may now find matrices ${\bf A}_{\mathcal{D}_1}, {\bf A}_{\mathcal{D}_2}, {\bf A}_{\mathcal{D}_3}, {\bf B}_{\mathcal{D}_1}, {\bf B}_{\mathcal{D}_2}, {\bf B}_{\mathcal{D}_3}$ (where ${\bf A}_{\mathcal{D}_i}$ has $k$ rows and ${\bf B}_{\mathcal{D}_i}$ has $\delta$ rows) such that 
\begin{enumerate}
\item $({\bf A}_{\mathcal{D}_1}; {\bf B}_{\mathcal{D}_1}) \in \mathbb{F}_q^{(k+\delta)\times \delta_1}$ is a basis of $\langle ({\bf A}_\mathcal{D}; {\bf B}_\mathcal{D}) \rangle \cap \langle ({\bf A}_{\mathcal{D}^c}; {\bf B}_{\mathcal{D}^c}) \rangle$,
\item $({\bf A}_{\mathcal{D}_2}; {\bf B}_{\mathcal{D}_2}) \in \mathbb{F}_q^{(k+\delta)\times \delta_2}$ and $( ({\bf I}_{k\times k}; {\bf 0}_{\delta\times k}), ({\bf A}_{\mathcal{D}_1}; {\bf B}_{\mathcal{D}_1}),  ({\bf A}_{\mathcal{D}_2}; {\bf B}_{\mathcal{D}_2}) )$ is a basis of $\langle {\bf A}_\mathcal{D}; {\bf B}_\mathcal{D} \rangle$,
\item $({\bf A}_{\mathcal{D}_3}; {\bf B}_{\mathcal{D}_3}) \in \mathbb{F}_q^{(k+\delta)\times \delta_3}$ and $( ({\bf A}_{\mathcal{D}_1}; {\bf B}_{\mathcal{D}_1}),  ({\bf A}_{\mathcal{D}_3}; {\bf B}_{\mathcal{D}_3}) )$ is a basis of $\langle {\bf A}_{\mathcal{D}^c}; {\bf B}_{\mathcal{D}^c} \rangle$,
\item $\Big( ({\bf I}_{k\times k}; {\bf 0}_{\delta\times k}), ({\bf A}_{\mathcal{D}_1}; {\bf B}_{\mathcal{D}_1}),  ({\bf A}_{\mathcal{D}_2}; {\bf B}_{\mathcal{D}_2}), ({\bf A}_{\mathcal{D}_3}; {\bf B}_{\mathcal{D}_3}) \Big)$ is a basis of $\langle {\bf A}; {\bf B} \rangle ~=\langle {\bf I}_{(k+\delta)\times (k+\delta) }\rangle$ and $({\bf B}_{\mathcal{D}_1}, {\bf B}_{\mathcal{D}_2}, {\bf B}_{\mathcal{D}_3})$ is a basis of $\langle {\bf B} \rangle  ~=\langle {\bf I}_{ \delta \times \delta } \rangle  $.
\end{enumerate}
Then define
\begin{eqnarray}
&& \bm{b}_1' = \bm{a} {\bf A}_{\mathcal{D}_1} + \bm{b} {\bf B}_{\mathcal{D}_1} \in \mathbb{F}_q^{1\times \delta_1}, \\
&& \bm{b}_2' = \bm{a} {\bf A}_{\mathcal{D}_2} + \bm{b} {\bf B}_{\mathcal{D}_2} \in \mathbb{F}_q^{1\times \delta_2}, \\
&& \bm{b}_3' = \bm{a} {\bf A}_{\mathcal{D}_3} + \bm{b} {\bf B}_{\mathcal{D}_3} \in \mathbb{F}_q^{1\times \delta_3}.
\end{eqnarray}

As $({\bf B}_{\mathcal{D}_1}, {\bf B}_{\mathcal{D}_2}, {\bf B}_{\mathcal{D}_3})$ is a basis of $\langle {\bf I}_{ \delta \times \delta} \rangle$,  the product $\bm{b} ({\bf B}_{\mathcal{D}_1}, {\bf B}_{\mathcal{D}_2}, {\bf B}_{\mathcal{D}_3})$ is invertible to $\bm{b} \in \mathbb{F}_q^{1\times \delta}$, and then
\begin{eqnarray}
\bm{b}' \triangleq (\bm{b}_1', \bm{b}_2', \bm{b}_3')~\mbox{is invertible to}~\bm{b} \in \mathbb{F}_q^{1\times \delta} ~\mbox{for any fixed}~ \bm{a}. \label{enc:inv}
\end{eqnarray}
We are now ready to proceed to decoding. The state for $R Q_1 \cdots Q_N = R \mathcal{D} \mathcal{D}^c$ is expressed as follows. 

\begin{eqnarray}
&& \sum_{\bm{a}} \sqrt{p_{\bm{a}}}  \ket{\bm{a}} \frac{1}{\sqrt{q^\delta}} \sum_{\bm{b}}  \ket{\bm{a} {\bf A} + \bm{b} {\bf B}} \\
&=&  \sum_{\bm{a}} \sqrt{p_{\bm{a}}}  \ket{\bm{a}} \frac{1}{\sqrt{q^\delta}} \sum_{\bm{b}}  \ket{\bm{a} {\bf A}_\mathcal{D} + \bm{b} {\bf B}_\mathcal{D}} \ket{\bm{a} {\bf A}_{\mathcal{D}^c} + \bm{b} {\bf B}_{\mathcal{D}^c}} \\
&\rightsquigarrow&  \sum_{\bm{a}} \sqrt{p_{\bm{a}}}  \ket{\bm{a}} \frac{1}{\sqrt{q^\delta}} \sum_{\bm{b}} \ket{\bm{a}}  \ket{\bm{b}_1'}  \ket{\bm{b}_2'}  \ket{l(\bm{b}_1', \bm{b}_2')}  
\ket{l(\bm{b}_1', \bm{b}_3')}  \label{eq:dec1} \\
&=&  \sum_{\bm{a}} \sqrt{p_{\bm{a}}}  \ket{\bm{a}} \frac{1}{\sqrt{q^\delta}} \sum_{\bm{b}'} \ket{\bm{a}}  \ket{\bm{b}_1'}  \ket{\bm{b}_2'}  \ket{l(\bm{b}_1', \bm{b}_2')}  
\ket{l(\bm{b}_1', \bm{b}_3')} \label{eq:dec2} \\
&=&  \sum_{\bm{a}} \sqrt{p_{\bm{a}}}  \ket{\bm{a}} \ket{\bm{a}} \frac{1}{\sqrt{q^\delta}} \sum_{\bm{b}'} \ket{\bm{b}_1'}  \ket{\bm{b}_2'}  \ket{l(\bm{b}_1', \bm{b}_2')}   
\ket{l(\bm{b}_1', \bm{b}_3')}  \\
&=& \ket{\varphi}_{R \widehat{Q}_0} ~\otimes~ \left( \frac{1}{\sqrt{q^\delta}} \sum_{\bm{b}'} \ket{\bm{b}_1'}  \ket{\bm{b}_2'}  \ket{l(\bm{b}_1', \bm{b}_2')}   
\ket{l(\bm{b}_1', \bm{b}_3')} \right),
\end{eqnarray}
where (\ref{eq:dec1}) involves a change of basis operation following space decomposition using ${\bf A}_{\mathcal{D}_i}, {\bf B}_{\mathcal{D}_i}$ presented above and change of basis is unitary; 
$l(\bm{b}_1', \bm{b}_2')$ represents the remaining $\kappa \Lambda(\mathcal{D}) - (k+\delta_1+\delta_2)$ columns of $\bm{a}{\bf A}_\mathcal{D} + \bm{b}{\bf B}_\mathcal{D}$ that can be written as linear combinations of $\bm{b}_1', \bm{b}_2'$; 
and $\bm{a}{\bf A}_{\mathcal{D}^c} + \bm{b}{\bf B}_{\mathcal{D}^c}$ is written as $l(\bm{b}_1', \bm{b}_3')$ (so nothing is changed and this is required for decoding as $\mathcal{D}^c$ is not in the decoding set). 
(\ref{eq:dec2}) follows from the observation that for any fixed $\bm{a}$, as we sum over all possible values of $\bm{b}$ over $\mathbb{F}_q^{1\times \delta}$, $\bm{b}'$ also takes all values over $\mathbb{F}_q^{1\times \delta}$ (because of the invertible map in (\ref{enc:inv})). In the last step, we have unentangled $R \widehat{Q}_0$  from the remaining  qudits and $\ket{\varphi}_{R \widehat{Q}_0} = \ket{\varphi}_{R {Q}_0}$, so decoding is successful.

\paragraph{\it Rate Achieved:} $\lambda_0 = k/\kappa$ so the proof of the lower bound on capacity in Theorem \ref{thm:css} is complete.

\subsection{Proof of Corollary \ref{cor:css}}\label{sec:csscor}
Noting that $\bm{a}, \bm{b}$ consist of i.i.d. uniform elements in $\mathbb{F}_q$, we may express the entropy and mutual information terms in (\ref{css:dec}), (\ref{css:sec}) as rank conditions of the associated matrices and then Corollary \ref{cor:css} will follow.

First, consider (\ref{css:rank1}).
\begin{eqnarray}
0 &\overset{(\ref{css:dec})}{=}& H(\bm{a}\mid \mathcal{Y}_{\mathcal{D}(e)}) \\
&=& H(\bm{a}, \mathcal{Y}_{\mathcal{D}(e)}) - H(\mathcal{Y}_{\mathcal{D}(e)}) \\
&=& H(\bm{a}) + H( \mathcal{Y}_{\mathcal{D}(e)} \mid \bm{a}) - H( (\bm{a}, \bm{b}) ({\bf A}_{\mathcal{D}(e)}; {\bf B}_{\mathcal{D}(e)})) \\
&=& H(\bm{a}) + H(\bm{b} {\bf B}_{\mathcal{D}(e)} \mid \bm{a} ) - H( (\bm{a}, \bm{b}) ({\bf A}_{\mathcal{D}(e)}; {\bf B}_{\mathcal{D}(e)})) \\
&=& k + \mbox{rank}({\bf B}_{\mathcal{D}(e)}) - \mbox{rank}({\bf A}_{\mathcal{D}(e)}; {\bf B}_{\mathcal{D}(e)}),
\end{eqnarray}
where the last step follows from the fact that $\bm{a}, \bm{b}$ are i.i.d. uniform and (\ref{css:rank1}) is proved. 

(\ref{css:space1}) follows from (\ref{css:rank1}) because the rank of ${\bf A}_{\mathcal{D}(e)}$ is at most $k$ and $\mbox{rank}({\bf A}_{\mathcal{D}(e)}; {\bf B}_{\mathcal{D}(e)})$ $-$ $\mbox{rank}({\bf 0}; {\bf B}_{\mathcal{D}(e)})$ $= k$ so that $\langle {\bf I}_{k\times k}; {\bf 0}_{\delta\times k} \rangle$ must be a subspace of $\langle {\bf A}_{\mathcal{D}(e)}; {\bf B}_{\mathcal{D}(e)} \rangle$.

Second, consider (\ref{css:rank2}).
\begin{eqnarray}
0 &\overset{(\ref{css:sec})}{=}& I(\bm{a};\mathcal{Y}_{{\mathcal{D}^c(e)}}) \\
&=& H(\mathcal{Y}_{{\mathcal{D}^c(e)}}) - H(\mathcal{Y}_{{\mathcal{D}^c(e)}} \mid \bm{a}) \\
&=& H( (\bm{a}, \bm{b}) ({\bf A}_{{\mathcal{D}^c(e)}}; {\bf B}_{{\mathcal{D}^c(e)}})) - H( \bm{b} {\bf B}_{{\mathcal{D}^c(e)}} \mid \bm{a}) \\
&=&  \mbox{rank}({\bf A}_{{\mathcal{D}^c(e)}}; {\bf B}_{{\mathcal{D}^c(e)}}) - \mbox{rank}({\bf B}_{{\mathcal{D}^c(e)}})
\end{eqnarray}
and (\ref{css:rank2}) is proved.

(\ref{css:space2}) follows from (\ref{css:rank2}) because $\mbox{rank}({\bf A}_{{\mathcal{D}^c(e)}}; {\bf B}_{{\mathcal{D}^c(e)}}) = \mbox{rank}({\bf 0}; {\bf B}_{{\mathcal{D}^c(e)}})$ so that $\langle {\bf I}_{k\times k}; {\bf 0}_{\delta\times k} \rangle$ must have no intersection with $\langle {\bf A}_{{\mathcal{D}^c(e)}}; {\bf B}_{{\mathcal{D}^c(e)}} \rangle$.

\subsection{Proof of Corollary \ref{thm:small}}\label{sec:small}
\subsubsection{If and only if condition for $C(\mathcal{G}) = 0$}
The `if' direction is immediate, i.e., when $e_i \cap e_j = \emptyset$, by the intersection bound (\ref{up:intersection}), we have $C(\mathcal{G}) \leq 0$.

We then consider the `only if' direction, i.e., we show that as long as for any $\mathcal{D}(e_i), \mathcal{D}(e_j), e_i, e_j \in \mathcal{E}$, $e_i \cap e_j \neq \emptyset$, then there exists a feasible quantum storage code such that $C(\mathcal{G}) > 0$. We now present such a (simple) code. We devote no effort to improving the efficiency but note that this is possible, e.g., by following existing work in related topics such as (quantum) secret sharing \cite{Beimel}.

For the quantum code, we resort to the classical code in Theorem \ref{thm:css} whose performance translates into the quantum case.

For the classical secure storage problem with constraints (\ref{css:dec}) and (\ref{css:sec}), we consider $k=1$ uniform binary classical secret symbol ${a} \in \mathbb{F}_2$. Then consider each decoding set $\mathcal{D}(e_1), \cdots, \mathcal{D}(e_{|\mathcal{E}|}), e_i \in \mathcal{E}$ sequentially. For decoding set $\mathcal{D}(e_i) = \{Q_{i_1}, \cdots, Q_{i_{|e_i|}}\}, i \in [|\mathcal{E}|]$, we set
\begin{eqnarray}
&& Y_{Q_{i_1}} \overset{\mbox{\scriptsize include}}{\longleftarrow} b_{i_1}, Y_{Q_{i_2}} \overset{\mbox{\scriptsize include}}{\longleftarrow} b_{i_2}, \cdots, Y_{Q_{i_{|e_i|} - 1}} \overset{\mbox{\scriptsize include}}{\longleftarrow} b_{i_{|e_i|-1}},  \notag\\
&& Y_{Q_{i_{|e_i|} }}  \overset{\mbox{\scriptsize include}}{\longleftarrow} a - b_{i_1} - \cdots -  b_{i_{|e_i|-1}}, \label{feas:scheme}
\end{eqnarray}
where the $b$ symbols are all i.i.d. and uniform over $\mathbb{F}_2$ and in total $\sum_{i \in [|\mathcal{E}|]} (|e_i| - 1)$ uniform bits $b$ are used;  the notation $Y \overset{\mbox{\scriptsize include}}{\longleftarrow} b$ means $b$ is appended as the last element in vector $Y$, i.e., for any $\mathcal{D}(e_i), i \in [|\mathcal{E}|]$, one more bit is included in $Y_{Q}, Q \in \mathcal{D}(e_i)$.

The classical decoding constraint (\ref{css:dec}) is satisfied as from (\ref{feas:scheme}), $a = \sum_{j \in [|e_i|]} Y_{Q_{i_j}}$ for decoding set $\mathcal{D}(e_i)$. The classical security constraint (\ref{css:sec}) is satisfied because any two decoding sets are intersecting such that for any decoding set $\mathcal{D}(e_k), k \in [|\mathcal{E}|]$, its complement $\mathcal{D}^c(e_k)$ is missing at least one $b_{i}$ noise bit from any other decoding set $\mathcal{D}(e_i), i \neq k, i \in [|\mathcal{E}|]$ (note that in (\ref{feas:scheme}), any $|e_i| - 1$ $Y_{Q}$ see independent $b_i$ noise bits) and thus nothing about $a$ is revealed. 
$C(\mathcal{G}) \geq k/\kappa > 0$, where $\kappa$ is chosen so that the size constraint of each storage node is not exceeded, so the proof of the only if direction (and the overall proof) is now complete.

\subsubsection{Capacity of small graphs}

The graphs where $C(\mathcal{G}) = 0$ have been covered and we only need to consider graphs where $C(\mathcal{G}) > 0$. Note that we assume $\mathcal{G}$ has no redundant storage nodes (every storage node appears in some decoding set) and no redundant decoding sets (only minimal decoding sets are considered).

Let us start from the case when $|\mathcal{E}| \leq 3$. The case $|\mathcal{E}| = 1$ is trivial. When $|\mathcal{E}| = 2$, say $\mathcal{E} = \{e_1, e_2\}$, we have $C(\mathcal{G}) = \Lambda(\mathcal{D}(e_1) \cap \mathcal{D}(e_2))$; converse follows from the intersection bound (\ref{up:intersection}) and achievability follows from storing the qudits directly in the storage nodes that are in both $\mathcal{D}(e_1)$ and $\mathcal{D}(e_2)$.
When $|\mathcal{E}| = 3$, say $\mathcal{E} = \{e_1, e_2, e_3\}$, we may partition $\mathcal{Q}(\mathcal{G})$ into the following $7$ disjoint sets.
\begin{eqnarray}
&& \mathcal{Q}_{123} \triangleq \mathcal{D}(e_1) \cap \mathcal{D}(e_2) \cap \mathcal{D}(e_3), \\
&& \mathcal{Q}_{123^c} \triangleq \mathcal{D}(e_1) \cap \mathcal{D}(e_2) \cap \mathcal{D}^c(e_3), ~
\mathcal{Q}_{12^c3} \triangleq \mathcal{D}(e_1) \cap \mathcal{D}^c(e_2) \cap \mathcal{D}(e_3), \notag\\
&&~~~ \mathcal{Q}_{1^c23} \triangleq \mathcal{D}^c(e_1) \cap \mathcal{D}(e_2) \cap \mathcal{D}(e_3), \\
&& \mathcal{Q}_{12^c3^c} \triangleq \mathcal{D}(e_1) \cap \mathcal{D}^c(e_2) \cap \mathcal{D}^c(e_3), ~ 
\mathcal{Q}_{1^c23^c} \triangleq \mathcal{D}^c(e_1) \cap \mathcal{D}(e_2) \cap \mathcal{D}^c(e_3), \notag\\
&&~~~\mathcal{Q}_{1^c2^c3} \triangleq \mathcal{D}^c(e_1) \cap \mathcal{D}^c(e_2) \cap \mathcal{D}(e_3).
\end{eqnarray}
From the intersection bound (\ref{up:intersection}), we have
\begin{eqnarray}
C(\mathcal{G}) &\leq& \min\Big( \Lambda\big( \mathcal{D}(e_1) \cap \mathcal{D}(e_2) \big) , \Lambda\big( \mathcal{D}(e_1) \cap \mathcal{D}(e_3) \big) , \Lambda\big( \mathcal{D}(e_2) \cap \mathcal{D}(e_3) \big) \Big) \\
&=& \Lambda(\mathcal{Q}_{123}) + \min\Big( \Lambda( \mathcal{Q}_{123^c} ), \Lambda( \mathcal{Q}_{12^c3} ), \Lambda( \mathcal{Q}_{1^c23} )  \Big),
\end{eqnarray}
which can be achieved by dividing the quantum message $Q_0$ into two parts - the first part has $\Lambda(\mathcal{Q}_{123})$ qudits and is directly stored in $\mathcal{Q}_{123}$; the second part has $\min( \Lambda( \mathcal{Q}_{123^c} ), \Lambda( \mathcal{Q}_{12^c3} ), \Lambda( \mathcal{Q}_{1^c23} )  )$ qudits and is stored over the MDS storage graph with three node sets $\mathcal{Q}_{123^c} , \mathcal{Q}_{12^c3}, \mathcal{Q}_{1^c23} $ and any two node sets form a decoding set (invoking the MDS storage graph $\mathcal{M}_{3,2}$ result from Theorem \ref{thm:mds}).

Next we consider $N \leq 4$ settings, among which $|\mathcal{E}| \leq 3$ cases have been covered above and the only remaining cases are
\begin{enumerate}
\item the wheel graph $\mathcal{W}_4$ that is covered by Theorem \ref{thm:wheel},
\item the MDS storage graph $\mathcal{M}_{4,3}$ that is covered by Theorem \ref{thm:mds}.
\end{enumerate}
So all small graph cases are settled.

\subsection{Proof of Theorem \ref{thm:mds}}\label{sec:mds} 

The converse $C(\mathcal{M}_{N,K}) \leq \lambda_1 + \cdots + \lambda_{2K-N}$ follows from the intersection bound (\ref{up:intersection}) by setting $e_1 = [K], e_2 = \{1,2,\cdots, 2K-N, K+1, \cdots, N\}$ so that $e_1 \cap e_2 = [2K-N]$ and $\Lambda(\mathcal{D}(e_1) \cap \mathcal{D}(e_2)) = \lambda_1 + \cdots + \lambda_{2K-N}$, leading to $C_u(\mathcal{M}_{N,K}) \leq 2K-N$ in the uniform storage case where $\lambda_1 = \cdots = \lambda_N = 1$.

We now prove the achievability and start from the uniform storage case. We show that $C_u(\mathcal{M}_{N,K})$ $=$ $2K - N$ is achievable. Set $q$~as a prime power where $q \geq N$, and
\begin{eqnarray}
&&  k = 2K - N, ~\bm{a} = (a_1, \cdots, a_k), \\
&& \delta = N-K, ~\bm{b} = (b_1, \cdots, b_\delta), \\
&& \kappa = 1, n = N,  ({\bf A}; {\bf B}) = [(\alpha_j)^{k+\delta-i}]_{i,j} \in \mathbb{F}_q^{(k+\delta) \times n}, \\
&& (Y_{Q_1}, \cdots, Y_{Q_N}) = \bm{a} {\bf A} + \bm{b} {\bf B},
\end{eqnarray}
where $({\bf A};{\bf B})$ is a Vandermonde matrix with the element in the $i^{th}$ row and $j^{th}$ column being $\alpha_j^{k+\delta-i}, i \in [k+\delta], j \in [n]$ and $\alpha_1, \cdots, \alpha_n$ are distinct elements in $\mathbb{F}_q$. Then the classical decoding constraint (\ref{css:dec}) is satisfied as we may verify the equivalent rank constraint (\ref{css:rank1}) as follows. For any $e \in \mathcal{E} = \binom{[N]}{K}$,
\begin{eqnarray}
\mbox{rank}({\bf A}_{\mathcal{D}(e)}; {\bf B}_{\mathcal{D}(e)}) - \mbox{rank}({\bf B}_{\mathcal{D}(e)}) = (k+\delta) - \delta = k,
\end{eqnarray}
where $({\bf A}_{\mathcal{D}(e)}; {\bf B}_{\mathcal{D}(e)})$ is a sub-matrix of the Vandermonde matrix $({\bf A};{\bf B})$ with $|\mathcal{D}(e)| = K = k + \delta$ columns and is also square Vandermonde thus has full rank; $({\bf B}_{\mathcal{D}(e)})$ is a sub-matrix of the Vandermonde matrix $({\bf A}_{\mathcal{D}(e)}; {\bf B}_{\mathcal{D}(e)})$ with the last $\delta$ rows and is also Vandermonde and has full rank (note that in $({\bf A};{\bf B})$, we order the exponents of $\alpha_j$ increasingly from the bottom to the top). The classical security constraint (\ref{css:sec}) is satisfied as we may verify the equivalent rank constraint (\ref{css:rank2}) as follows. For any $e \in \mathcal{E}$,
\begin{eqnarray}
\mbox{rank}({\bf A}_{\mathcal{D}^c(e)}; {\bf B}_{\mathcal{D}^c(e)}) - \mbox{rank}({\bf B}_{\mathcal{D}^c(e)}) = \delta - \delta = 0,
\end{eqnarray}
where $({\bf A}_{\mathcal{D}^c(e)}; {\bf B}_{\mathcal{D}^c(e)})$ is a sub-matrix of $({\bf A};{\bf B})$ with $|\mathcal{D}^c(e)| = N - K = \delta$ columns and is Vandermonde and has full rank; $({\bf B}_{\mathcal{D}^c(e)})$ is a sub-matrix of $({\bf A}_{\mathcal{D}^c(e)}; {\bf B}_{\mathcal{D}^c(e)})$ with last $\delta$ rows and is square Vandermonde and has full rank. Thus we may apply Theorem \ref{thm:css} to obtain a quantum code and $C_u(\mathcal{M}_{N,K}) \geq k/\kappa = 2K- N$.

\bigskip
Next consider the non-uniform storage case and we show that $C(\mathcal{M}_{N,K}) = \lambda_1 + \lambda_2 + \cdots + \lambda_{2K-N}$ is achievable. We will combine (space-share over) uniform storage MDS graphs as follows. Set $q \geq N$ as a prime power.
For any $\epsilon>0$, there exist rational $\lambda_i'$ such that $\lambda_i' \leq \lambda_i, |\lambda_i' - \lambda_i| < \epsilon$ for all $i \in [N]$ as rationals are dense over the reals. For rationals $\lambda_1', \cdots, \lambda_N'$, one can choose an integer scaling factor $\kappa$ so that $\kappa \lambda_1', \cdots, \kappa \lambda_N'$ are integers.
\begin{eqnarray}
&& \mbox{Store $\lambda_0(1) = (2K-N) \kappa \lambda_1'$ qudits of $Q_0$ in $\mathcal{M}_{N,K}$} \notag\\
&&~~~ \mbox{ by utilizing $\lambda_i (1) = \kappa \lambda_1'$ qudits of storage for $Q_i, i \in [1:N]$}. \\
&& \mbox{Store $\lambda_0(2) = (2K-N-1) \kappa (\lambda_2' - \lambda_1')$ qudits of $Q_0$ in $\mathcal{M}_{N-1,K-1}$} \notag\\
&&~~~ \mbox{by utilizing $\lambda_i (2) = \kappa (\lambda'_2 - \lambda'_1)$ qudits of storage for $Q_i, i \in [2:N]$}. \\
&& \mbox{Store $\lambda_0(3) = (2K-N-2) \kappa (\lambda_3' - \lambda_2')$ qudits of $Q_0$ in $\mathcal{M}_{N-2,K-2}$} \notag\\
&&~~~ \mbox{by utilizing $\lambda_i (3) = \kappa (\lambda'_3 - \lambda'_2) $ qudits of storage for $Q_i, i \in [3:N]$}. \\
&& ~~~~~~~~~~~~~~~~~~~~~~\vdots \notag\\
&& \mbox{Store $\lambda_0(j) = (2K-N-j+1) \kappa (\lambda_j' - \lambda_{j-1}')$ qudits of $Q_0$ in $\mathcal{M}_{N-j+1,K-j+1}$} \notag\\
&&~~~ \mbox{by utilizing $\lambda_i (j) = \kappa (\lambda_j' - \lambda_{j-1}') $ qudits of storage for $Q_i, i \in [j:N]$}. \\
&& ~~~~~~~~~~~~~~~~~~~~~~\vdots \notag\\
&& \mbox{Store $\lambda_0(2K-N) =  1\times \kappa (\lambda_{2K-N}' - \lambda_{2K-N-1}')$ qudits of $Q_0$ in $\mathcal{M}_{2N-2K+1,N-K+1}$} \notag\\
&&~~~ \mbox{by utilizing $\lambda_i (2K-N) = \kappa( \lambda_{2K-N}' - \lambda_{2K-N-1}') $ qudits of storage for $Q_i, i \in [2K-N : N]$}. \notag\\
\end{eqnarray}
For each $j \in [2K-N]$, we are using the capacity achieving code for uniform storage MDS graph $\mathcal{M}_{N-j+1,K-j+1}$ where $C \geq 2K-N-j+1 > 0$. Further the component code for each $j \in [2K-N]$ satisfies the original $K$ out of $N$ MDS constraint because 
\begin{eqnarray}
\underbrace{(j-1)}_{\mbox{\scriptsize unused $Q_i$}} + (K-j+1) = K.
\end{eqnarray}
In total the number of qudits from the quantum message stored and the number of qudits used in the storage nodes are
\begin{eqnarray}
&& \lambda_0(1) + \cdots + \lambda_0(2K-N) = \kappa ( \lambda_1' + \cdots + \lambda_{2K-N}') \triangleq \kappa \lambda_0, \\
&& \lambda_1(1)  = \kappa \lambda_1' \leq \kappa \lambda_1, \\
&& \lambda_2(1) + \lambda_2(2) = \kappa \lambda_2' \leq \kappa \lambda_2, \\
&& ~~~~~~~~~~\vdots \\
&& \lambda_{2K-N}(1) + \cdots + \lambda_{2K-N}(2K-N) = \kappa \lambda_{2K-N}' \leq \kappa \lambda_{2K-N},  \\
&& \kappa \lambda_{2K-N}' \leq \kappa \lambda_{i} , i \in [2K-N+1 : N].
\end{eqnarray}
Thus, $C(\mathcal{M}_{N,K}) \geq \sup \lambda_0 = \sup \lambda_1' + \cdots + \lambda_{2K-N}' \overset{\epsilon \rightarrow 0}{=} \lambda_1 + \cdots + \lambda_{2K-N}$, as desired.

\subsection{Proof of Theorem \ref{thm:wheel}}\label{sec:wheel_proof}
Consider the upper bound $C(\mathcal{W}_N) \leq \min\left( \lambda_1, \lambda_2, ~\frac{\lambda_1 + \lambda_2}{3},~\frac{\lambda_1 + \lambda_2 + \lambda_3}{5},\cdots, \frac{\lambda_1+\lambda_2+\cdots+\lambda_{N-2}}{2N-5}\right)$. The first two terms follow from the intersection bound (\ref{up:intersection}), 
\begin{eqnarray}
e_1 = \{1,2\}, e_2 = \{1,3\} &\overset{(\ref{up:intersection})}{\implies}& C(\mathcal{W}_N) \leq \Lambda\Big(\mathcal{D}(e_1) \cap \mathcal{D}(e_2)\Big) = \lambda_1, \\
e_1 = \{1,2\}, e_N = \{2,3,\cdots,N\} &\overset{(\ref{up:intersection})}{\implies}& C(\mathcal{W}_N) \leq \Lambda\Big(\mathcal{D}(e_1) \cap \mathcal{D}(e_N)\Big) = \lambda_2.
\end{eqnarray}
The remaining terms follow from the wheel bound (\ref{up:wheel}) by setting $n = N$, $\mathcal{Q}_i = \{Q_i\}$ for all $i \in [n]$.
Next, we prove the achievability of the upper bound for the following two cases. 

With a similar proof as that for the MDS graph in Section \ref{sec:mds}, we  assume without loss of generality that $\lambda_i, i \in [N]$ and the number of qudits used in the schemes (i.e., $\frac{2\lambda_1 - \lambda_2}{2N-5}$ in (\ref{eq:qd2}), $\frac{(N-2)\lambda_2-\lambda_1}{2N-5}$ in (\ref{eq:qd3}), $\frac{2\lambda_2 - \lambda_1}{3}$ in (\ref{eq:qd4}), $\frac{2\lambda_1 - \lambda_2}{3}$ in (\ref{eq:qd5}), $\frac{2(\lambda_1+\lambda_2) - 3\lambda_3}{5}$ in (\ref{eq:qd6}), $\frac{4\lambda_2 - (\lambda_1+\lambda_3)}{5} $ in (\ref{eq:qd7})) are integers (through approximating reals by rationals and scaling rationals to integers).

\subsubsection{$\lambda_2 = \lambda_3 = \cdots = \lambda_N$}
We show that $C(\mathcal{W}_N) = \min(\lambda_1, \lambda_2, \frac{\lambda_1 + (N-3)\lambda_2}{2N-5})$. Note that the uniform storage case, where $\lambda_1 = \lambda_2 = \cdots = \lambda_N = 1$ and $C_u(\mathcal{W}_N) = \frac{N-2}{2N-5}$, is covered and thus needs no proof. The proof in this section is a generalization of the proof of $C(\mathcal{W}_4)$ presented in Section \ref{ex:wheel}.

We first present two component codes.
\begin{enumerate}
\item $\lambda_0 = 1$ code for storage graph $\mathcal{W}_N$ with $(\lambda_1, \lambda_2, \cdots, \lambda_N) = (N-2, 1, \cdots, 1)$.

Set $\kappa = 1$ and $q \geq N-1$ is a prime power. Use the following classical (generic linear) secure storage code,
\begin{eqnarray}
&& Y_{Q_1} = (b_1, b_2, \cdots, b_{N-2}),\\
&& (Y_{Q_2}, Y_{Q_3}, \cdots, Y_{Q_N}) = (a, b_1, \cdots, b_{N-2}) \times [\alpha_j^{N-1-i}]_{i,j},
\end{eqnarray}
where $a, b_1, \cdots, b_{N-2}$ are i.i.d. uniform elements in $\mathbb{F}_q$, $[\alpha_j^{N-1-i}]_{i,j} \in \mathbb{F}_q^{(N-1) \times (N-1)}, i, j \in [N-1]$ is a Vandermonde matrix and $\alpha_j$ are distinct elements in $\mathbb{F}_q$. The classical constraints (\ref{css:dec}), (\ref{css:sec}) are satisfied and Theorem \ref{thm:css} produces the desired quantum code.

\item $\lambda_0 = 1$ code for storage graph $\mathcal{W}_N$ with $(\lambda_1, \lambda_2, \cdots, \lambda_N) = (1, 2, \cdots, 2)$.

Set $\kappa = 1$ and $q \geq N-1$ is a prime power. Use the following classical (structured) secure storage code. $Y_{Q_i}, i \in [2:N]$ contains two symbols $Y_{Q_i} = (Y_{Q_i}(1), Y_{Q_i}(2))$.
\begin{eqnarray}
&& Y_{Q_1} = b_1, Y_{Q_i}(1) = a + b_1, i \in [2:N], \\
&& (Y_{Q_2}(2), Y_{Q_3}(2), \cdots, Y_{Q_N}(2)) = (a, b_2, \cdots, b_{N-1}) \times [\alpha_j^{N-1-i}]_{i,j},
\end{eqnarray}
where $a, b_1, \cdots, b_{N-1}$ are i.i.d. uniform elements in $\mathbb{F}_q$, $[\alpha_j^{N-1-i}]_{i,j} \in \mathbb{F}_q^{(N-1) \times (N-1)}, i, j \in [N-1]$ is Vandermonde and $\alpha_j$ are distinct in $\mathbb{F}_q$. The classical constraints (\ref{css:dec}), (\ref{css:sec}) are satisfied and Theorem \ref{thm:css} produces the desired quantum code.
\end{enumerate}

Next we use the above two component codes to achieve $C(\mathcal{W}_N) = \min\left( \lambda_1, \lambda_2 , \frac{\lambda_1 + (N-3)\lambda_2}{2N-5} \right)$. There are three cases. Choose a prime power $q \geq N-1$.
\begin{enumerate}
\item When $\lambda_1 \leq \lambda_2/2$, $\min\left( \lambda_1, \lambda_2 , \frac{\lambda_1 + (N-3)\lambda_2}{2N-5} \right) = \lambda_1$. 
Use the second component code to store $\lambda_0 = \lambda_1$ qudits of $Q_0$ in $(Q_1, Q_2, \cdots, Q_N)$ while utilizing storage in the amounts  $(\lambda_1, 2\lambda_1, \cdots, 2\lambda_1)$ qudits, respectively. The scheme works as $\lambda_1 \leq \lambda_2/2$ and $Q_2, \cdots, Q_N$ might have extra unused qudits.
\item When $ \lambda_2/2 \leq \lambda_1 \leq (N-2) \lambda_2$, $\min\left( \lambda_1, \lambda_2 , \frac{ \lambda_1 + (N-3) \lambda_2}{2N-5} \right) = \frac{ \lambda_1 + (N-3) \lambda_2}{2N-5}$.
\begin{eqnarray}
&& \mbox{Use the first component code to store $\lambda_0(1) = \frac{2\lambda_1 - \lambda_2}{2N-5} \geq 0$ qudits of $Q_0$} \notag\\
&& ~~~~\mbox{in $(Q_1, Q_2, \cdots, Q_N)$ by utilizing $((N-2)\lambda_0(1), \lambda_0(1), \cdots, \lambda_0(1))$ qudits of storage.} \notag\\
\label{eq:qd2}\\
&& \mbox{Use the second component code to store $\lambda_0(2) = \frac{(N-2)\lambda_2-\lambda_1}{2N-5} \geq 0$ qudits of $Q_0$} \notag\\
&& ~~~~\mbox{in $(Q_1, Q_2, \cdots, Q_N)$ by utilizing $(\lambda_0(2), 2\lambda_0(2), \cdots, 2\lambda_0(2))$ qudits of storage.} \label{eq:qd3}
\end{eqnarray}
The scheme works because we have stored the desired number of qudits of $Q_0$, $\lambda_0(1) + \lambda_0(2) = \frac{\lambda_1 + (N-3)\lambda_2}{2N-5}$, in $Q_1, Q_2, \cdots, Q_N$ and all storage qudits are fully used, $\lambda_1 = (N-2)\lambda_0(1) + \lambda_0(2), \lambda_2 = \lambda_0(1) + 2\lambda_0(2)$.
\item When $(N-2) \lambda_2 \leq \lambda_1$, $\min\left( \lambda_1, \lambda_2 , \frac{\lambda_1 + (N-3)\lambda_2}{2N-5} \right) = \lambda_2$. 
Use the first component code to store $\lambda_0 = \lambda_2$ qudits of $Q_0$ in $(Q_1, Q_2, \cdots, Q_N)$ by utilizing storage in the amounts $((N-2) \lambda_2, \lambda_2, \cdots, \lambda_2)$ qudits, respectively. The scheme works as $(N-2) \lambda_2 \leq \lambda_1$ and $Q_1$ might have extra unused qudits.
\end{enumerate}

The results of this section generalize (slightly) to cases where $\lambda_{N-1}, \lambda_N$ are unconstrained (so not necessarily equal to $\lambda_2$ and $\lambda_2 \leq \lambda_{N-1} \leq \lambda_N$) as they do not influence the upper bound (\ref{cap:wheel}) and the above code construction.

\subsubsection{$N=5$}
Note that the $N=4$ case is covered in Section \ref{ex:wheel} and we only need to consider $N=5$. We show that $C(\mathcal{W}_5) = \min(\lambda_1, \lambda_2, \frac{\lambda_1+\lambda_2}{3}, \frac{\lambda_1+\lambda_2+\lambda_3}{5})$ is achievable. The idea is similar to that in the above section, i.e., carefully combining component codes, so repetitive details are omitted and only differences are highlighted.

We first present three component codes, where the first two are identical to above.
\begin{enumerate}
\item $\lambda_0 = 1$ code for storage graph $\mathcal{W}_5$ with $(\lambda_1, \lambda_2, \lambda_3, \lambda_4, \lambda_5) = (3, 1, 1, 1, 1)$.
\item $\lambda_0 = 1$ code for storage graph $\mathcal{W}_5$ with $(\lambda_1, \lambda_2, \lambda_3, \lambda_4, \lambda_5) = (1, 2, 2, 2, 2)$.
\item $\lambda_0 = 1$ code for storage graph $\mathcal{W}_5$ with $(\lambda_1, \lambda_2, \lambda_3, \lambda_4, \lambda_5) = (2, 1, 2, 2, 2)$.

Set $\kappa = 1$ and $q$ is any prime power. Use the following classical (structured) secure storage code.
\begin{eqnarray}
&& Y_{Q_1} = (b_1, b_4), Y_{Q_2} = (a+b_1), Y_{Q_3} = (b_2, a+b_4), \\
&& Y_{Q_4} = (b_3, a+b_4), Y_{Q_5} = (a+b_2+b_3, a+b_4),
\end{eqnarray}
where $a, b_1, b_2, b_3, b_4$ are i.i.d. uniform elements in $\mathbb{F}_q$. Interestingly, a different alignment structure is applied here.
The classical constraints (\ref{css:dec}), (\ref{css:sec}) are satisfied and Theorem \ref{thm:css} produces the desired quantum code.
\end{enumerate}

Next we combine the above three component codes to achieve $C(\mathcal{W}_5) = \min(\lambda_1, \lambda_2, \frac{\lambda_1+\lambda_2}{3}, \frac{\lambda_1+\lambda_2+\lambda_3}{5})$. Depending on which term is the minimum, we have four cases. Choose a prime power $q \geq 4$.
Note that $\lambda_2 \leq \lambda_3 \leq \lambda_4 \leq \lambda_5$.
\begin{enumerate}
\item When $\min(\lambda_1, \lambda_2, \frac{\lambda_1+\lambda_2}{3}, \frac{\lambda_1+\lambda_2+\lambda_3}{5}) = \lambda_1$, we have $\lambda_1 \leq \lambda_2/2 \leq (\lambda_2+\lambda_3)/4$.
Use the second component code to store $\lambda_0 = \lambda_1$ qudits of $Q_0$ in $(Q_1, Q_2, Q_3, Q_4, Q_5)$ by utilizing storage in the amounts $(\lambda_1, 2\lambda_1, 2\lambda_1, 2\lambda_1, 2\lambda_1)$ qudits, respectively.
\item When $\min(\lambda_1, \lambda_2, \frac{\lambda_1+\lambda_2}{3}, \frac{\lambda_1+\lambda_2+\lambda_3}{5}) = \lambda_2$, we have $\lambda_2 \leq \lambda_1/2$ and $\lambda_2 \leq (\lambda_1+\lambda_3)/4$. There are two sub-cases here.
\begin{enumerate}
\item $\lambda_1 \leq \lambda_3$ or $2\lambda_2 \leq \lambda_3 \leq \lambda_1$: Use the third component code to store $\lambda_0 = \lambda_2$ qudits of $Q_0$ in $(Q_1, Q_2, Q_3, Q_4, Q_5)$ by utilizing storage in the amounts $(2\lambda_2, \lambda_2, 2\lambda_2, 2\lambda_2, 2\lambda_2)$ qudits, respectively, where $2\lambda_2 \leq \lambda_1$ and $2\lambda_2 \leq \lambda_3 \leq \lambda_4 \leq \lambda_5$. 
\item $\lambda_3 \leq \lambda_1$ and $\lambda_3 \leq 2\lambda_2$:
\begin{eqnarray}
&& \mbox{Use the first component code to store $\lambda_0(1) = 2\lambda_2 - \lambda_3 \geq 0$ qudits of $Q_0$} \notag\\
&& ~~~~\mbox{in $Q_1, Q_2, Q_3, Q_4, Q_5$ by utilizing $(3\lambda_0(1), \lambda_0(1),  \lambda_0(1),  \lambda_0(1), \lambda_0(1))$ qudits of storage.} \notag\\
\\
&& \mbox{Use the third component code to store $\lambda_0(2) = \lambda_3 - \lambda_2 \geq 0$ qudits of $Q_0$} \notag\\
&& ~~~~\mbox{in $Q_1, Q_2, Q_3, Q_4, Q_5$ by utilizing $(2\lambda_0(2), \lambda_0(2), 2\lambda_0(2), 2\lambda_0(2), 2\lambda_0(2))$ qudits of storage.}\notag\\
\end{eqnarray}
The scheme works because the number of stored qudits of $Q_0$ is $\lambda_0 = \lambda_0(1) + \lambda_0(2) = \lambda_2$, and the number of used storage qudits are
\begin{eqnarray}
Q_1: && 3\lambda_0(1) + 2\lambda_0(2) = 4\lambda_2 - \lambda_3 \leq \lambda_1, \\
Q_2: && \lambda_0(1) + \lambda_0(2) = \lambda_2, \\
Q_3, Q_4, Q_5: && \lambda_0(1) + 2\lambda_0(2) = \lambda_3 \leq \lambda_4 \leq \lambda_5.
\end{eqnarray}
\end{enumerate}
\item When $\min(\lambda_1, \lambda_2, \frac{\lambda_1+\lambda_2}{3}, \frac{\lambda_1+\lambda_2+\lambda_3}{5}) = \frac{\lambda_1+\lambda_2}{3}$, we have $2\lambda_1 \geq \lambda_2, 2\lambda_2 \geq \lambda_1, 3\lambda_3 \geq 2(\lambda_1+\lambda_2)$
and use\footnote{$(\lambda_1, \lambda_2, \lambda_3, \lambda_4, \lambda_5) \times \lambda_0$ denotes the coding scheme that stores $\lambda_0$ qudits of $Q_0$ in $(Q_1, Q_2, Q_3, Q_4, Q_5)$ by utilizing storage in the amounts $(\lambda_0\lambda_1, \lambda_0\lambda_2, \lambda_0\lambda_3, \lambda_0\lambda_4, \lambda_0\lambda_5)$ qudits, respectively.}
\begin{eqnarray}
&& \mbox{second component code}~(1, 2, 2, 2, 2) \times \frac{2\lambda_2 - \lambda_1}{3} ~(\geq 0), \label{eq:qd4} \\
&& \mbox{third component code}~(2, 1, 2, 2, 2) \times \frac{2\lambda_1 - \lambda_2}{3} ~(\geq 0). \label{eq:qd5}
\end{eqnarray}
The scheme works because
\begin{eqnarray}
Q_0: && \lambda_0 = \frac{2\lambda_2 - \lambda_1}{3} + \frac{2\lambda_1 - \lambda_2}{3} = \frac{\lambda_1+\lambda_2}{3}, \\
Q_1: && 1\times \frac{2\lambda_2 - \lambda_1}{3} + 2\times \frac{2\lambda_1 - \lambda_2}{3} = \lambda_1, \\
Q_2: && 2\times \frac{2\lambda_2 - \lambda_1}{3} + 1\times \frac{2\lambda_1 - \lambda_2}{3}  = \lambda_2, \\
Q_3, Q_4, Q_5: && 2\times \frac{2\lambda_2 - \lambda_1}{3} + 2\times \frac{2\lambda_1 - \lambda_2}{3}  = \frac{2(\lambda_1+\lambda_2)}{3} \leq \lambda_3 \leq \lambda_4 \leq \lambda_5.
\end{eqnarray}
\item When $\min(\lambda_1, \lambda_2, \frac{\lambda_1+\lambda_2}{3}, \frac{\lambda_1+\lambda_2+\lambda_3}{5}) = \frac{\lambda_1+\lambda_2+\lambda_3}{5}$, we have
\begin{eqnarray}
4\lambda_1 &\geq& \lambda_2 + \lambda_3, \label{eq:l1}\\
4\lambda_2 &\geq& \lambda_1 + \lambda_3, \label{eq:l2}\\
2(\lambda_1+\lambda_2) &\geq& 3\lambda_3 \label{eq:l3}
\end{eqnarray}
and use
\begin{eqnarray}
&& \mbox{first component code}~(3, 1, 1, 1, 1) \times \beta_1, ~\mbox{where}~\beta_1 = \frac{2(\lambda_1+\lambda_2) - 3\lambda_3}{5} \geq 0~\mbox{due to (\ref{eq:l3})}, \notag\\
\label{eq:qd6} \\
&& \mbox{second component code}~(1, 2, 2, 2, 2) \times \beta_2, ~\mbox{where}~\beta_2 = \frac{4\lambda_2 - (\lambda_1+\lambda_3)}{5} ~\geq 0~\mbox{due to (\ref{eq:l2})}, \notag\\
\label{eq:qd7}\\
&& \mbox{third component code}~(2, 1, 2, 2, 2) \times \beta_3,~\mbox{where}~\beta_3 =  (\lambda_3 - \lambda_2) ~\geq 0.
\end{eqnarray}
The scheme works because
\begin{eqnarray}
Q_0: && \lambda_0 = \beta_1 + \beta_2 + \beta_3,
= \frac{\lambda_1+\lambda_2+\lambda_3}{5}, \\
Q_1: && 3\beta_1 + \beta_2 + 2\beta_3 = \lambda_1, \\
Q_2: && \beta_1 + 2\beta_2 + \beta_3 = \lambda_2, \\
Q_3, Q_4, Q_5: &&  \beta_1 + 2\beta_2 + 2\beta_3 = \lambda_3 \leq \lambda_4 \leq \lambda_5.
\end{eqnarray}
\end{enumerate}

\subsection{Proof of Theorem \ref{thm:int}}\label{sec:int}
For uniform storage, $\lambda_i = 1, \forall i$. The intersection bound (\ref{up:intersection}) gives the converse as follows. 
\begin{eqnarray}
C_u(\sqcap_{\Delta,m}) &\leq& \Lambda( \mathcal{D}(e_1) \cap \mathcal{D}(e_2)) = \Lambda\left(\left\{Q_{\overline{\mathcal{S}}}: 1 \in \mathcal{S}, 2 \in \mathcal{S}, \mathcal{S} \in \binom{[\Delta]}{m} \right\} \right) \\
&=& \binom{\Delta-2}{m-2}.
\end{eqnarray}

For the achievability, we use Theorem \ref{thm:css} and translate from the following classical (structured) secure storage code. Set $\kappa = 1$ and $q$ as a prime power such that $q > \binom{\Delta-1}{m-1} + \Delta \binom{\Delta - 2}{m-2}$, 
and $k = \binom{\Delta - 2}{m-2}$, $\delta = \binom{\Delta-2}{m-1}$ so that $\bm{a} \in \mathbb{F}_q^{1 \times k}$ and $\bm{b} \in \mathbb{F}_q^{1\times \delta}$ (with i.i.d. elements in $\mathbb{F}_q$). Note that $k+\delta = \binom{\Delta-1}{m-1}$.
For the design of $Y_{Q}$, consider first $\mathcal{D}(e_1)$.
\begin{eqnarray}
\left(Y_{Q_{\overline{\mathcal{S}}}}: 1 \in \mathcal{S}, \mathcal{S} \in \binom{[\Delta]}{m} \right) = (\bm{a}, \bm{b}) \times \left( {\bf A}_{\mathcal{D}(e_1)} ; {\bf B}_{\mathcal{D}(e_1)} \right)  \in \mathbb{F}_q^{1 \times (k+\delta)}, \label{int:enc1}
\end{eqnarray}
where $( {\bf A}_{\mathcal{D}(e_1)} ; {\bf B}_{\mathcal{D}(e_1)} ) \in \mathbb{F}_q^{(k+\delta) \times (k+\delta)}$ is a square matrix that needs to satisfy some rank constraints so as to guarantee (\ref{css:dec}), (\ref{css:sec}) (refer to Lemma \ref{lemma:matrix}). For now, it suffices to imagine it as a sufficiently generic matrix. In particular, if we randomly generate each of its element over a large field, then it will work with high probability.

Consider next $\mathcal{D}^c(e_1)$. For any $Q_{\overline{\mathcal{S}}}, 1 \notin \mathcal{S}$, denote
\begin{eqnarray}
\mathcal{S} = \{i_1, i_2, \cdots, i_m\}, 1 < i_1 < i_2 < \cdots < i_m
\end{eqnarray}
and define the set where $i_j, j \in [m]$ in $\mathcal{S}$ is replaced by $1$ as $\mathcal{S}_j$, 
\begin{eqnarray}
\mathcal{S}_j \triangleq \{i_1, \cdots, i_{j-1}, 1, i_{j+1}, \cdots, i_m\}, j \in [m]. 
\end{eqnarray}
Then for any $\mathcal{S}$ such that $1 \notin \mathcal{S}, \mathcal{S} \in \binom{[\Delta]}{m}$, set 
\begin{eqnarray}
Y_{Q_{\overline{\mathcal{S}}}} = \sum_{j \in [m]} (-1)^{j} Y_{{Q_{\overline{\mathcal{S}_j}}}}. \label{int:enc}
\end{eqnarray}
The encoding is now complete. Next we show that this encoding satisfies some useful properties in the following lemma and the proof is deferred to Section \ref{sec:enc}.

\begin{lemma}\label{lemma:enc}
For the encoding in (\ref{int:enc}), the following three properties hold.
\begin{enumerate}
\item For any $i \in [2:\Delta]$, 
\begin{eqnarray}
\left(Y_{Q_{\overline{\mathcal{S}}}}: i \in \mathcal{S}, \mathcal{S} \in \binom{[\Delta]}{m} \right) \overset{\mbox{\scriptsize invertible}}{\longleftrightarrow} \left(Y_{Q_{\overline{\mathcal{S}}}}: 1 \in \mathcal{S}, \mathcal{S} \in \binom{[\Delta]}{m} \right). \label{enc:dec}
\end{eqnarray}
\item For any $i \in [2:\Delta]$, 
\begin{eqnarray}
\left(Y_{Q_{\overline{\mathcal{S}}}}: i \notin \mathcal{S}, \mathcal{S} \in \binom{[\Delta]}{m} \right) ~\mbox{is a function of}~\left(Y_{Q_{\overline{\mathcal{S}}}}: 1 \in \mathcal{S}, i \notin \mathcal{S}, \mathcal{S} \in \binom{[\Delta]}{m} \right). \label{enc:sec1}
\end{eqnarray}
\item 
\begin{eqnarray}
\left(Y_{Q_{\overline{\mathcal{S}}}}: 1 \notin \mathcal{S}, \mathcal{S} \in \binom{[\Delta]}{m} \right) ~\mbox{is a function of}~\left(Y_{Q_{\overline{\mathcal{S}}}}: 1 \notin \mathcal{S}, 2 \in \mathcal{S}, \mathcal{S} \in \binom{[\Delta]}{m} \right). \label{enc:sec2}
\end{eqnarray}
\end{enumerate}
\end{lemma}

\begin{remark}
Lemma \ref{lemma:enc} is intuitively described as follows. In (\ref{enc:dec}), the LHS is all classical shares in $\mathcal{D}(e_i), i \neq 1$ and the RHS is all classical shares in $\mathcal{D}(e_1)$; their invertibility means that if $\mathcal{D}(e_1)$ can decode the classical secret, then so can all $\mathcal{D}(e_i)$ (see (\ref{int:dec}) below). Referring to $\sqcap_{5,3}$ in Figure \ref{fig:intgraphs} for an example, $\mathcal{D}(e_2)$ contains classical shares $(z_1, z_4, z_5, z_4-z_1+z_2, z_5-z_1+z_3, z_5-z_4+z_6)$, which are invertible to $(z_1, z_2, z_3, z_4, z_5, z_6)$, all classical shares in $\mathcal{D}(e_1)$. In (\ref{enc:sec1}), the LHS is all classical shares in $\mathcal{D}^c(e_i), i \neq 1$ and the RHS is a subset where $1$ belongs to the label. 
The property that LHS is a function of the RHS means that for security, we need to show the noise ($b_i$ terms) in the RHS spans the full space thus leaking no information (see (\ref{int:sec1}) below). Referring to $\sqcap_{5,3}$ in Figure \ref{fig:intgraphs}, $\mathcal{D}^c(e_2)$ contains classical shares $(z_2, z_3, z_6, z_6-z_3+z_2)$ where $z_6+z_3+z_2$ is a function of the remaining three (RHS). Regarding alignment, $z_6-z_3+z_2$ is aligned in the space spanned by $z_2, z_3, z_6$ so that it cannot contribute additional information. In (\ref{enc:sec2}), the LHS is all classical shares in $\mathcal{D}^c(e_1)$, which are a function of the RHS, a subset of LHS.
Thus for security, it suffices to show the noise in the RHS has full rank (see (\ref{int:sec2}) below). Referring to $\sqcap_{5,3}$ in Figure \ref{fig:intgraphs}, $\mathcal{D}^c(e_1)$ contains $(z_4-z_1+z_2, z_5-z_1+z_3, z_5-z_4+z_6, z_6-z_3+z_2)$ where the last element $z_6-z_3+z_2$ is a function of the first three elements (alignment), i.e., $(z_4-z_1+z_2) - (z_5-z_1+z_3) + (z_5-z_4+z_6)$. Thanks to such alignment, the dimension in the complement of a decoding set, $\binom{\Delta-1}{m}$ is reduced to the dimension of noise, $\binom{\Delta-2}{m-1}$.
\end{remark}

Equipped with the above properties, we proceed to consider the decoding constraint (\ref{css:dec}) and the security constraint (\ref{css:sec}).

First, consider (\ref{css:dec}). For any decoding set $\mathcal{D}(e_i), i \in [\Delta]$, from (\ref{int:enc1})  and (\ref{enc:dec}), (\ref{css:dec}) is equivalent to that
\begin{eqnarray}
\mbox{square matrix}~\left( {\bf A}_{\mathcal{D}(e_1)}; {\bf B}_{\mathcal{D}(e_1)} \right) ~\mbox{has full rank}~k+\delta \label{int:dec}
\end{eqnarray}
because for $\mathcal{D}(e_1)$, if the square matrix $( {\bf A}_{\mathcal{D}(e_1)}; {\bf B}_{\mathcal{D}(e_1)} )$ has full rank, then by  (\ref{int:enc1}), we can decode all $\bm{a}, \bm{b}$ from $\mathcal{Y}_{\mathcal{D}(e_1)}$ and for $\mathcal{D}(e_i), i \in [2:\Delta]$, by (\ref{enc:dec}),  $\mathcal{Y}_{\mathcal{D}(e_i)}$ is invertible to $\mathcal{Y}_{\mathcal{D}(e_1)}$ so that the full rank of $( {\bf A}_{\mathcal{D}(e_1)}; {\bf B}_{\mathcal{D}(e_1)} )$ also suffices. 
 
Second, consider (\ref{css:sec}) for $\mathcal{D}(e_i), i \in [2:\Delta]$. From (\ref{enc:sec1}), we know that $\mathcal{Y}_{\mathcal{D}^c(e_i)}$ is a function of $\left(Y_{Q_{\overline{\mathcal{S}}}}: 1 \in \mathcal{S}, i \notin \mathcal{S}, \mathcal{S} \in \binom{\Delta}{m} \right)$ and denote
\begin{eqnarray}
\left(Y_{Q_{\overline{\mathcal{S}}}}: 1 \in \mathcal{S}, i \notin \mathcal{S}, \mathcal{S} \in \binom{[\Delta]}{m} \right) \triangleq (\bm{a}, \bm{b}) \times ({\bf A}_{\mathcal{D}^c_{i}}; {\bf B}_{\mathcal{D}^c_{i}}),
\end{eqnarray}
where $ ({\bf A}_{\mathcal{D}^c_{i}}; {\bf B}_{\mathcal{D}^c_{i}}) \in \mathbb{F}_q^{(k+\delta) \times \delta}$ is a sub-matrix of $ ({\bf A}_{\mathcal{D}(e_1)}; {\bf B}_{\mathcal{D}(e_1)}) $ (with corresponding columns that produce the above $Y_{Q_{\overline{\mathcal{S}}}}$). Then (\ref{css:sec}) and equivalently (\ref{css:rank2}) hold if 
\begin{eqnarray}
\mbox{rank} \left( {\bf A}_{\mathcal{D}^c_{i}}; {\bf B}_{\mathcal{D}^c_i} \right) = \mbox{rank}\left( {\bf B}_{\mathcal{D}^c_i}  \right), \label{int:sec1r}
\end{eqnarray}
which is further equivalent to 
\begin{eqnarray}
~\mbox{square matrix}~{\bf B}_{\mathcal{D}^c_i}  \in \mathbb{F}_q^{\delta \times \delta}~\mbox{has full rank}, i \in [2:\Delta], \label{int:sec1}
\end{eqnarray}
as $({\bf A}_{\mathcal{D}^c_i}  ;  {\bf B}_{\mathcal{D}^c_i}  )$ has $\delta$ columns and its rank is no smaller than the rank of ${\bf B}_{\mathcal{D}^c_i} $. 

Third, consider (\ref{css:sec}) for $\mathcal{D}(e_1)$. From (\ref{enc:sec2}), we know that $\mathcal{Y}_{\mathcal{D}^c(e_1)}$ is a function of 
\begin{eqnarray}
\left(Y_{Q_{\overline{\mathcal{S}}}}: 1 \notin \mathcal{S}, 2 \in \mathcal{S}, \mathcal{S} \in \binom{[\Delta]}{m} \right) \triangleq (\bm{a}, \bm{b}) \times \left( {\bf A}_{\mathcal{D}^c_1} ; {\bf B}_{\mathcal{D}^c_1}  \right),
\end{eqnarray}
where $ ( {\bf A}_{\mathcal{D}^c_1}; {\bf B}_{\mathcal{D}^c_1} ) \in \mathbb{F}_q^{(k+\delta) \times \delta}$ can be written as linear combinations of the columns of $({\bf A}_{\mathcal{D}(e_1)}; {\bf B}_{\mathcal{D}(e_1)})$ because from (\ref{int:enc}), $Y_{Q_{\overline{\mathcal{S}}}}, 1 \notin \mathcal{S}$ is coded through $Y_{Q_{\overline{\mathcal{S}}}}, 1 \in \mathcal{S}$, i.e., $( {\bf A}_{\mathcal{D}(e_1)} ; {\bf B}_{\mathcal{D}(e_1)} )$. 
Then (\ref{css:sec}) and equivalently (\ref{css:rank2}) hold if 
\begin{eqnarray}
~\mbox{square matrix}~ {\bf B}_{\mathcal{D}^c_1} \in \mathbb{F}_q^{\delta \times \delta}~\mbox{has full rank}. \label{int:sec2}
\end{eqnarray}

We have now established that (\ref{css:dec}), (\ref{css:sec}) hold if (\ref{int:dec}), (\ref{int:sec1}), (\ref{int:sec2}) hold, which can be guaranteed according to the following lemma. So the proof is complete.

\begin{lemma}\label{lemma:matrix}
When the field size $q > (k+\delta) + \Delta \delta = \binom{\Delta-1}{m-1} + \Delta \binom{\Delta - 2}{m-2}$, there exists an assignment of matrix $({\bf A}_{\mathcal{D}(e_1)}; {\bf B}_{\mathcal{D}(e_1)})$ such that (\ref{int:dec}), (\ref{int:sec1}), (\ref{int:sec2}) are satisfied.
\end{lemma}

{\it Proof:} Take the product of the determinants of the square matrices in (\ref{int:dec}), (\ref{int:sec1}), (\ref{int:sec2}) and view it as a polynomial, denoted as $\Pi$, whose variables are the $(k+\delta)^2$ elements of the square matrix $({\bf A}_{\mathcal{D}(e_1)}; {\bf B}_{\mathcal{D}(e_1)})$. $\Pi$ has degree $(k+\delta) + \Delta \delta = \binom{\Delta-1}{m-1} + \Delta \binom{\Delta - 2}{m-2} < q$. 

The polynomial $\Pi$ is not the zero polynomial because there exists a realization of the $(k+\delta)^2$ variables such that any single matrix in (\ref{int:dec}), (\ref{int:sec1}), (\ref{int:sec2}) has full rank 
(for (\ref{int:sec2}), note that each $Y_{Q_{\overline{\mathcal{S}}}}$ where $1 \in \mathcal{S}, 2\notin \mathcal{S}$ appears in one distinct $Y_{Q_{\overline{\mathcal{S}'}}}$ where $1 \notin \mathcal{S}', 2\in \mathcal{S}'$). 
Sample the $(k+\delta)^2$ variables independently and uniformly from $\mathbb{F}_q$ so that by the Schwartz–Zippel lemma, the probability of $\Pi$ being zero is no greater than the degree of $\Pi$ over $q$, which is strictly smaller than $1$. Therefore there exists an assignment of $({\bf A}_{\mathcal{D}(e_1)}; {\bf B}_{\mathcal{D}(e_1)})$ such that the matrices in (\ref{int:dec}), (\ref{int:sec1}), (\ref{int:sec2}) all have full rank.

\hfill\qed

\subsubsection{Proof of Lemma \ref{lemma:enc}}\label{sec:enc}
First, we prove (\ref{enc:dec}). Consider any $i \in [2:\Delta]$.
\begin{eqnarray}
&& \left(Y_{Q_{\overline{\mathcal{S}}}}: i \in \mathcal{S}, \mathcal{S} \in \binom{[\Delta]}{m} \right) \notag \\
&\overset{\mbox{\scriptsize invertible}}{\longleftrightarrow}& \left(Y_{Q_{\overline{\mathcal{S}}}}: 1 \in \mathcal{S}, i \in \mathcal{S}, \mathcal{S} \in \binom{[\Delta]}{m}, Y_{Q_{\overline{\mathcal{S}}}}: 1 \notin \mathcal{S}, i \in \mathcal{S}, \mathcal{S} \in \binom{[\Delta]}{m} \right). \label{enc:1}
\end{eqnarray}
Note that from (\ref{int:enc}), for any $\mathcal{S} = \{i_1, \cdots, i_m\}$ such that $1 \notin \mathcal{S}, i \in \mathcal{S}$, 
\begin{eqnarray}
Y_{Q_{\overline{\mathcal{S}}}} = \sum_{j \in [m], i_j \neq i} (-1)^{j} Y_{Q_{\overline{\mathcal{S}_j}}} + \sum_{j \in [m], i_j = i} (-1)^{j} Y_{Q_{\overline{\mathcal{S}_j}}},
\end{eqnarray}
where the first sum term contains $m-1$ terms and each term, $(-1)^{j} Y_{Q_{\overline{\mathcal{S}_j}}}$ satisfies that $1 \in \mathcal{S}_j, i \in \mathcal{S}_j$; and the second sum term contains only $1$ term, $(-1)^j Y_{Q_{\overline{\mathcal{S}_j}}}, i_j = i$ so that $\mathcal{S}_j = \{i_1, i_2, \cdots, i_{j-1}, 1, i_{j+1}, \cdots, i_m\}$ and $1 \in \mathcal{S}_j, i \notin \mathcal{S}_j$. Plugging this property in the second term of the right-hand-side (RHS) of (\ref{enc:1}), we eliminate the contribution of the first term of RHS of (\ref{enc:1}) in the second term,
\begin{eqnarray}
&& \left(Y_{Q_{\overline{\mathcal{S}}}}: i \in \mathcal{S}, \mathcal{S} \in \binom{[\Delta]}{m} \right) \notag\\
&\overset{\mbox{\scriptsize invertible}}{\longleftrightarrow}& \left(Y_{Q_{\overline{\mathcal{S}}}}: 1 \in \mathcal{S}, i \in \mathcal{S}, \mathcal{S} \in \binom{[\Delta]}{m}, Y_{Q_{\overline{\mathcal{S}_j}}}: 1 \in \mathcal{S}_j, i \notin \mathcal{S}_j, \mathcal{S} \in \binom{[\Delta]}{m} \right)\\
&=& \left(Y_{Q_{\overline{\mathcal{S}}}}: 1 \in \mathcal{S}, i \in \mathcal{S}, \mathcal{S} \in \binom{[\Delta]}{m}, Y_{Q_{\overline{\mathcal{S}_j}}}: 1 \in \mathcal{S}_j, i \notin \mathcal{S}_j, \mathcal{S}_j \in \binom{[\Delta]}{m} \right) \label{enc:2}\\
&\overset{\mbox{\scriptsize invertible}}{\longleftrightarrow}&  \left(Y_{Q_{\overline{\mathcal{S}}}}: 1 \in \mathcal{S}, \mathcal{S} \in \binom{[\Delta]}{m} \right),
\end{eqnarray}
where (\ref{enc:2}) follows from the fact that for distinct $\mathcal{S}$, $\mathcal{S}_j$ are also distinct, i.e., if $\mathcal{S}' \neq \mathcal{S}$, then $\mathcal{S}_j'$ is obtained from $\mathcal{S}'$ by replacing $i$ by $1$ and is not equal to $\mathcal{S}_j$ which is obtained from $\mathcal{S}$ by replacing $i$ by $1$.

Second, we prove (\ref{enc:sec1}) as a simple consequence of (\ref{int:enc}). Consider any $i \in [2:\Delta]$.
\begin{eqnarray}
&& \left(Y_{Q_{\overline{\mathcal{S}}}}: i \notin \mathcal{S}, \mathcal{S} \in \binom{[\Delta]}{m} \right) \notag\\
&\overset{\mbox{\scriptsize invertible}}{\longleftrightarrow}& \left( Y_{Q_{\overline{\mathcal{S}}}}: 1 \in \mathcal{S}, i \notin \mathcal{S}, \mathcal{S} \in \binom{[\Delta]}{m}, Y_{Q_{\overline{\mathcal{S}}}}: 1 \notin \mathcal{S}, i \notin \mathcal{S}, \mathcal{S} \in \binom{[\Delta]}{m} \right), \label{enc:3}
\end{eqnarray}
where the second term of RHS of (\ref{enc:3}) is a function of the first term due to the encoding (\ref{int:enc}), so $ \left(Y_{Q_{\overline{\mathcal{S}}}}: i \notin \mathcal{S}, \mathcal{S} \in \binom{\Delta}{m} \right) $ is a function of the first term of RHS of (\ref{enc:3}) and this is our desired claim.

Third, we prove (\ref{enc:sec2}). It suffices to show that
\begin{eqnarray}
\left(Y_{Q_{\overline{\mathcal{S}}}}: 1 \notin \mathcal{S}, 2 \notin \mathcal{S}, \mathcal{S} \in \binom{[\Delta]}{m} \right) ~\mbox{is a function of}~\left(Y_{Q_{\overline{\mathcal{S}}}}: 1 \notin \mathcal{S}, 2 \in \mathcal{S}, \mathcal{S} \in \binom{[\Delta]}{m} \right).
\end{eqnarray}
To this end, consider any $\mathcal{S} = \{i_1, \cdots, i_m\}$ such that $1 \notin \mathcal{S}, 2 \notin \mathcal{S}, \mathcal{S} \in \binom{[\Delta]}{m}$. From (\ref{int:enc}), we have
\begin{eqnarray}
Y_{Q_{\overline{\mathcal{S}}}} = \sum_{j \in [m]} (-1)^{j} Y_{Q_{\overline{\mathcal{S}_j}}},
\end{eqnarray}
where $\mathcal{S}_j = \{i_1, \cdots, i_{j-1}, 1, i_{j+1}, \cdots, i_m\}$. Define $\mathcal{S}'_j = \{i_1, \cdots, i_{j-1}, 2, i_{j+1}, \cdots, i_m\}$ and it turns out that
\begin{eqnarray}
\sum_{j \in [m]} (-1)^{j} Y_{Q_{\overline{\mathcal{S}_j}}} = - \sum_{j \in [m]} (-1)^{j} Y_{Q_{\overline{\mathcal{S}'_j}}} \label{enc:5}
\end{eqnarray}
and the desired claim follows. So it remains to prove (\ref{enc:5}). We plug in the encoding function (\ref{int:enc}) to $Y_{Q_{\overline{\mathcal{S}'_j}}}$, where $\mathcal{S}'_j = \{2, i_1, \cdots, i_{j-1}, i_{j+1}, \cdots, i_m\}$ (note that $2 < i_1 < \cdots < i_{j-1} < i_{j+1} < \cdots i_m$).
\begin{eqnarray}
Y_{Q_{\overline{\mathcal{S}'_j}}} &=& - Y_{Q_{\overline{\{1, i_1, \cdots, i_{j-1}, i_{j+1}, \cdots, i_m\}}}} + Y_{Q_{\overline{\{2, 1, \cdots, i_{j-1}, i_{j+1}, \cdots, i_m\}}}} - \cdots \notag\\
&&~+(-1)^{j} Y_{Q_{\overline{\{2, i_1, \cdots, 1, i_{j+1}, \cdots, i_m\}}}} +(-1)^{j+1} Y_{Q_{\overline{\{2, i_1, \cdots, i_j, 1, \cdots, i_m\}}}} + \cdots \notag\\
&&~+(-1)^m Y_{Q_{\overline{\{2, i_1, \cdots, i_{j-1}, i_{j+1}, \cdots, 1\}}}} \label{enc:6}
\end{eqnarray}
and we show that the left-hand-side (LHS) of (\ref{enc:5}) is equal to RHS of (\ref{enc:5}). After expanding through (\ref{enc:6}), RHS of (\ref{enc:5}) contains more $Y_{Q_{\overline{\mathcal{S}}}}, 1 \in \mathcal{S}$ terms than those in LHS of (\ref{enc:5}). So we show that for the $Y_{Q_{\overline{\mathcal{S}}}}, 1 \in \mathcal{S}$ terms that exist in LHS, RHS has the same term and for the remaining $Y_{Q_{\overline{\mathcal{S}}}}, 1 \in \mathcal{S}$ terms that do not exist in LHS, they will cancel in RHS.

Consider any $Y_{Q_{\overline{\mathcal{T}}}}$ that is in LHS of (\ref{enc:5}), i.e., $\mathcal{T} =  \{i_1, \cdots, i_{j-1}, 1, i_{j+1}, \cdots, i_m\}$ and it is equal to $(-1)^{j} Y_{Q_{\mathcal{T}}}$ in LHS. In RHS, we have the same term because it appears as the first term in (\ref{enc:6}), i.e., $-(-1)^j \times (-Y_{Q_{\overline{\mathcal{T}}}})$. 

Consider any remaining $Y_{Q_{\overline{\mathcal{T}}}}$, i.e., $1 \in \mathcal{T}, 2 \in \mathcal{T}$ and $m-2$ elements from $i_1, \cdots, i_{j-1}, i_{j+1}, \cdots, i_m$ are in $\mathcal{T}$. For any $m-2$ such elements, $Y_{Q_{\overline{\mathcal{T}}}}$ will cancel in RHS of (\ref{enc:5}). We have two cases.
\begin{enumerate}
\item The $m-2$ elements are $i_1, \cdots, i_{t-1}, i_{t+1}, \cdots, i_{j-1}, i_{j+1}, \cdots, i_m$, i.e., $i_t$ is missing and $1 \leq t \leq j-1$.
$Y_{Q_{\overline{\mathcal{T}}}}, \mathcal{T} = \{1,2, i_1, \cdots, i_{t-1}, i_{t+1}, \cdots, i_{j-1}, i_{j+1}, \cdots, i_m\}$ will cancel in RHS of (\ref{enc:5}) because it appears twice - for the first time $\mathcal{S}'_j = \{i_1, \cdots, i_t, \cdots i_{j-1}, 2, i_{j+1}, \cdots, i_m\}$ and $i_t$ is replaced by $1$ to become $\mathcal{T}$ in (\ref{enc:6}), i.e., $(-1)^j \times (-1)^{t+1} Y_{Q_{\overline{\mathcal{T}}}}$; for the second time, $\mathcal{S}_t' = \{i_1, \cdots, i_{t-1}, 2, i_{t+1}, \cdots, i_{j-1}, i_j, i_{j+1}, \cdots, i_m\}$ and $i_j$ is replaced by $1$ to become $\mathcal{T}$ in (\ref{enc:6}), i.e., $(-1)^t \times (-1)^j Y_{Q_{\overline{\mathcal{T}}}}$; thus the two $Y_{Q_{\overline{\mathcal{T}}}}$ terms have different signs and cancel.
\item The $m-2$ elements are $i_1, \cdots, i_{j-1}, i_{j+1}, \cdots, i_{t-1}, i_{t+1}, \cdots, i_m$, i.e., $i_t$ is missing and $j+1 \leq t \leq m$.
$Y_{Q_{\overline{\mathcal{T}}}}, \mathcal{T} = \{1,2, i_1, \cdots, i_{j-1}, i_{j+1}, \cdots, i_{t-1}, i_{t+1}, \cdots, i_m\}$ will cancel in RHS of (\ref{enc:5}) because it appears twice - for the first time $\mathcal{S}'_j = \{i_1, \cdots i_{j-1}, 2, i_{j+1}, \cdots, i_t, \cdots, i_m\}$ and $i_t$ is replaced by $1$ to become $\mathcal{T}$ in (\ref{enc:6}), i.e., $(-1)^j \times (-1)^{t} Y_{Q_{\overline{\mathcal{T}}}}$; for the second time, $\mathcal{S}_t' = \{i_1, \cdots, i_{j-1}, i_j, i_{j+1}, \cdots, i_{t-1}, 2, i_{t+1}, \cdots, i_m\}$ and $i_j$ is replaced by $1$ to become $\mathcal{T}$ in (\ref{enc:6}), i.e., $(-1)^t \times (-1)^{j+1} Y_{Q_{\overline{\mathcal{T}}}}$; thus the two $Y_{Q_{\overline{\mathcal{T}}}}$ terms have different signs and cancel.
\end{enumerate} 

\subsection{Proof of Theorem \ref{thm:strong}}\label{sec:strong}

First, consider the `if' direction. We show that if the condition in Theorem \ref{thm:strong} is satisfied for $C(\mathcal{G})$, $\mathcal{G}=((\lambda_1, \cdots, \lambda_N),\mathcal{E})$ , then for any $\mathcal{G}'=((\lambda_1', \cdots, \lambda'_{N'}),\mathcal{E}')$
 such that $\mathcal{G}$ is a proper subgraph of $\mathcal{G}'$ we have $C(\mathcal{G}') = 0$. We have two cases for the proper subgraph.
\begin{enumerate}
\item $N = N'$ and $\mathcal{E}'$ contains at least one more hyperedge that is not in $\mathcal{E}$, denoted as $e'$. Then by the condition in Theorem \ref{thm:strong}, there exists one $e \in \mathcal{E}$ such that $\mathcal{D}(e') \cap \mathcal{D}(e) \in \{\emptyset, \mathcal{D}(e) \}$.  
\begin{enumerate}
\item $ \mathcal{D}(e') \cap  \mathcal{D}(e) = \emptyset$. By the intersection bound (\ref{up:intersection}), we have $C(\mathcal{G}') \leq \Lambda( \mathcal{D}(e') \cap  \mathcal{D}(e)) = 0$, as desired.
\item $ \mathcal{D}(e') \cap  \mathcal{D}(e) =  \mathcal{D}(e)$. Then $ \mathcal{D}(e')$ is redundant for $\mathcal{G}'$ as a strict subset of $ \mathcal{D}(e')$, $\mathcal{D}(e)$ is also a decoding set. Note that in our definition of storage graphs, we do not include redundant decoding sets so that this case violates the assumption that $\mathcal{G}$ is a proper subgraph of $\mathcal{G}'$ and cannot happen.
\end{enumerate}
\item $N' > N$, i.e., $\mathcal{G}'$ contains at least one more storage node that is not in $\mathcal{G}$, say $Q_{N+1}$. As we do not include redundant storage nodes in our definition of storage graphs, $\mathcal{G}'$ must contain one decoding set that includes $Q_{N+1}$, denoted as $\mathcal{D}(e') = \{\mathcal{S}, Q_{N+1}\}$ where $\mathcal{S} \subset \{Q_1, \cdots, Q_N\}$. Then by the condition in Theorem \ref{thm:strong}, there exists one $e \in \mathcal{E}$ such that $\mathcal{S} \cap \mathcal{D}(e) \in \{\emptyset, \mathcal{D}(e)\}$.
\begin{enumerate}
\item $\mathcal{S} \cap \mathcal{D}(e) = \emptyset$. By the intersection bound (\ref{up:intersection}), we have $C(\mathcal{G}') \leq \Lambda( \mathcal{D}(e') \cap \mathcal{D}(e) ) = 0$.
\item $\mathcal{S} \cap \mathcal{D}(e) = \mathcal{D}(e)$. Then $\mathcal{D}(e')$ is a redundant decoding set for $\mathcal{G}'$ and $\mathcal{G}$ is not a proper subgraph of $\mathcal{G}'$.
\end{enumerate}   
\end{enumerate}

Second, consider the `only if' direction. We show that if the condition in Theorem \ref{thm:strong} is not satisfied for $C(\mathcal{G})$, then there exists $\mathcal{G}'$ such that $\mathcal{G}$ is a proper subgraph of $\mathcal{G}'$ and $C(\mathcal{G}') > 0$. 

When the condition in Theorem \ref{thm:strong} is not satisfied, there exists a set of storage nodes, denoted as $\mathcal{S} \subset \{Q_1,\cdots,Q_N\}$ so that for any decoding set $\mathcal{D}(e), e \in \mathcal{E}$, we have $\mathcal{S} \cap \mathcal{D}(e) \notin \{\emptyset, \mathcal{D}(e)\}$, i.e., $\mathcal{S}$ intersects with every decoding set $\mathcal{D}(e)$ and is not equal to any $\mathcal{D}(e)$. As a result, we can include $\mathcal{S}$ as an additional non-redundant decoding set for $\mathcal{G}'$ and $C(\mathcal{G}')>0$ as $\mathcal{G}'$ satisfies the condition for $C > 0$ in Corollary \ref{thm:small} (i.e., any two distinct decoding sets are intersecting).

\subsection{Proof of Corollary \ref{cor:max}}\label{sec:max}
\begin{enumerate}
\item MDS graph $\mathcal{M}_{N, K}$. First, consider strong maximality. Suppose $2K- N = 1$ and consider any $\mathcal{S} \subset \{Q_1,\cdots, Q_N\}$. When $|\mathcal{S}| < K$, $\mathcal{S}^c$ must contain a decoding set as a subset, thus $\mathcal{S}$ is completely excluded by this decoding set; when $|\mathcal{S}| \geq K$, $\mathcal{S}$ must include a decoding set as a subset, i.e., this decoding set is completely included by $\mathcal{S}$. So the condition of Theorem \ref{thm:strong} is satisfied and $\mathcal{M}_{2K-1, K}$ is strongly maximal. Suppose $2K - N > 1$ and consider $\mathcal{S} = \{Q_1, \cdots, Q_{K-1}\}$. Note that $\mathcal{S}$ intersects with any decoding set $\mathcal{D}(e)$, where $|\mathcal{D}(e)| = K$ as $K-1 + K - N > 0$, so there is no decoding set that is completely included or excluded by $\mathcal{S}$, i.e., the condition in Theorem \ref{thm:strong} is not satisfied and $\mathcal{M}_{N,K}, 2K-N>1$ is not strongly maximal. To summarize, we have proved that $\mathcal{M}_{N,K}$ is strongly maximal if and only if $2K-N = 1$.

Second, consider weak maximality. By definition, strongly maximal storage graphs are weakly maximal, so we only need to consider the cases not covered above, i.e., when $2K-N > 1$. For all $\mathcal{G}' = ( (\lambda'_1, \cdots, \lambda_N'), \mathcal{E}')$ 
such that $\mathcal{G} = ( \{\lambda_1,\cdots, \lambda_N\}, \mathcal{E}) = \mathcal{M}_{N,K}$ is a proper subgraph of $\mathcal{G}'$, there must exist (non-redundant) $e' \in \mathcal{E}'$ such that for all $e \in \mathcal{E}$, $\mathcal{D}(e') \cap \mathcal{D}(e) \notin \{\emptyset, \mathcal{D}(e)\}$.
\begin{enumerate}
\item $\mathcal{D}(e') \subset \{Q_1, \cdots, Q_N\}$.  Then $|\mathcal{D}(e')| < K$ and $\mathcal{D}(e')$ intersects with every $\mathcal{D}(e), e \in \mathcal{E}$. Consider $\mathcal{D}(e)$ such that $|\mathcal{D}(e') \cap \mathcal{D}(e)|$ is the smallest among all $e \in \mathcal{E}$. By the intersection bound (\ref{up:intersection}), $C(\mathcal{G}') \leq |\mathcal{D}(e') \cap \mathcal{D}(e)| < 2K-N = C(\mathcal{G}) = C(\mathcal{M}_{N,K})$.
\item $\mathcal{D}(e') \not\subset \{Q_1, \cdots, Q_N\}$. Then $\mathcal{D}(e')$ contains a subset $\mathcal{S}$ such that $\mathcal{S} \subset \{Q_1,\cdots, Q_N\}$, $\mathcal{S} \cap \mathcal{D}(e) \notin \{\emptyset, \mathcal{D}(e)\}$ and then $|\mathcal{S}| < K$. The same argument as above shows that $C(\mathcal{G}') < C(\mathcal{G})$.
\end{enumerate}
So the definition of weak maximality is satisfied.

\item Wheel graph $\mathcal{W}_N$. We show that $\mathcal{W}_N, N \geq 4$ is strongly maximal by verifying the condition in Theorem \ref{thm:strong}. Consider any $\mathcal{S} \subset \mathcal{Q} = \{Q_1, \cdots, Q_N\}$.
\begin{enumerate}
\item When $Q_1 \in \mathcal{S}$, then $\mathcal{S}$ must be decoding set $\{Q_1, Q_i\}$ for some $i \in [2:N]$ or must include decoding set $\{Q_1, Q_i\}$ for some $i \in [2:N]$. Thus there exists a decoding set that is completely included by $\mathcal{S}$.
\item When $Q_1 \notin \mathcal{S}$, then $\mathcal{S}$ is either equal to decoding set $\{Q_2, \cdots, Q_N\}$ or is completely excluded by decoding set $\{Q_1, Q_i\}$ for some $i \in [2:N]$.
\end{enumerate}

\item Fano graph $\mathcal{F}_7$. We show that $\mathcal{F}_7$ is strongly maximal by verifying the condition in Theorem \ref{thm:strong}. As $\mathcal{F}_7$ only contains $7$ storage nodes, we may verify through checking all $\mathcal{S} \subset \mathcal{Q} = \{Q_1, \cdots, Q_7\}$. When $|\mathcal{S}| \geq 4$, $\mathcal{S}$ either contains a decoding set or is equal to the complement of a decoding set. When $|\mathcal{S}| \leq 3$, $\mathcal{S}$ is either a decoding set or is a subset of the complement of some decoding set.

\item Intersection graph $\sqcap_{\Delta, m}$. As strongly maximal storage graphs are by definition also weakly maximal storage graphs, we only need to prove the if direction for strong maximality (when $\Delta = 3, m = 2$, $\sqcap_{3,2}$ is isomorphic to MDS graph $\mathcal{M}_{3,2}$, so this has been proved above) and the only if direction for weak maximality, presented next. We show that we may include a non-redundant decoding set to $\sqcap_{\Delta, m}$ that intersects with every existing decoding sets while the capacity remains the same so $\sqcap_{\Delta, m}, (\Delta, m) \neq (3,2), \Delta > m$ is not weakly maximal. Specifically, for $\mathcal{G} =  ((\lambda_1,\cdots,\lambda_N), \mathcal{E}) = \sqcap_{\Delta, m}$, we set $\mathcal{G}' = ((\lambda_1,\cdots,\lambda_N), \mathcal{E}')$ where $\mathcal{E}' = \mathcal{E} \cup e^*$ and
\begin{eqnarray}
\mathcal{D}(e^*) = \{ \mathcal{D}(e_2) \setminus Q_{\overline{\{2, \Delta - m + 2, \cdots, \Delta - 1, \Delta \}}} \} \cup \left\{ Q_{\overline{ \{i, \Delta - m +2, \cdots, \Delta - 1, \Delta \}}} : i \in \{ 1, 3, \cdots, \Delta - m + 1 \} \right\}, \label{key} \notag\\
\end{eqnarray}
i.e., the additional decoding set is chosen as deleting one element in $\mathcal{D}(e_2)$ and including $\Delta - m$ elements (specifically, deleting the last one in lexicographic order and including through replacing $2$ by each of $\{1, 3, \cdots, \Delta - m + 1\}$ in the label). $\mathcal{D}(e^*) \cap \mathcal{D}(e) \notin \{\emptyset, \mathcal{D}(e)\}$ for all $e \in \mathcal{E}$ and $C(\mathcal{G}') = C(\mathcal{G})$ because the same code for $\sqcap_{\Delta,m}$ in Section \ref{sec:int} continues to work.
The decoding and security constraint of the corresponding classical code for decoding set $\mathcal{D}(e^*)$ is guaranteed by proving similar arguments to (\ref{enc:dec}), (\ref{enc:sec1}).
\end{enumerate}

\section{Discussion and Open Problems}
In this work, we have explored the capacity of storing a  quantum message over a number of storage nodes of certain specified relative sizes such that from certain specified subsets of the storage nodes, the message must be perfectly recovered. The problem is surprisingly rich as it is intimately related to a corresponding classical secure storage problem. This stands in sharp contrast to the problem of storing a \emph{classical} message (over either classical or quantum storage nodes) where the cut-set bound (Holevo bound for quantum storage nodes) is tight; such a significant difference stems from the distinct properties of quantum systems, in particular the no-cloning theorem that introduces additional security constraints. The quantum storage problem turns out to be quite challenging as the related (via CSS codes) classical secure storage problem can be equivalently formulated as a secure network coding problem, which is known to be hard \cite{Huang_Ho_Langberg_Kliewer}. In light of this connection, we have applied an interference alignment perspective to obtain various code constructions and provided tight converse bounds through quantum information inequalities. As an initial step towards an important capacity problem that has not been considered in the literature, there are many interesting future research avenues and open problems and we list a few in the following.

\begin{enumerate}
\item Is the achievable scheme in Theorem \ref{thm:css} always tight? Equivalently, we may view the classical secure storage problem (\ref{css:dec}), (\ref{css:sec}) as a stand-alone problem and ask if the capacity of this stand-alone classical secure storage problem is always equal to the capacity of the quantum storage problem, $C(\mathcal{G})$? For the cases settled in this work, the answer is yes while the general case is open and interesting, because if the general answer is yes, then the quantum storage problem fully reduces to a classical secure storage problem.

\item What is the capacity of the wheel graph $C(\mathcal{W}_N)$ in general, particularly $N \geq 6$? What is the capacity of the intersection graph $C(\sqcap_{\Delta, m})$ in general?

\item Extremal rate values for uniform storage: if $C(\mathcal{G})\neq0$, then what is the smallest (across all feasible storage graphs $\mathcal{G}$ with $N$ storage nodes) capacity value $C_u(\mathcal{G})$ as a function of $N$? In other words, what is the worst storage graph in terms of storage capacity for a fixed number of storage nodes $N$? Generalizing Csirmaz's lower bound \cite{Csirmaz_Large} from classical secret sharing to quantum storage, we can find a graph with polynomially small (in $N$) storage capacity, but can the capacity be exponentially small?

\item Additivity of combined storage: given storage graphs $\mathcal{G}_1=( (\alpha_1, \cdots, \alpha_{N_1}), \mathcal{E}_1)$ and $\mathcal{G}_2=( (\beta_1, \cdots, \beta_{N_2}), \mathcal{E}_2)$, for distinct storage systems (the storage nodes are disjoint), the combined storage graph can be defined as $\mathcal{G}_{12}\triangleq ( (\alpha_1, \cdots, \alpha_{N_1}, \beta_1, \cdots, \beta_{N_2}), \mathcal{E}_{12})$ where 
\begin{align}
& \mbox{$\mathcal{G}_{12}$ has $N_1+N_2$ storage nodes,}\\
& \mathcal{E}_{12}=\{e_1\cup e_2 \mid e_1\in \mathcal{E}_1, e_2\in \mathcal{E}_2\}.
\end{align}
Clearly $C(\mathcal{G}_{12})\geq C(\mathcal{G}_1)+C(\mathcal{G}_2)$ since quantum information can be divided into two parts with the parts stored separately in $\mathcal{G}_1,\mathcal{G}_2$. The question is, can the combined capacity ever be strictly larger than the sum of capacities? In other words, is storage capacity always \emph{additive} or can it be \emph{super-additive}? 
A  special case worth considering is the combination of a zero storage capacity system with any other system, i.e., if $C(\mathcal{G}_1)=0$, then is $C(\mathcal{G}_{12})=C(\mathcal{G}_2)$? Or even more specifically, what if $\mathcal{G}_1=( (1,1), \{\{1\},\{2\}\})$?
\end{enumerate}

\bibliography{Thesis}
\end{document}